\documentclass{jfm}

\usepackage[]{amsmath}
\usepackage[dvipsnames]{xcolor}
\usepackage[]{subfigure}
\usepackage{upgreek}
\usepackage{overpic}
\usepackage[]{multirow}
\usepackage[]{url}
\usepackage[]{nicefrac}
\usepackage{bm}
\usepackage{enumerate}
\usepackage[normalem]{ulem} 
\usepackage{algorithm}
\usepackage{algorithmicx}  
\usepackage{algpseudocode}  

\begin{document}

\newtheorem{lemma}{Lemma}
\newtheorem{corollary}{Corollary}

\shorttitle{Cluster-based hierarchical network model} 
\shortauthor{N.~Deng, B.~R.~Noack, M.~Morzy\'nski and L.~R.~Pastur} 

\title{Cluster-based hierarchical network model of the fluidic pinball --- Cartographing transient and post-transient, multi-frequency, multi-attractor behaviour}

\author{Nan Deng\aff{1,2,3}, 
	Bernd R.~Noack\aff{1,4}\corresp{\email{bernd.noack@hit.edu.cn}},
	Marek Morzy\'nski\aff{5}
	\and Luc~R.~Pastur\aff{2}\corresp{\email{luc.pastur@ensta-paris.fr}} 
}
\affiliation
{
\aff{1}
School of Mechanical Engineering and Automation, 
Harbin Institute of Technology, 
Shenzhen 518055, 
People's Republic of China
\aff{2}
Institute of Mechanical Sciences and Industrial Applications, 
ENSTA-Paris, Institut Polytechnique de Paris, 
828 Bd des Mar\'echaux, 
F-91120 Palaiseau, France. 
\aff{3}
Universit\'e Paris-Saclay, CNRS, Laboratoire Interdisciplinaire des Sciences du Num\'erique, 
F-91400 Orsay, France
\aff{4}
Institut f\"ur Str\"omungsmechanik und Technische Akustik (ISTA),
Technische Universit\"at Berlin,
M\"uller-Breslau-Stra{\ss}e 8,
D-10623 Berlin, Germany
\aff{5}
Department of Virtual Engineering, 
Pozna\'n University of Technology,
Jana Pawla II 24, 
PL 60-965 Pozna\'n, Poland
}

\maketitle


\begin{abstract} 
We propose a self-supervised cluster-based hierarchical reduced-order modelling methodology to model and analyse the complex dynamics arising from a sequence of bifurcations for a two-dimensional incompressible flow of the unforced fluidic pinball.
The hierarchy is guided by a triple decomposition separating a slowly varying base flow, dominant shedding and secondary flow structures.
All these flow components are kinematically resolved by a hierarchy of clusters, starting with the base flow in the first layer, resolving the vortex shedding in the second layer and distilling the secondary flow structures in the third layer.
The transition dynamics between these clusters is described by a directed network, called the cluster-based hierarchical network model (HiCNM) in the sequel.
Three consecutive Reynolds number regimes for different dynamics are considered:
(i) periodic shedding at $\Rey=80$,
(ii) quasi-periodic shedding at $\Rey=105$, and
(iii) chaotic shedding at $\Rey=130$,
involving three unstable fixed points, three limit cycles, two quasi-periodic attractors and a chaotic attractor.
The HiCNM enables identifying the transient and post-transient dynamics between multiple invariant sets in a self-supervised manner.
Both the global trends and the local structures during the transition are well resolved by a moderate number of hierarchical clusters.
The proposed reduced-order modelling provides a visual representation of transient and post-transient, multi-frequency, multi-attractor behaviour 
and may automate the identification and analysis of complex dynamics 
with multiple scales and multiple invariant sets.
\end{abstract}

\section{Introduction}
\label{Sec:Introduction}

Fluid flows generally involve complex, high-dimensional and nonlinear dynamics, which makes them hard to understand. 
However, even at high Reynolds numbers, the flow dynamics keeps trace of the instabilities undergone at increasing Reynolds number \citep{huerre1990ARFM}. 
Stationary laminar flows are generally stable with respect to infinitesimal perturbations at sufficiently low Reynolds number. 
This steady state becomes unstable when the Reynolds number increases beyond a critical value $Re_c$, where a bifurcation occurs.
On the way towards a fully turbulent regime, the flow may undergo a succession of bifurcations with increasing Reynolds number. 
\citet{Ruelle1971nature} shows that the flow can reach a chaotic regime after a small number of bifurcations. 
The complex flow dynamics can be seen as the result of the interactions between the fundamental structures of different instabilities \citep{chomaz2005ARFM, bagheri2009JFM}.
A reduced-order model incorporating the underlying mechanisms is always the promising solution for flow analysis \citep{amsallem2008AIAA, legresley2000AIAA} and control \citep{choi2008ARFM, bagheri2009AMR, barbagallo2009JFM}.

Numerous reduced-order models (ROMs) have been developed and applied \citep{taira_AIAA2017}.
The classical method starts with projecting the full system into a low-dimensional subspace, where the high-dimensional dynamics can be approximated with the optimal basis.
This process is so-called Galerkin projection, which leads to a Galerkin system describing the dynamics in reduced-order ordinary differential equations (ODEs).
According to the dimensionality reduction techniques and the model selection strategies, there exist many different projection-based ROMs.
Proper orthogonal decomposition (POD) \citep{berkooz1993arfm, holmes2012turbulence} is the most popular one, which has many empirical variations, for example, balanced POD \citep{rowley2005bpod} with balanced truncation.
The POD-Galerkin method can be optimized and extended with incorporating the pressure term \citep{bergmann2009enablers}, with numerical stabilizationis \citep{iollo2000stability}, with variational multiscale method \citep{iliescu2014variational} and with closure modelling strategies \citep{wang2012proper}.
Based on first principles, the mean-field theory of \citet{Landau1944} and \citet{Stuart1958jfm} is the lowest dimensional mean-field model to account for a supercritical Hopf bifurcation.
Weakly nonlinear mean-field analysis has also been applied to more complex situations in which the flow has undergone two successive bifurcations, such as in the wake of axisymmetric bodies \citep{Fabre2008pof}, the wake of a disk \citep{Meliga2009jfm} or the wake of the fluidic pinball \citep{deng2020jfm}. 
\citet{Gomez2016jfm, Rigas2017aiaa} included mean-field considerations in their resolvent analysis, decomposing the flow in time-resolved linear dynamics and a feedback term with the quadratic nonlinearity.

Alternatively, data-driven strategies show their advantage in pattern and system recognition without prior knowledge about flow dynamics \citep{brunton2020ARFM}, like 
Koopman analysis \citep{Schmid2010jfm, mezic2013ARFM} using dynamic mode decomposition (DMD) \citep{tu2013DMD, kutz2016DMD}, 
data-driven Galerkin modelling \citep{Noack2016jfm} using recursive DMD, and 
multiscale Proper Orthogonal Decomposition (mPOD) \citep{mendez2019jfm} using a matrix factorization framework to enhance feature detection capabilities.
Above mentioned methods still start with a modal decomposition of the original flow fields.
The advances in machine-learning algorithms provide huge potential for data-driven ROMs, for example,
using artificial neural network (ANN) to stabilize projection-based ROMs \citep{san2018extreme} or to build the ANN ROMs \citep{san2019artificial}, turbulence modelling with deep neural networks \citep{kutz2017deep}, feature-based manifold modelling \citep{loiseau2018sparse} with sparse identification \citep{brunton2016discovering}.

Inspired with centroidal Voronoi tessellation ROMs in \citet{burkardt2006pod},
\citet{kaiser2014jfm} proposed the cluster-based reduced-order modelling (CROM) method to partition the flow data into clusters and analyze the flow dynamics with a cluster-based Markov model (CMM).
CROM provides us with a novel modelling strategy, liberating us from the issue of choosing a low-dimensional space of the traditional projection method.
\citet{nair2019JFM} applied CROM to the nonlinear feedback flow control and introduced the directed network \citep{newman2018book} for the dynamical modelling.
With the clusters being the nodes and the transitions between clusters being the edges, an extended Markov model with a directed network was built, emphasizing the non-trivial transitions between clusters.
\citet{fernex2020cnm} and \citet{li2020CNM} further proposed the cluster-based network model (CNM) for time-resolved data by introducing local interpolations between clusters with the pre-specified transition times.
The CNM can be seen as an extension of the traditional CMM, using the network model instead of the standard Markov model to describe the transient dynamics.
Networks of complex dynamical systems have attracted a great deal of interest, forming an increasingly important interdisciplinary field known as network science \citep{watts1998collective, albert2002statistical, barabasi2013network}.
The network-based approaches have been used in fluid mechanics to describe the interactions among vortical elements \citep{nair2015jfm}, detect the Lagrangian vortex motion \citep{hadjighasem2016pre}, and model and analyze turbulent flows \citep{taira2016jfm, yeh2021jfm}. 
Together with the clustering approaches, networks have been also used to extract key features of complex flows \citep{bollt2001ijbc, schlueter2017jfm, murayama2018pre, krueger2019prsa}. 
The critical structures modifying the flow can be identified by the intra- and inter-cluster interactions using community detection \citep{meena2018pre, meena2021jfm}.
Theories and techniques in the field of network science may play a crucial role in the modelling, analysis and control of fluid systems.

The accuracy of the cluster-based model depends on the number of clusters.
However, too many clusters will increase the complexity of the Markov/network model. 
A high level of human experience is required to achieve a good compromise between resolution and a simple model. 
The focus of this paper is to optimise the data-driven cluster analysis by introducing a hierarchical structure and a systematic self-supervised way to model the transient and post-transient flows in the case of multiple unstable solutions and multiple attractors. 
The hierarchical modelling strategy shows good consistency with the Reynolds decomposition from the mathematical foundation. The systematic data treatment process shows its great potential for multiscale and multi-frequency modelling.

Inspired by the hierarchical Markov model \citep{fine1998ml}, we apply a scale-dependent hierarchical clustering to the classic network modelling under the mean-field consideration.
The time-scales of the different flow components provide a good indicator for figuring out the typical structures in multiscale flows, and enable the hierarchical model to address the complex dynamics of multiscale problems. 
The resulting cluster-based hierarchical network model (HiCNM) can systematically identify complex dynamics involved in the case of multiple attractors. 
Both the global trends and the local structure during the transition can be well preserved by a fewer number of clusters in the hierarchical structure, which leads to a better understanding of the physical mechanisms involved in the flow dynamics.

\begin{figure}
 \centerline{
 \includegraphics[width = .95\linewidth]{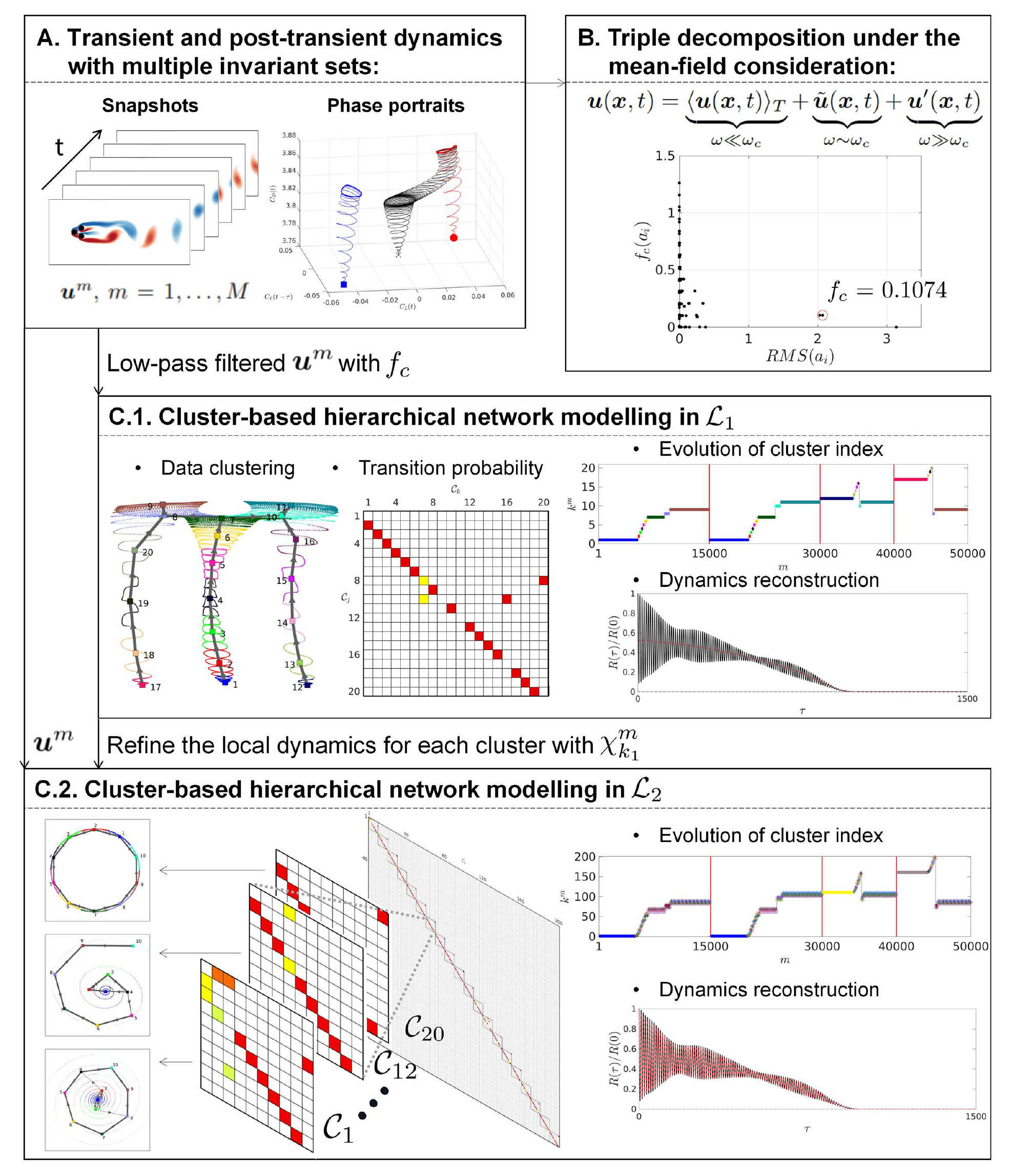}
 }
\caption{Overview of the cluster-based hierarchical network modelling framework exemplified at $\Rey=80$.
(A) The flow dynamics involves six invariant sets associated with three unstable fixed points, three limit cycles, as shown in the 3D phase portrait of the drag and lift forces.
(B) Under the mean-field consideration, the flow can be decomposed into a slowly-varying mean flow, the coherent and incoherent components, separated by the dominant frequency of the coherent part.
The non-coherent fluctuating component is weak in this case, and the third term of the triple decomposition can be ignored.
(C) Therefore, a HiCNM with two layers is enough to extract the global trend and the local dynamics of the varying mean-flow field.
The transient and post-transient dynamics, characterized by multiple frequencies and multiple invariant sets, are introduced in \S~\ref{Sec:FlowFeatures}.
The hierarchical network modelling strategy is discussed in \ref{SubSubSec:HiModelling} under the mean-filed consideration in \S~\ref{SubSubSec:FlowDecomposition}.
The dynamics reconstruction of the resulting hierarchical network model is given in \S~\ref{SubSubSec:Validation}.
}
\label{Fig:Sketch}
\end{figure}

We consider the two-dimensional incompressible flow configuration of \citet{bansal2017experimental}, defined as the unforced ``fluidic pinball'' in \citet{deng2020jfm}. 
With increasing Reynolds number, the wake undergoes a first instability leading to a periodic vortex shedding, then a static symmetry breaking, and finally a transition to a quasi-periodic regime before transiting to a chaotic regime. 
HiCNMs are built for these flow regimes, which have multiple invariant sets and exhibit different transient dynamics. 
We provide a principle sketch of our HiCNM framework in figure~\ref{Fig:Sketch}.

The manuscript is organised as follows:
\S~\ref{Sec:configuration} describes the numerical plant of the fluidic pinball and the flow features at different Reynolds number.
\S~\ref{Sec:HiCROM} discusses the different perspectives on the cluster-based hierarchical network modelling strategy.
In \S~\ref{Sec:HiCROM_PINBALL}, we discuss the HiCNMs applied to the transient and post-transient dynamics of a flow configuration involving six invariant sets, for three different Reynolds numbers, respectively associated with a periodic, a quasi-periodic and a chaotic dynamics.
\S~\ref{Sec:Conclusions} summarises the main findings and gives some suggestions for improvement and future directions.

\section{Flow configuration and flow features}
\label{Sec:configuration}

We consider two-dimensional incompressible flows in the fluidic pinball \citep{Noack2017put} as the benchmark configuration for our hierarchical modelling strategy.
The flow configuration and the direct Navier-Stokes solver are described in \S~\ref{Sec:Pinball}. 
The transient and post-transient dynamics at different Reynolds numbers are illustrated in \S~\ref{Sec:FlowFeatures}.

\subsection{Flow configuration and direct Navier-Stokes solver}
\label{Sec:Pinball}
\begin{figure}
 \centerline{
 \includegraphics[width = .6\linewidth]{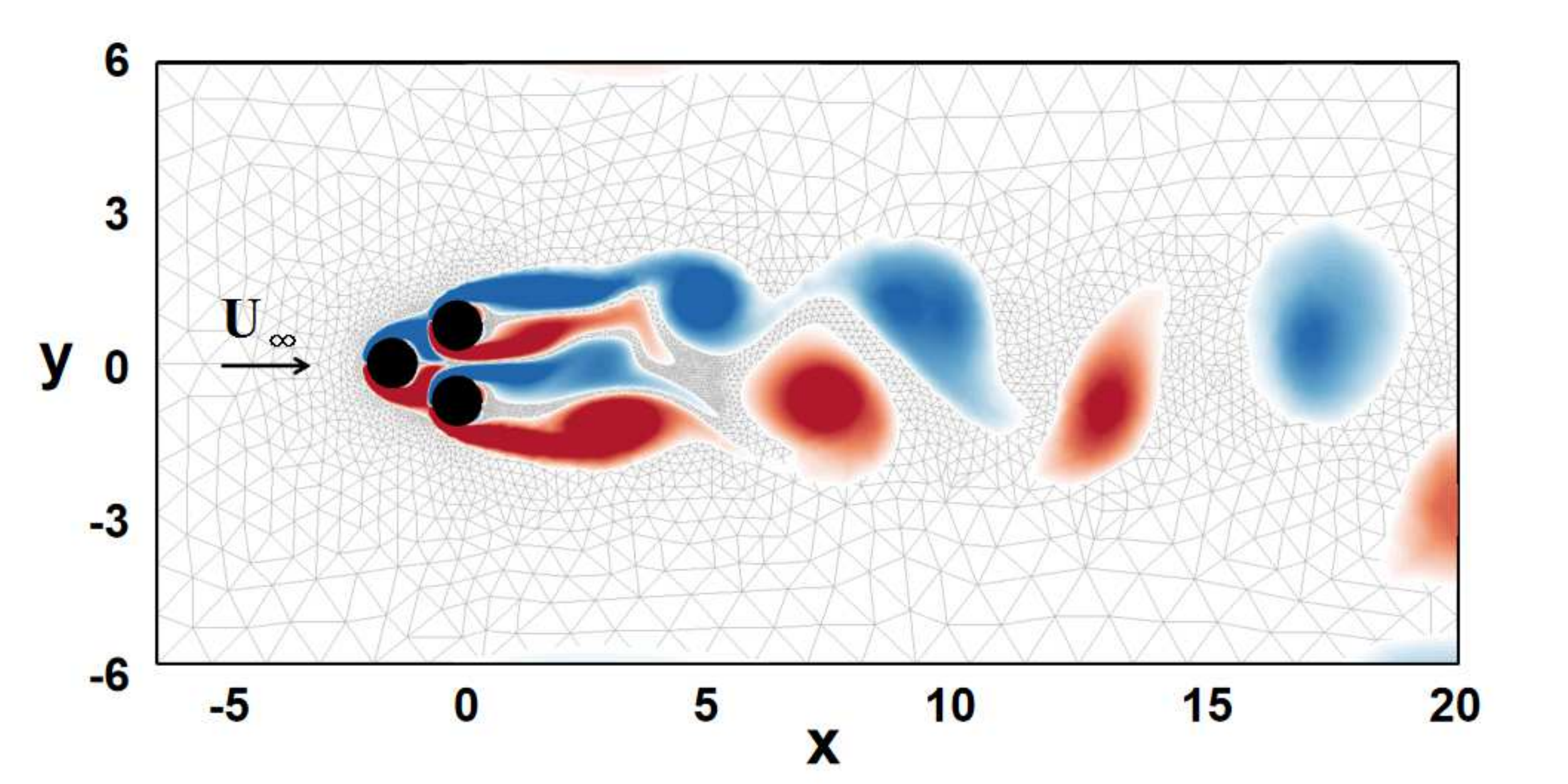}
 }
\caption{Configuration of the fluidic pinball and computational grid for the simulated domain.
The upstream velocity is denoted $ U_\infty $. An example vorticity field at $\Rey=130$ is colour-coded in the range $[-1.5, 1.5]$ from blue to red.}
\label{Fig:pinball}
\end{figure}
Figure~\ref{Fig:pinball} shows the geometric configuration of the fluidic pinball, consisting of three fixed cylinders of unit diameter $ D $.
Their axes are placed on the vertices of an equilateral triangle of side $ 3D/2 $ in the $ (x, y) $ plane.
The upstream flow is in the $x$-axis direction with a uniform velocity $ U_\infty $ at the inlet of the domain.
The computational domain $\Omega$ is bounded by a rectangular box of size $ [-6D,+20D] \times [-6D,+6D] $.
A Cartesian coordinate system is used for description, and its origin is placed in the middle of the back two cylinders considering the symmetry of this configuration.
Since no external force is applied to these three cylinders, a no-slip condition is applied on the cylinders, and the velocity in the far wake is assumed to be $ U_\infty $.
The Reynolds number is defined as $ Re = U_\infty D/\nu $, where $ \nu $ is the kinematic viscosity of the fluid.
A no-stress condition is applied at the outlet of the domain. 

The fluid flow is governed by the non-dimensionalized incompressible Navier-Stokes equations in scales with the cylinder diameter $D$ and the velocity $ U_\infty $, which read
\begin{equation}
\label{Eqn:NSE}
\partial_t \bm{u} + \nabla \cdot \bm{u} \otimes \bm{u} = \nu \triangle \bm{u} - \nabla p, \quad \nabla \cdot \bm{u}=0,
\end{equation}
where $p$ and $\bm{u}$ are respectively the pressure and velocity flow fields and $\nu = 1/Re$.
The advection time scale is $ D/U_\infty$ and the pressure scale is $\rho U^2_\infty $, where $\rho$ is the unit fluid density for the incompressible flow.
It is assumed that there exists a solution $(\bm{u}_s,p_s)$ satisfying the steady Navier-Stokes equations
\begin{equation}
\label{Eqn:NSE:Steady}
 \nabla \cdot \bm{u}_s \otimes \bm{u}_s = \nu \triangle \bm{u}_s - \nabla p_s, \quad \nabla \cdot \bm{u}_s=0.
\end{equation}
The inner product of two square-integrable velocity fields $\bm{u} (\bm{x})$ and $\bm{v} (\bm{x})$  in the computational domain $\Omega$ reads
\begin{equation}
\label{Eqn:InnerProduct}
\left( \bm{u} , \bm{v} \right)_{\Omega} := \int\limits_{\Omega} \!\! d\bm{x} \> \bm{u}(\bm{x}) \cdot \bm{v}(\bm{x}).
\end{equation}
The associated norm of the velocity field $\bm{u}(\bm{x})$ is defined as
\begin{equation}
\label{Eqn:Norm}
||\bm{u}||_{\Omega}:= \sqrt{\left( \bm{u} , \bm{u} \right)_{\Omega}}.
\end{equation}
The direct numerical simulation (DNS) of the Navier-Stokes equations \eqref{Eqn:NSE} is based on a second-order finite-element discretization method of the Taylor-Hood type \citep{taylor1973ef}, on an unstructured grid of 4\,225 triangles and 8\,633 vertices, and an implicit integration of the third-order in time. 
The unsteady flow field is calculated by an unsteady solver with Newton-Raphson iteration until the residual is less than a prescribed tolerance. 
This approach is also employed to calculate the steady solution by a steady solver for the steady Navier-Stokes equations \eqref{Eqn:NSE:Steady}.
The direct Navier-Stokes solver used herein has been validated in \citet{noack2003jfm, deng2020jfm}, and the grid used for the simulations provides a consistent flow dynamics comparing to a refined grid.
A relevant numerical investigation for this kind of equilateral-triangle configuration can also be found in \citet{chen2020jfm}.
The data-driven HiCNM method is exemplified on this benchmark configuration with a blockage ratio $B=0.21$, defined as the ratio of the cross-section length of the cluster of three cylinders $ 5D/2 $ to the width of the computational domain $12D$, which mimics the experimental setups in \citet{raibaudo2020pof}.
The numerical results with the current computational domain remain similar compared with a larger domain with $B=0.025$, as detailed in appendix~\ref{Sec:BlockageRatio}.

\subsection{Flow features}
\label{Sec:FlowFeatures}

 \begin{figure}
\centering
 \raisebox{-0.5\height}{\includegraphics[width=.9\linewidth]{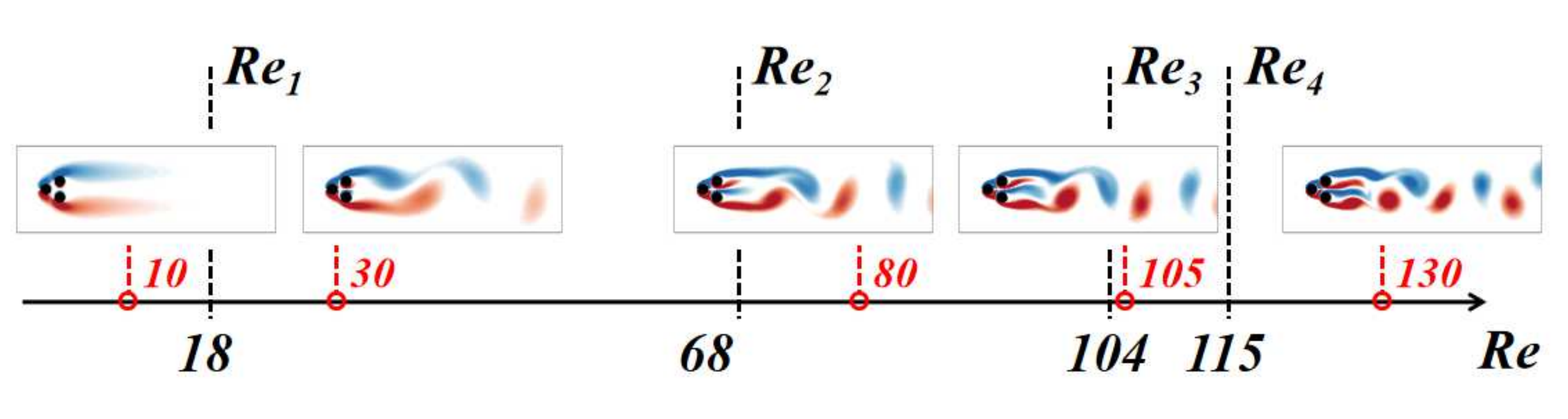}}
\caption{Post-transient flow state for different flow regimes at the Reynolds numbers marked in red.
The critical values of the supercritical Hopf bifurcation $Re_1$, the supercritical pitchfork bifurcation $Re_2$, and the Neimark-S\"acker bifurcation $Re_3$ before the system entering into chaos at $Re_4$ are marked in black on the $Re$-axis.}
\label{Fig:FlowRe}
\end{figure}

As shown in figure~\ref{Fig:FlowRe},
the flow undergoes a supercritical Hopf bifurcation at $Re_1\approx 18$, a supercritical pitchfork bifurcation at $Re_2\approx 68$ and a Neimark-S\"acker bifurcation at $Re_3\approx 105$, before entering the chaotic regime beyond $Re_4\approx 115$ with increasing the Reynolds number \citep{deng2020jfm}. 
Depending on the Reynolds number, the wake flow may present rich transient dynamics due to multiple exact solutions of the Navier-Stokes equations, assciated with the co-existing invariant sets in the state space. 
For instance, at $Re_1<Re<Re_2$, the symmetric steady solution $\bm{u}_s$ is the only fixed point of the system. This exact solution of the Navier-Stokes equations is unstable. 
The only attractor in the state space is a symmetric limit cycle, associated with the cyclic release of vortices in the wake of the cylinders, forming a von K\'arm\'an street of regular vortices.
At $Re>Re_2$, three fixed points are solutions of the steady Navier-Stokes equations, one symmetric $\bm{u}_s$ and two asymmetric steady solutions $\bm{u}_s^{\pm}$, and all three points are unstable. 
Meanwhile, the unsteady Navier-Stokes equations have three periodic solutions.
The symmetric limit cycle, associated with symmetric vortex shedding, is unstable.
The two mirror-conjugated asymmetric limit cycles, associated with asymmetric vortex sheddings, co-exist as attractors of the flow dynamics in the state space. 
For $Re_3<Re<Re_4$, the two attracting asymmetric limit cycles thicken into torii by introducing an additional low frequency, which modulates the vortex shedding quasi-periodically. 
Beyond $Re_4$, the vortex shedding dynamics is chaotic. The interested reader can find more details on the route to chaos in the fluidic pinball in \citet{deng2020jfm}.

The flow features can be illustrated by the forces exerted on the body. 
The drag $F_D$ and lift $F_L$ forces are the projection on $\bm{e}_x$ and $\bm{e}_y$ of the resultant force $\bm{F} =F_D\,\bm{e}_x + F_L\,\bm{e}_y$, obtained by integrating the viscous and pressure forces over the cylinder surfaces.
The flow dynamics is analyzed with the lift coefficient $C_L$,
\begin{equation}
\label{Eqn:LiftCoef}
C_L (t)= \frac{2F_L (t)}{\rho\, U_\infty^2}.
\end{equation}

\begin{figure}
\centering
\begin{tabular}{cc}
 (a) & \raisebox{-0.5\height}{\includegraphics[width=.65\linewidth]{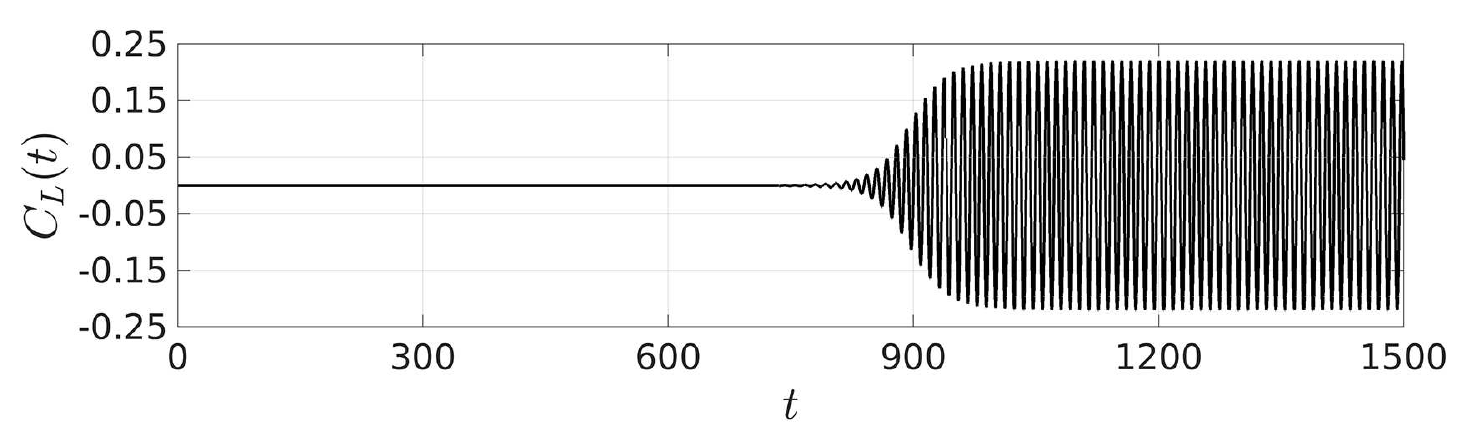} }\\
 (b) & \raisebox{-0.5\height}{\includegraphics[width=.65\linewidth]{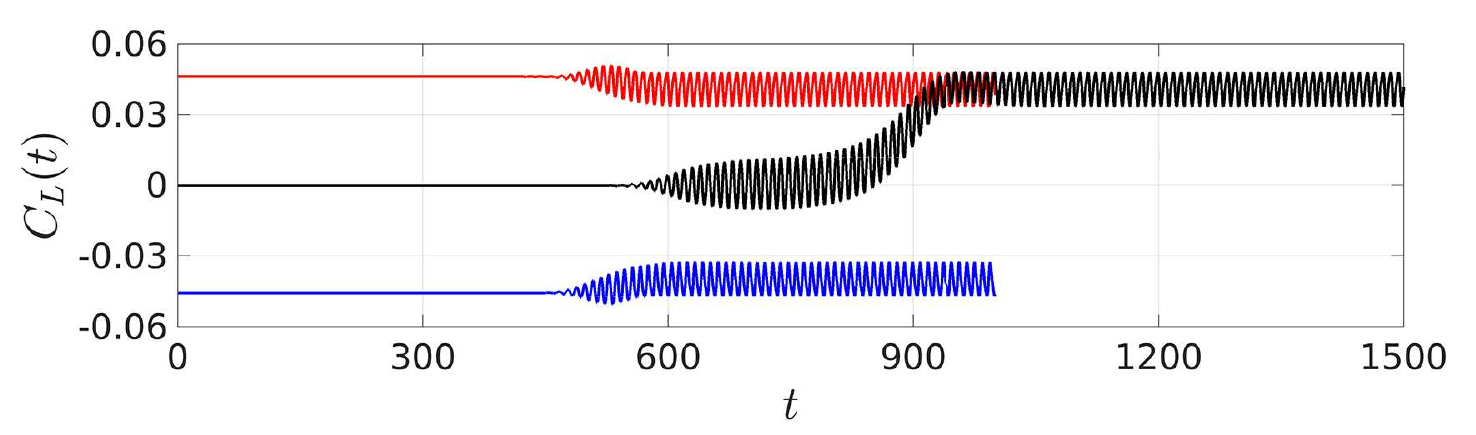} }\\
 (c) & \raisebox{-0.5\height}{\includegraphics[width=.65\linewidth]{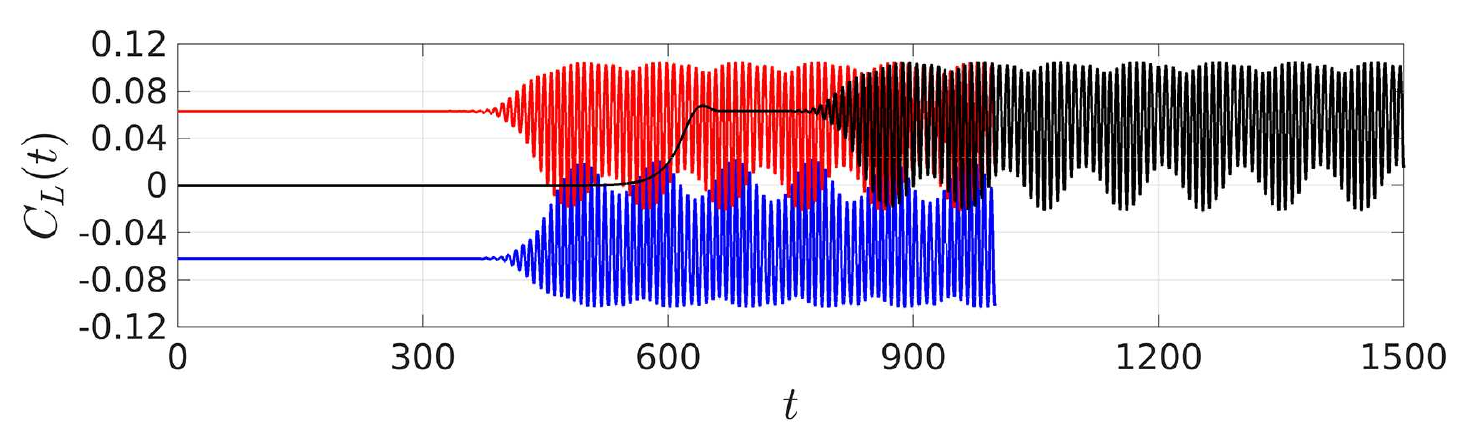} }\\
 (d) & \raisebox{-0.5\height}{\includegraphics[width=.65\linewidth]{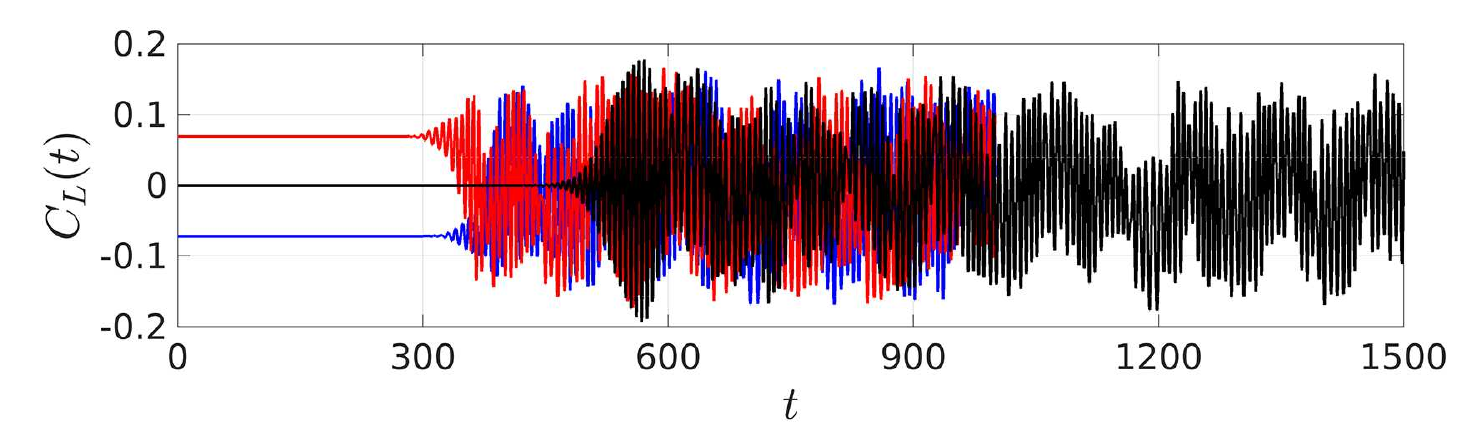} }\\
\end{tabular}
\caption{Transient and post-transient dynamics starting with different steady solutions, illustrated with the time evolution of the lift coefficient $C_L$ at $Re = 30$ (a), $80$ (b), $105$ (c), $130$ (d).}
\label{Fig:CL}
\end{figure}

We apply the DNSs at $Re = 30$, $80$, $105$ and $130$ respectively, starting close to the symmetric steady solution (for $Re>Re_1$) until $t=1\,500$ and the asymmetric steady solutions (for $Re>Re_2$) until $t=1\,000$.
The time evolutions of the lift coefficient $ C_L $ are shown in figure~\ref{Fig:CL}, where different transient dynamics are observed, from the steady solutions to the asymptotic regimes.

At $Re = 30$, as shown in figure~\ref{Fig:CL}(a), the lift coefficient $ C_L $ starts to oscillate visibly at $t \approx 800$, indicating that the flow leaves the neighborhood of the symmetric steady solution. 
Then, $ C_L $ oscillates around a vanishing value with increasing amplitude until converging to a fixed amplitude.
This state refers to a symmetric vortex shedding, as the instantaneous flow is oscillating around a geometrical symmetric mean-flow field.

At $Re = 80$, as shown in figure~\ref{Fig:CL}(b), the primary transition is the same as at $\Rey=30$.
Next, the slowly-varying mean lift coefficient $ \langle C_L\rangle _T $, averaged over the oscillation period $T$, leaves from $0$ to $0.04$.
This indicates that the oscillatory dynamics in the permanent regime has lost the statistical symmetry, and the flow state refers to an asymmetric vortex shedding.
Starting nearby either one of the two asymmetric steady solutions, $ C_L $ directly evolves to the asymmetric vortex shedding regime thats shares the same asymmetry.

At $Re = 105$, $C_L$ visibly increases at $t \approx 580$ starting with the vanishing value of the symmetric steady solution, as illustrated with the black curve in figure~\ref{Fig:CL}(c).
However, the initial transition reaches a non-oscillating value equal to the initial value of the red curve, which refers to one of the two asymmetric steady solutions. 
It eventually enters a quasi-periodic state, the vortex shedding oscillations being modulated at a low frequency.
Starting from the other two asymmetric steady solutions will directly evolve into the permanent quasi-periodic state with the same asymmetry.

At $Re = 130$, the initial transition of the black curve in figure~\ref{Fig:CL}(d) is similar to the initial transition at $Re = 80$, but the dynamics enters a chaotic regime shortly after the symmetric vortex shedding has started. Simulations converge to the same chaotic attracting set, starting with all the three different steady solutions.

\section{Cluster-based hierarchical reduced-order modelling}
\label{Sec:HiCROM}

In this section, the general approach of the cluster-based hierarchical reduced-order modelling is described and discussed. 
In \S~\ref{Sec:Background}, we present the relevant background on the flow decomposition and the standard cluster-based reduced-order model (CROM).
The cluster-based reduced-order modelling with hierarchical structure is described in \S~\ref{Sec:HiModelling}, as well as the relevant analysis of the cluster-based hierarchical network model (HiCNM).

\subsection{Background}
\label{Sec:Background}
The standard CROM is obtained in two steps:
the snapshots are first clustered into coarse-grained representative states before building either a Markov or a network model for the analysis of the dynamics. 
Clustering all transients and post-transients at once can suffer from the inability to accurately capture the dynamics at different scales.
Under the mean-field consideration, we introduce a hierarchical structure of clusters for the flow dynamics of different time scales, resulting in a cluster-based hierarchical reduced-order model (HiCROM).

\subsubsection{Flow decomposition with mean-field consideration}
\label{SubSubSec:FlowDecomposition}

The starting point of the HiCROM is the triple decomposition of the flow field similar to \citet{Reynolds1972jfm}
\begin{equation}
\label{Eqn:ReynoldsDecomposition}
\bm{u} (\bm{x}, t)
= \underbrace{ \langle \bm{u} (\bm{x},t)\rangle _T }_{\omega \ll \omega_c}
+  \underbrace{ \tilde{\bm{u}} ( \bm{x}, t) }_{\omega \sim \omega_c}
+  \underbrace{ \bm{u}^{\prime} ( \bm{x}, t) }_{\omega \gg \omega_c},
\end{equation}
where the dominant angular frequency $\omega_c$ is defined as the dominant peak in the Fourier spectrum of the velocity field. 
Here, the velocity field is decomposed into a slowly-varying mean-flow field $\langle \bm{u}\rangle _T$, a coherent component on time-scales of order $2\pi /\omega_c$, involving coherent structures $\tilde{\bm{u}}$, and the remaining non-coherent small scale fluctuations $\bm{u}^{\prime}$. 
This kind of decomposition can also be found in the low-order Galerkin models of \cite{Tadmor2011ptrsa} and the weakly nonlinear modelling of \cite{rigas2017jfm}. 

The slowly-varying mean-flow field $\langle \bm{u}\rangle _T$ can be defined as the average of the velocity field $\bm{u}$ over one local period $T\approx 2\pi /\omega_c$ of the coherent structures, \begin{equation}
\label{Eqn:PhaseAverage}
 \langle \bm{u} ( \bm{x}, t ) \rangle _T  
:= \frac{1}{T} \int\limits_{t-T/2}^{t+T/2} \!\!\! d\tau  \> \bm{u} ( \bm{x}, \tau ),
\end{equation}
which eliminates both the coherent contribution from $\tilde{\bm{u}}$ and the non-coherent contribution from $\bm{u}^{\prime}$. Unlike the mean-flow field defined by the post-transient limit,
\begin{equation}
\label{Eqn:TimeAverage}
\bm{\bar{u}} (\bm{x})
= \lim_{T\rightarrow \infty }\frac{1}{T}\int\limits_{0}^{T} \bm{u} (\bm{x}, \tau) d\tau,
\end{equation}
the finite-time averaged-flow field considered in this study owns a slowly varying dynamics. From the mean-field theory of \cite{Stuart1958jfm}, the slowly-varying mean-flow field evolves out of the steady solution under the action of the Reynolds stress associated with the most unstable eigenmode(s). The mean-flow field deformation $\bm{u}_{\Delta}$ is used to describe the difference between the slowly-varying mean-flow field and the invariant steady solution $\bm{u}_s (\bm{x})$, which reads
\begin{equation}
\label{Eqn:MFAnsatz}
 \langle \bm{u} (\bm{x},t)\rangle _T =  \bm{u}_s (\bm{x})  + \bm{u}_{\Delta} (\bm{x},t).
\end{equation}

\subsubsection{Clustering algorithm}
\label{SubSubSec:kmeans}
We consider the state vectors, for instance, the velocity fields $\bm{u} (\bm{x},t)$ in the computational domain $\Omega$, which is sampled at times $t^m = m\Delta t$ with a time step $\Delta t$, where the superscript $m=1,\ldots,M$ is the snapshot index.
The clustering process aims at partitioning the $M$ time-discrete states (snapshots) $\bm{u}^m = \bm{u} (\bm{x},t^m) $ into $K$ clusters $\mathcal{C}_k$, $k = 1,\,\ldots,\,K$.
Snapshots of a given cluster share similar attributes featured by its cluster centroid $\bm{c}_k$. 
The distance between the snapshot $\bm{u}^m$ and the centroid $\bm{c}_k$ is defined as 
\begin{equation}
\label{Eqn:Dmn_velocity}
D^{m}_{k}:= ||\bm{u}^m - \bm{c}_k||_{\Omega}.
\end{equation}
Each snapshot is partitioned to the cluster of the closest centroid by $\underset{k}{ \hbox{argmin}} \> D^{m}_{k}$, and the characteristic function is defined as
\begin{equation}
 \chi_{k}^{m} := \begin{cases}
                1, &  \mbox{if }\bm{u}^m\in \mathcal{C}_k,\\
                0, &  \mbox{otherwise}.
               \end{cases}
 \label{Eqn:CharacteristicFunction}
\end{equation}
A cluster index $k^m$, $m=1,\ldots,M$, indicates the cluster assignment of the corresponding snapshot with $\bm{u}^m \in \mathcal{C}_k$, and records the visited clusters consecutively.
The number of snapshots $n_k$ in cluster $k$ is given by
\begin{equation}
 n_k = \sum\limits_{m=1}^M\, \chi_{k}^{m}.
 \label{Eqn:Countingnumber}
\end{equation}
The cluster centroids $\bm{c}_k$ are defined as the average of the snapshots belonging to the cluster $\mathcal{C}_k$: 
\begin{equation}
\label{Eqn:DefCentroids}
\bm{c}_k
= \frac{1}{n_k} \sum\limits_{m=1}^M  \> \chi_{k}^{m} \bm{u}^m.
\end{equation}
The performance of clustering is judged by the within-cluster variances: 
\begin{equation}
\label{Eqn:TotalClusterVariance}
 J \left(\bm{c}_1,\ldots,\bm{c}_K \right)
= \sum\limits_{k=1}^K \,\sum\limits_{m=1}^M \, 
\chi_{k}^{m}\,\left\Vert\bm{u}^m - 
 \bm{c}_k \right\Vert_{\Omega}^2.
\end{equation}
The clustering algorithm minimizes $J$ and determines the optimal centroid positions,
\begin{equation}
\label{Eqn:OptimalTotalClusterVariance}
\bm{c}_1^{\rm{opt}}, \ldots,\bm{c}_K^{\rm{opt}} = 
\underset{\bm{c}_1,\ldots,\bm{c}_K}{ \hbox{argmin}} \> J \left(\bm{c}_1,\ldots,\bm{c}_K \right),
\end{equation}
by iteratively updating the characteristic function and the centroid positions.

To solve the optimization problem \eqref{Eqn:OptimalTotalClusterVariance}, we use the $k$-means$++$ algorithm \citep{arthur2006k}. 
Comparing to the traditional $k$-means algorithm, the $k$-means$++$ algorithm selects the initial centroids as far away as possible to avoid any bias from the initial conditions. 
The remaining steps of the two algorithms are the same. 
At each iteration, the snapshots are divided into clusters of the nearest newly determined centroids. The optimal centroids are obtained by iterating until either convergence or when the maximum number of iterations is reached. 

\subsubsection{Cluster-based network model}
\label{SubSubSec:CNM}

Based on the clustering result, \citet{kaiser2014jfm} derived a cluster-based Markov model (CMM), which provides a probabilistic representation of the system using a Markov process, with the assumption that the fluid system is memoryless.
\citet{nair2019JFM} removed the transitions residing in the same cluster and emphasized the non-trivial transitions between two different clusters.
In these two works, the transitions are only characterised by probabilities. 
The cluster-based network model (CNM) proposed in \citet{fernex2020cnm} and \citet{li2020CNM}, inherited the idea of focusing on the non-trivial transitions, and further introduced time-scale characteristics by recording the transition times.
We here briefly review some concepts of the CNM, as they will be used in our benchmark of HiCNM.

The $M$ consecutive snapshots define $M-1$ transitions, containing trivial transitions staying in the same cluster and non-trivial transitions between two different clusters.
The number of transitions from $\mathcal{C}_j$ to $\mathcal{C}_i$ reads
\begin{equation}
n_{ij} = \sum\limits_{m=1}^{M-1}\, \chi_{j}^{m} \chi_{i}^{m+1}.
\end{equation}
Considering the non-trivial transitions, ${n_{j}}$ is the total number of departing snapshots from $\mathcal{C}_j$, with $ n_j = \sum\limits_{i=1}^K\, (1-\delta_{ij}) n_{ij}$.
The direct transition probability $P_{ij}$ reads
\begin{equation}
 P_{ij}= \frac{(1-\delta_{ij}) n_{ij}}{n_{j}},   \>   \> i, j = 1, \cdots, K
\end{equation}
where the non-migrating transition ${n_{jj}}$ is eliminated. All the non-trivial transitions are identified with the direct transition matrix $\mathsfbi{P}$.

The residence time matrix $\mathsfbi{T}$ relies on the time information of the snapshots.
After clustering, each snapshot $\bm{u}^m= \bm{u}(t^m) $ is associated with the closest centroid with a cluster index $k^m$.
Assuming that $N$ $(N < M)$ non-trivial transitions occur along the trajectory, 
the moments of transition $t_n$, $n=1,\ldots,N$ --- including the initial time $t_0=t^1$ --- are defined as the time entering into a new cluster, 
\begin{equation}
t_n = t^m \> \> \mbox{if }\bm{u}^{m-1}\in \mathcal{C}_k \And \bm{u}^{m} \notin \mathcal{C}_k,
\end{equation}
with ascending order $t_0 < t_1 < \ldots < t_N$. The cluster index $k^m$ remains unchanged in a time range $[t_n, t_{n+1} )$. The sequence of visited clusters over time can be simplified with the first entering snapshot for each non-trivial transition with $k^m$ taken from  the  moments of transition $t_n$.
\begin{figure}
 \centerline{
 \includegraphics[width = .5\linewidth]{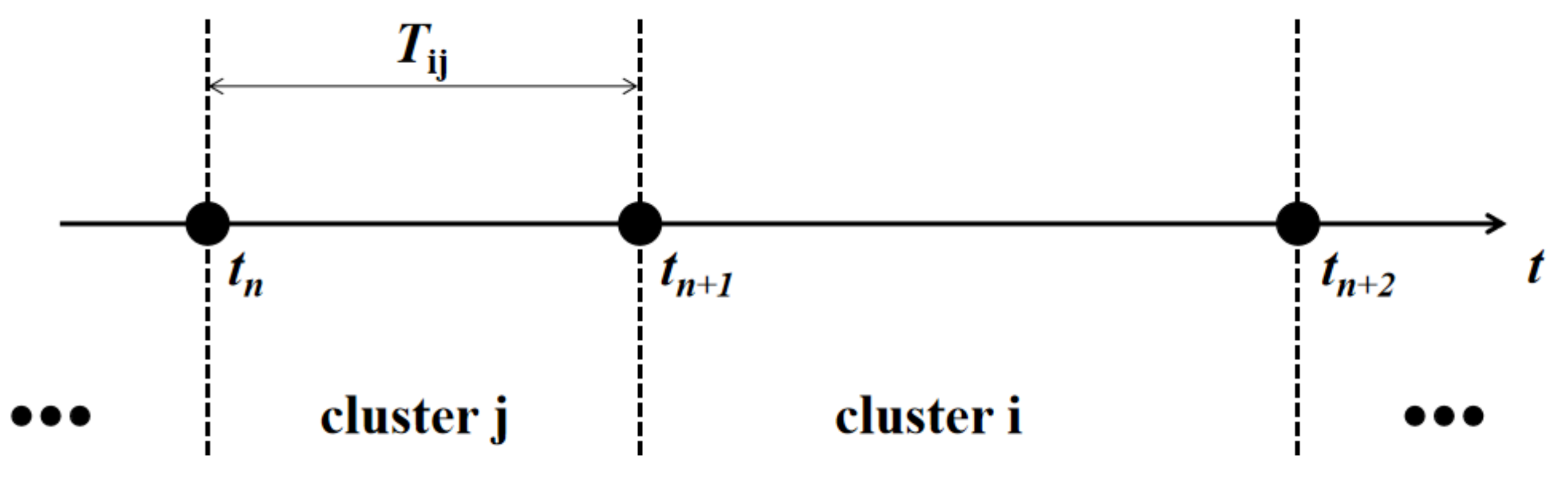}
 }
\caption{An illustration of the residence time in the cluster-based network model. $\bullet$ remarks the entering time into new clusters.}
\label{Fig:CNM}
\end{figure}

For a simple transition from $\mathcal{C}_j$ to $\mathcal{C}_i$ as illustrated in  figure \ref{Fig:CNM}, where the trajectory first enters in $\mathcal{C}_j$ at time $ t_{n}$, and leaves $\mathcal{C}_j$ for $\mathcal{C}_i$ at time $ t_{n+1}$, the residence time in $\mathcal{C}_j$ is defined as 
\begin{equation}
T_{ij}:= t_{n+1} - t_{n}.
\end{equation}
In the case of multiple trajectories of transition from $\mathcal{C}_j$ to $\mathcal{C}_i$, the residence time will be averaged according to the number of trajectories. 

At this point, the time-resolved snapshots $\bm{u}^m$ can be represented by cluster centroids $\bm{c}_{k^m}$ with the time evolution of the cluster index $k^m$.
The transient dynamics is described by both the direct transition matrix of non-migrating transitions $\mathsfbi{P}$ and the residence time matrix $\mathsfbi{T}$. 

\subsection{Hierarchical modelling with mean-field consideration}
\label{Sec:HiModelling}

As introduced in figure~\ref{Fig:Sketch}, the framework of cluster-based hierarchical network modelling contains the following three steps.
The hierarchical clustering with mean-field consideration is introduced in \S~\ref{SubSubSec:HiClustering}.
Based on the identified clusters, the network modelling with hierarchical structure is derived in \S~\ref{SubSubSec:HiModelling} for the mean-field model of Eq.~\eqref{Eqn:ReynoldsDecomposition} under a small number of general assumptions.
\S~\ref{SubSubSec:Validation} introduces the autocorrelation function and its root mean square error of the rebuilt flow for the validation of the HiCNM. 
An introductory example is introduced in Appendix~\ref{SubSubSec:example} to clarify the primary form of the HiCNM.

\subsubsection{Hierarchical clustering inspired by the triple decomposition}
\label{SubSubSec:HiClustering}

To better understand the global and local properties of the data, we present a novel hierarchical clustering algorithm inspired by the triple decomposition introduced in \S~\ref{SubSubSec:FlowDecomposition}.
The principle of hierarchical clustering is to divide the snapshots into layers of clusters. 
Snapshots belonging to clusters of the parent layer are further partitioned into clusters of the child layer. 
Hierarchical clustering algorithms are generally divided into two categories:
\begin{enumerate}[(a) ]
\item The agglomerating (``bottom-up'') hierarchical clustering begins with the smallest clusters at the bottom, each snapshot being an elementary cluster. The two closest clusters are merged to generate a new cluster according to certain criteria, introducing an additional layer from the bottom. This merging is repeated until all snapshots belong to one cluster at the top of the hierarchy.
\item The divisive (``top-down'') hierarchical clustering starts with only one cluster, which owns all the snapshots. In our case, the centroid of the top cluster would be the mean-flow field from by ensemble averaging. From the top to the bottom, the snapshots in each cluster of the parent layer are divided into multiple clusters in the child layer, according to certain criteria. The bottom layers will be associated with the small scale fluctuations of the flow field. The division can be continued until each snapshot is a cluster. 
\end{enumerate}
We employ a divisive hierarchical clustering to distil the different features in a hierarchy, which is consistent with the triple decomposition of Eq.~\eqref{Eqn:ReynoldsDecomposition}.
For instance, fluid flows characterised by multiple frequencies require only a finite number of layers to describe the different components bounded by frequency.

\begin{figure}
\centerline{
\includegraphics[width = .9\linewidth]{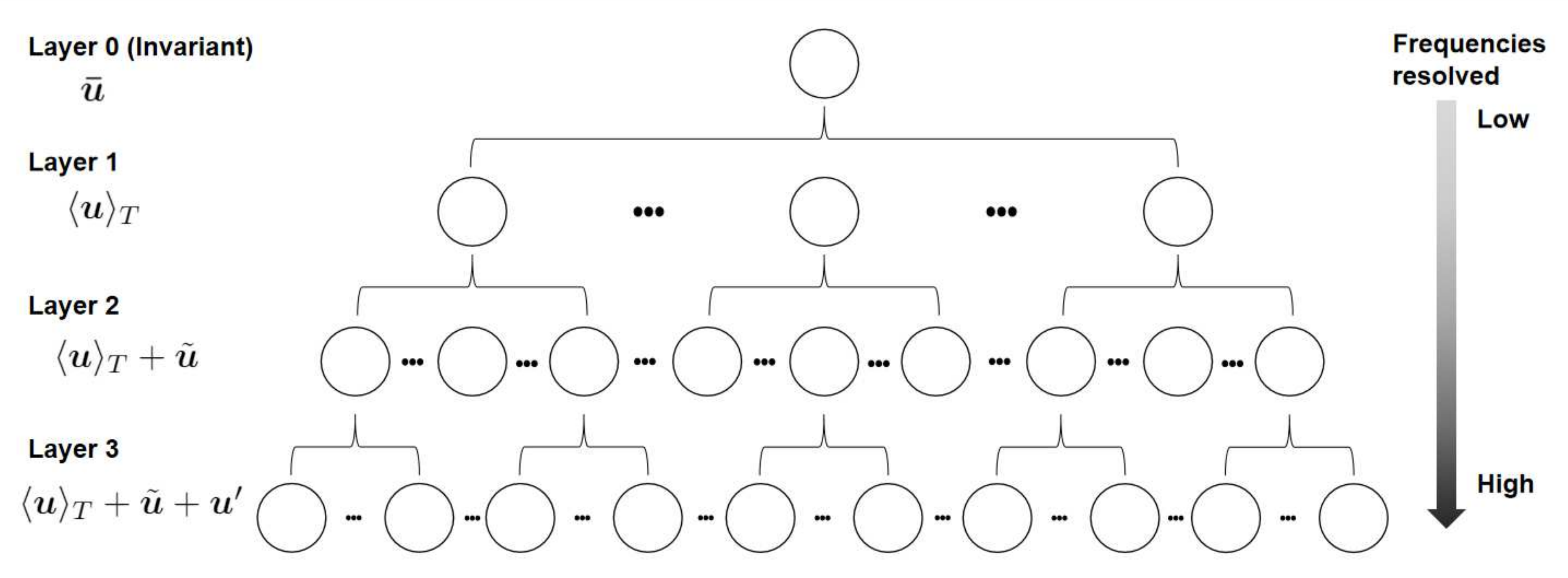}
}
\caption{
An illustration of the hierarchical structure with different scales in the triple flow decomposition in \S~\ref{SubSubSec:FlowDecomposition}.
Layer 0: The top layer is characterized by the invariant mean flow $\bm{\bar{u}}$. 
Layer 1: The global trend is described by the slowly-varying mean-flow field $\langle \bm{u}\rangle _T$.
Layer 2: The coherent part $\tilde{\bm{u}}$ is added for the local dynamics around the varying mean-flow field.
Layer 3: The non-coherent part $\bm{u}^{\prime}$ is considered in the case of turbulent flow.
}
\label{Fig:H_structure}
\end{figure}

Transient and post-transient dynamics are statistically non-homogeneous due to the existence of multiple invariant sets. If so, a scale subdivision of the flow-field decomposition like in Eq.~\eqref{Eqn:ReynoldsDecomposition} is used during the clustering process. 
Accounting for the Reynolds stress contribution, the slowly-varying mean-flow field $\langle \bm{u}\rangle _T$ is enough to describe the global trend.
Next, the local dynamics around $\langle \bm{u}\rangle _T$ can be zoomed in, considering the coherent structures involved in $\tilde{\bm{u}}$. 
This scale subdivision can still be extended to a hierarchical structure with more layers, which involve secondary frequencies in the case of quasi-periodic dynamics or turbulence from $\bm{u}^{\prime}$, as illustrated in figure~\ref{Fig:H_structure}.

In order to clearly describe the clusters in the hierarchy, we systematically name the clusters from top to bottom. 
The sole cluster on the top $\mathcal{L}_0$ contains the ensemble of input data, and we define it symbolically as $\mathcal{C}_0$. 
The sub-division of this cluster leads to $K_1$ subclusters in the first layer $\mathcal{L}_1$, named as $\mathcal{C}_{0,\, k_1}$, $k_1=1,\cdots, K_1$. The first subscript $k_0 = 0$ can be ignored because there is only one cluster in $\mathcal{L}_0$,  and the second subscript $k_1$ indicates the index of the subcluster in $\mathcal{L}_1$.
The second subdivision works on each cluster $\mathcal{C}_{k_1}$ separately, and generates refined $K_2$ subclusters for each of them.
The cluster index in the current layer $\mathcal{L}_2$ is presented by an additional subscript $k_2 = 1,\, \cdots,\, K_2$, which is written as $\mathcal{C}_{k_1,\, k_2}$.
For a higher layer number $\mathcal{L}_{L \in \mathbb{N}}$, $L \geqslant 3$, more subscripts $k_L$, $l = 1,\, \cdots,\, L$,  are needed to record the cluster index in each layer $\mathcal{L}_l$ from the top to the bottom, written as $\mathcal{C}_{k_1, \, \ldots , \, k_L}$. 
This naming method can clearly trace out all clusters in the hierarchy, and also works for other properties of clusters, e.g., the centorids $\bm{c}_{k_1, \, \ldots , \, k_L}$ and the characteristic function $\chi_{k_1, \, \ldots , \, k_L}^{m} $.
In this work, two or three layers ($L \leqslant 3 $) will be enough to extract the transient dynamics out of multiple invariant sets.

\subsubsection{Hierarchical network modelling}
\label{SubSubSec:HiModelling}

The starting point is the hierarchical Markov model of \citet{fine1998ml}, which introduces the hierarchical structure to describe the stochastic processes, comparing to the standard Markov model.
Each state of a Markov model in the parent layer is considered separately, and a new Markov model of the sub-states of a state is built in the child layer.
As the layer increases, the state is continuously subdivided.
Therefore, the hierarchical Markov model records a sequence of states in different layers.
In our case, each cluster is seen as a state.
When a cluster in the parent layer is activated, its subclusters in the child layer turns activated recursively. Meanwhile, the refined dynamics between the subclusters can be described by a Markov model.
Hence, the hierarchical Markov model can more effectively solve the problem of subsets.

In this work, we derive the HiCNM by replacing the Markov model by the network model. 
The hierarchical structure is identical, and the only change is the way to describe the transient dynamics between clusters.

\begin{figure}
 \centerline{
 \includegraphics[width = .4\linewidth]{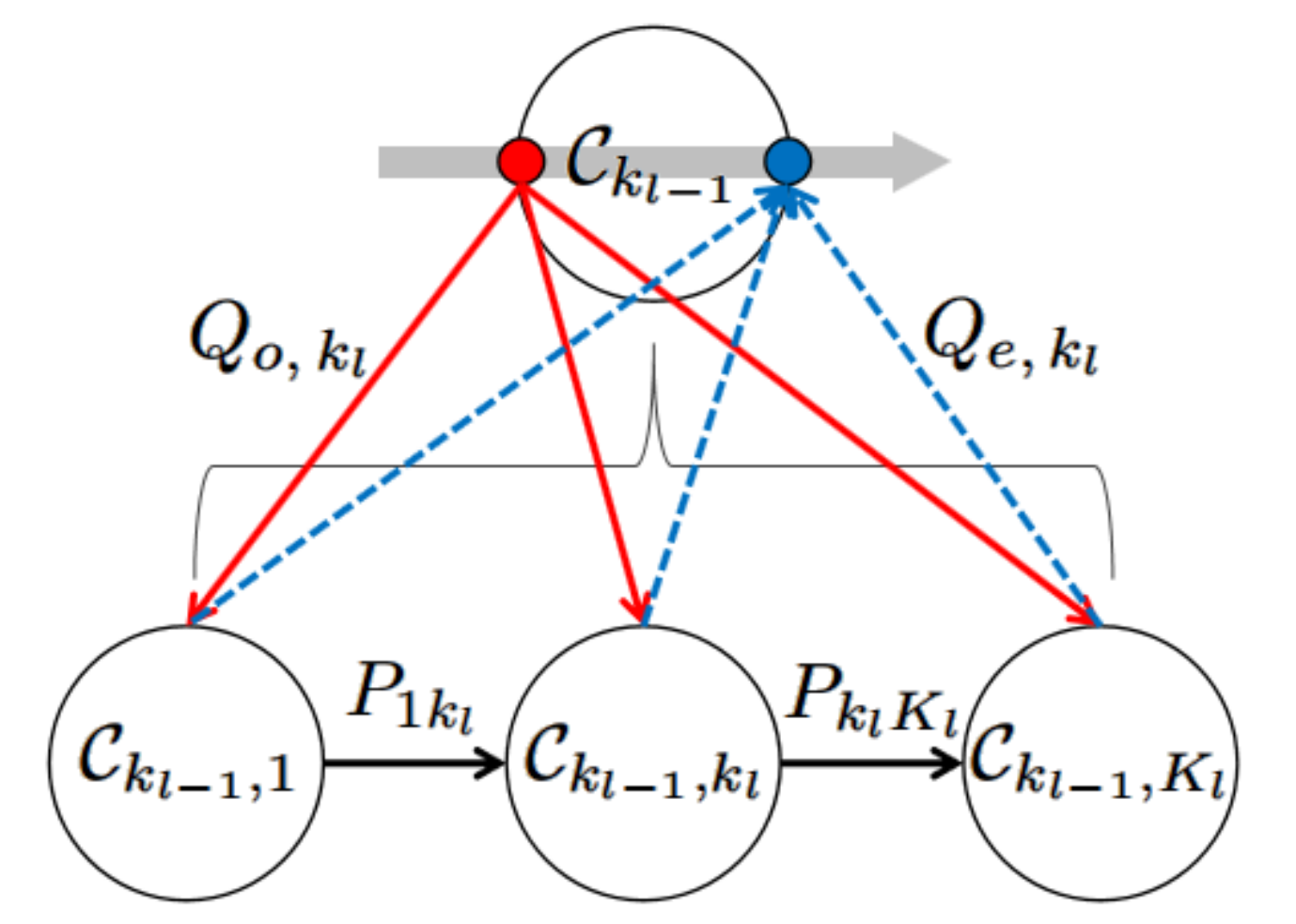}
 }
\caption{An illustration of the transitions between the parent and child layers in the hierarchical network model. The trajectories pass through the cluster in the parent layer: the entering and exiting snapshots are marked with red dot and the blue dot. After clustering, a classic network model is built between $N$ subclusters, with transition probability $P$. 
The vertical transitions indicates the ports of entry and exit of the subclusters with probability $Q_{o,j}$ and $Q_{e,j}$.}
\label{Fig:H_network_cell}
\end{figure}

A typical structure between the parent and child layers is shown in figure \ref{Fig:H_network_cell}. 
We start with cluster $\mathcal{C}_{k_1 , \, \ldots , \, k_{l-1} }$ in the parent layer $\mathcal{L}_{l-1}$, where the leading subscripts $k_1 , \, \ldots , \, k_{l-2}$ refers to the cluster number in each upper layer. 
For convenience, when the context will be unambiguous, the cluster will be only referenced by its number in the current layer, e.g. $\mathcal{C}_{k_{l-1}}$.
As indicated with the sequence of cluster numbers in its complete name, this cluster comes from the sub-division of the cluster $\mathcal{C}_{k_1 , \, \ldots , \, k_{l-2} }$ in the parent layer $\mathcal{L}_{l-2}$.
We suppose that $M$ snapshots $\bm{u}^{m}$, $m=1,\ldots,M$, exist in this cluster and are divided into $K_{l-1}$ subclusters by a sub-division clustering algorithm.
The sub-cluster $\mathcal{C}_{k_{l-1}}$ contains $n_{k_{l-1} }$ snapshots, calculated from Eq.~\eqref{Eqn:Countingnumber} with the characteristic function $\chi_{k_{l-1} }^{m}$. 
A standard network model for the cluster $\mathcal{C}_{k_1 , \, \ldots , \, k_{l-2} }$ can be derived with the direct transition matrix $\mathsfbi{P}_{k_1 , \, \ldots , \, k_{l-2} }$ and the residence time matrix $\mathsfbi{T}_{ k_1 , \, \ldots , \, k_{l-2} }$, as recorded in \S~\ref{SubSubSec:CNM}, which describe the dynamics between the subclusters $\mathcal{C}_{k_{l-1}}$.

In the following, we focus on the trajectories passing through cluster $\mathcal{C}_{k_{l-1}}$.
The snapshots entering and leaving from $\mathcal{C}_{k_{l-1}}$ are marked out for each trajectory, with the following characteristic function 
\begin{equation}
\begin{matrix}
 \chi_{o, k_{l-1}}^{m} := \begin{cases}
                1, &  \mbox{if }\bm{u}^{m-1} \notin \mathcal{C}_{k_{l-1}} \And \bm{u}^{m} \in \mathcal{C}_{k_{l-1}}, \\ 
                0, &  \mbox{otherwise}.
               \end{cases} \\
 \chi_{e, k_{l-1}}^{m} := \begin{cases}
                1, &  \mbox{if }\bm{u}^{m}\in \mathcal{C}_{k_{l-1}} \And \bm{u}^{m+1} \notin \mathcal{C}_{k_{l-1}}, \\ 
                0, &  \mbox{otherwise}.
               \end{cases}
\end{matrix}
 \label{Eqn:CharacteristicFunction2}
\end{equation}
The entering snapshots are denoted by the subscript ``$o$'', and the exiting snapshots by the subscript ``$e$''.  
The number of entering snapshots $n_o$ and of exiting snapshots $n_e$ read
\begin{equation}
n_o = \sum\limits_{m=1}^M\, \chi_{o,k_{l-1}}^{m}, \> \> \>
n_e = \sum\limits_{m=1}^M\, \chi_{e,k_{l-1}}^{m}.
\end{equation}

In the child layer $\mathcal{L}_{l}$, the $n_k$ snapshots in the cluster $\mathcal{C}_{k_{l-1}}$ have been divided into the subclusters $\mathcal{C}_{k_1 , \, \ldots , \, k_{l} }$ , $k_{l} =1,\ldots,k_L$. 
Without loss of generality, a standard network model for the cluster $\mathcal{C}_{k_{l-1}}$ can be built with its subclusters with the direct transition matrix $\mathsfbi{P}_{k_1 , \, \ldots , \, k_{l-1} }$ and the residence time matrix $\mathsfbi{T}_{ k_1 , \, \ldots , \, k_{l-1} }$. 

The snapshots $\bm{u}^m$ are approximated by the time evolution of the cluster centroids $\bm{c}_{k_1^m, \, \ldots , \, k_{l}^m }$ in $\mathcal{L}_{l}$,
\begin{equation}
\label{Eqn:HCNM}
\hat{\bm{u}}^m_{\mathcal{L}_{l}} = \bm{c}_{k_1^m, \, \ldots , \, k_{l}^m }.
\end{equation}
The residence time elements of $\mathsfbi{T}_{k_1 , \, \ldots , \, k_{l}}$ can be assembled in order to determine the moments of transition based on the sequence of visited clusters in $\mathcal{L}_{l}$. 

The entering and exiting snapshots defined in Eq.~\eqref{Eqn:CharacteristicFunction2} can be used to describe the vertical transitions.
Although they are not necessary to describe the dynamics of the fluidic pinball, the probability of the vertical transitions $Q_{o,j}$ and $Q_{e,j}$, described in appendix~\ref{Sec:VT_HiCNM}, completes all possible transitions in our hierachical structure and make it consistent with the classic hierarchical Markov model of \citet{fine1998ml}.

\subsubsection{Dynamics reconstruction of the hierarchical network model}
\label{SubSubSec:Validation}

The reconstructed flow in Eq.~\eqref{Eqn:HCNM} is a statistical representation of the original snapshot sequence by a few representative centroids, which is a highly discretized description compared to the full dynamics.
The approximations in the different layers provide different metrics for the flow dynamics.

The cluster-based hierarchical model uses the centroids $\bm{c}_{k_1^m, \, \ldots , \, k_{l}^m }$ in the original data space together with the time evolution of the cluster index $k_{l}^m$ to simplify the description of the original flow, which is more intuitive and closer to the original flow than the POD reconstruction.
For input data with $I$-dimensional state vectors of the velocity field and $M$ snapshots, a POD reconstruction truncated to $R$ modes will lead to a $I \times R$ matrix of POD modes and a $R \times M$ matrix of mode amplitudes.
A HiCNM with $K$ centroids leads to a $I \times K$ matrix of centroids and a sequence of the $N$ visited clusters of length $N \ll M$.
 In this sense, the compressive ability of HiCNM is more powerful for the large amount of continuously sampled data, as $I \times K < (I+M)\times R$. In addition, the hierarchical clustering works as a sparse sampling technique, extracting the representive states according to the clustering subspace.

We use the unbiased auto-correlation function \citep{protas2015jfm},
\begin{equation}
\label{Eqn: CorrelationFunction}
  R(\tau)=\frac{1}{T-\tau}\int\limits_{\tau}^{T}\left(\bm{u}(\bm{x}, t-\tau) - \bm{u}_s(\bm{x}),\bm{u}(\bm{x}, t)- \bm{u}_s(\bm{x})\right)_{\Omega} \mathrm{d}t, \quad \tau\in
  [0,T)\,,
\end{equation}
after normalization with respect to $R(0)$ to check the accuracy of the dynamics reconstruction of the HiCNM in Eq.~\eqref{Eqn:HCNM}.
The autocorrelation function without delay $R(0)$ is twice the time-averaging kinetic energy. 
The modeled autocorrelation function $\hat{R}_{\mathcal{L}_{l}}(\tau)$ in layer $\mathcal{L}_{l}$ is based on the rebuilt flow $\hat{\bm{u}}_{\mathcal{L}_{l}}$ in Eq.~\eqref{Eqn:HCNM} instead of $\bm{u}$ in Eq.~\eqref{Eqn: CorrelationFunction}.

For the discrete snapshots, the root mean-square error (RMSE) of the autocorrelation function $R(\tau)$ of the reference data and that of the model $\hat{R}_{\mathcal{L}_{l}}(\tau)$ is defined as
\begin{equation}
  \label{Eq:RMSE_R}
  R_{\rm rms}^l : = \sqrt{\frac{1}{M}\sum\limits_{m=1}^{M}\left(R(\tau)-\hat{R}_{\mathcal{L}_{l}}(\tau)\right)^2}\,,
\end{equation}
where $M$ is the number of snapshots $\bm{u}^m$ and $\hat{\bm{u}}^m_{\mathcal{L}_{l}}$ at $\tau = m \Delta t$.

\section{Hierarchical network modelling of the fluidic pinball}
\label{Sec:HiCROM_PINBALL}

In this section, we apply the hierarchical modelling strategy to the fluidic pinball at different Reynolds numbers. 
With increasing Reynolds number, the flow dynamics is undergoing successive instabilities and bifurcations, introducing multiple exact solutions of the Navier-Stokes equations and multiple invariant sets for the dynamics.
In \S~\ref{Sec:HC_Pinball}, the modelling strategy dealing with multiple invariant sets is introduced.
We derive the HiCNMs for the transient dynamics involving six invariant sets at $\Rey=80$ in \S~\ref{Sec:HNM80}, for the quasi-periodic regime at $\Rey=105$ in \S~\ref{Sec:HNM105}, and for the chaotic regime at $\Rey=130$ in \S~\ref{Sec:HNM130}.

\subsection{Hierarchical modelling with multiple invariant sets}
\label{Sec:HC_Pinball}

The flow field is computed with the direct numerical simulation (DNS) described in \S~\ref{Sec:Pinball}.
The resulting flow field is an ensemble of time-resolved snapshots starting with some given initial condition.
The transient and post-transient dynamics of the flow constitute a time-resolved trajectory sampled with a fixed time step.
Different invariant sets and multiple attractors can co-exist in the state space, only part of them being explored by each individual trajectory from the initial condition to the asymptotic regime. 
All the cases of interest in this paper are such that $\Rey>\Rey_2$, i.e. beyond the supercritical pitchfork bifurcation. 
The data set consists of the snapshots computed from four different trajectories: two mirror-conjugated trajectories starting in the vicinity of the symmetric steady solution, the two others starting from the two mirror-conjugated asymmetric steady solutions. 
The simulations are respectively run until $t=1\,500$ and $t=1\,000$ for the symmetric steady solution and the asymmetric steady solutions.

Sampled with time step $\Delta t = 0.1$, the input data basis is an ensemble of $M=50\,000$ snapshots ${\bm{u}^m(\bm{x})}$ from four transient trajectories, where the superscript $m$ is the snapshot index for the successive instants $t^m = m \Delta t$.
In order to distinguish the different trajectories, the snapshot index $m$ is sorted as:
\begin{enumerate}[ (a) ]
    \item $m=1,\ldots,15000$ and $m=15001,\ldots,30000$ for the two mirror-conjugated trajectories starting in the vicinity of the symmetric steady solution,
    \item $m=30001,\ldots,40000$ and $m=40001,\ldots,50000$ for the two others starting from the two mirror-conjugated asymmetric steady solutions. 
\end{enumerate}
The time continuity is critically important during the dynamical analysis.
Snapshots in each trajectory are time-resolved but have no time relationship in different trajectories.
For a trajectory of $M$ snapshots, it exists $M-1$ transitions as described in \S~\ref{SubSubSec:CNM}.
Hence, $M-4$ transitions occur in the four individual trajectories mentioned above.

\begin{figure}
 \centerline{
 \includegraphics[width = .95\linewidth]{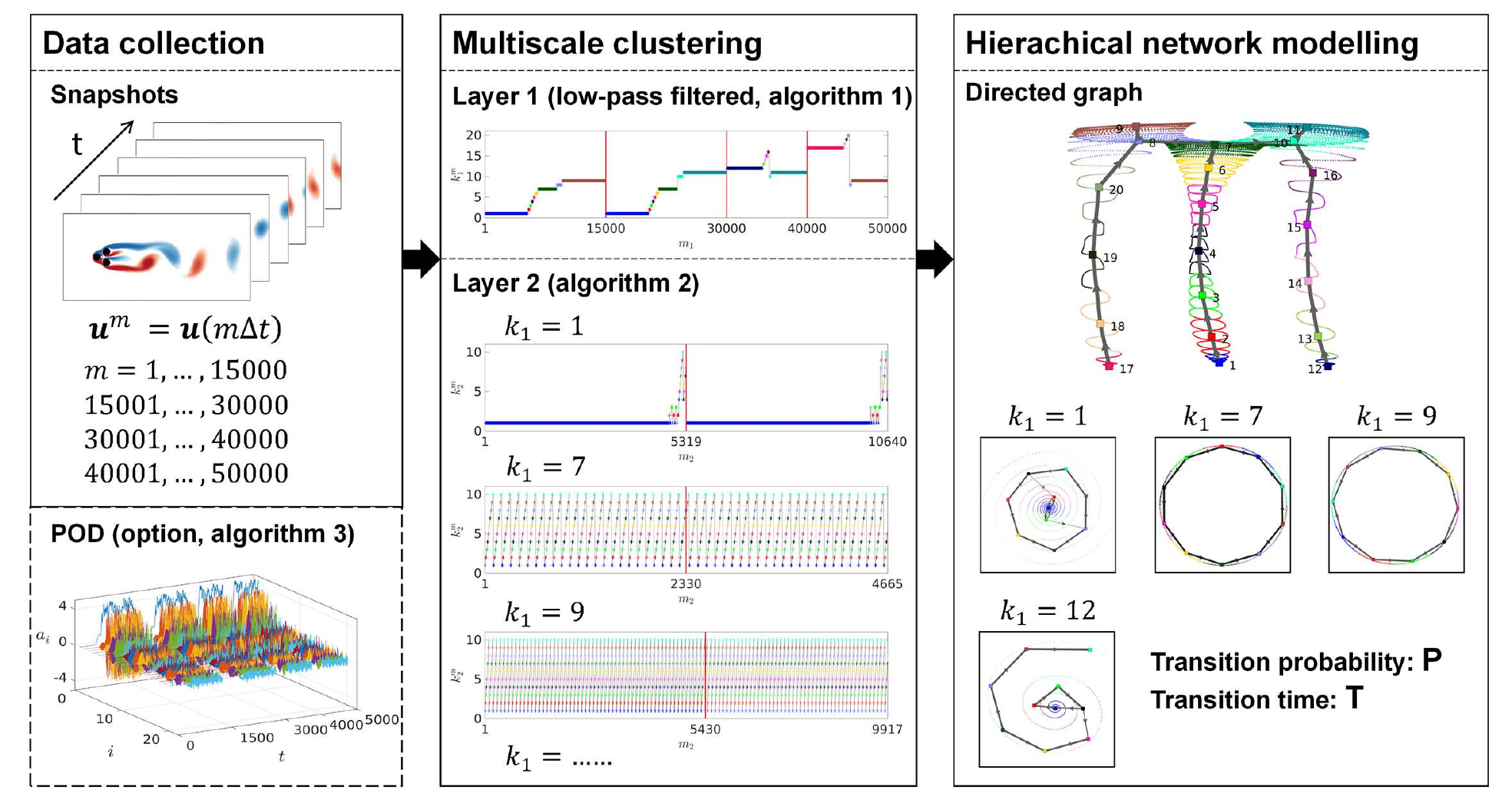}
 }
\caption{Sketch for HiCNM applied to the fluidic pinball at $\Rey=80$. See text for the details.}
\label{Fig:Cartoon_HiCNM}
\end{figure}

In our case, a standard network model with $200$ clusters are still not enough to distinguish the transitions starting from three different steady solutions, as shown in \S~\ref{Sec:CNM80}.
An optimal way to achieve correct classification is to hierarchically cluster the ensemble of snapshots.
The hierarchical clustering is performed with the slaving assumption under the mean-field consideration in \S~\ref{SubSubSec:FlowDecomposition}, 
by applying an unsupervised clustering algorithm (k-means$++$) for different time scales.
The clusters in the parent layer are split into subclusters in the child layer, where the clustering result in the parent layer works as a pre-classified indicator in the child layer.
The number of clusters is preset to 10 for each clustering algorithm, but can be adjusted to the minimal number for an accurate dynamics reconstruction, typically for the network model in the first layer.
The clustering algorithm in the first layer is meant to distinguish different invariant sets together with the transitions between them with a limited number of clusters. 
These clusters will be used to build a network model for the mean-field distortion, which will further supervise the clustering process in the second layer. 
A sketch for this process is shown in figure~\ref{Fig:Cartoon_HiCNM}.

From the mean-field consideration, the slowly-varying mean-flow field is the ideal candidate for the detection of several invariant sets. 
A fifth order Butterworth low-pass filter with cutoff frequency $0.2 f_c$ is applied to eliminate the coherent $\tilde{\bm{u}}$ and incoherent $\bm{u}^{\prime}$ components of $\bm{u}$ in the triple decomposition of Eq.\eqref{Eqn:ReynoldsDecomposition}. 
The clustering algorithm applied to the first layer is described in algorithm~\ref{alg:FilteredKmeans}. 
\begin{algorithm}[h]  
  \caption{Clustering algorithm with slowly varying mean flow}  
  \label{alg:FilteredKmeans}  
  \begin{algorithmic}[1]  
    \Require  
      $\bm{u}^m$: snapshots;  \quad $f_c$: frequency of coherent part; 
    \Statex \qquad
      $K_1$: number of clusters 
    \Ensure  
      $\chi_{k_1}^{m}$: characteristic function; \quad $k_1^m$: cluster indexes of snapshots $\bm{u}^m$; \quad 
    \Statex \qquad
      $\bm{c}_{k_1}$: optimal centroids
    \State compute the low-pass filtered $\bm{u}^m$ with cutoff frequency $0.2 f_c$, named $\bm{u}^m_{\rm LP}$;
    \State apply $k$-means$++$ algorithm with $K_1$ clusters to $\bm{u}^m_{\rm LP}$, and save the characteristic function $\chi_{k_1}^{m}$ and the cluster indexes $k_1^m$;
    \State compute and save the centroids in original data space: 
    \Statex $\bm{c}_{k_1} = { \sum\limits_{m=1}^M  \> \chi_{k_1}^{m} \bm{u}^m } / {\sum\limits_{m=1}^M  \> \chi_{k_1}^{m}}$.
  \end{algorithmic}  
\end{algorithm} 
The critical idea of the algorithm is to map the original data to the bounded low-frequency space, and then calculate the characteristic function in the low-frequency space. The resulting characteristic function $\chi_{k_1}^{m}$ is applied to the original data to achieve the clustering of the slowly-varying mean-flow field.

The divisive clustering algorithm of the clusters in the parent layer is described in algorithm~\ref{alg:KmeansChildLayer}, under the supervision of the characteristic function $\chi_{k_1}^{m}$ obtained from the parent layer. 
\begin{algorithm}[h]  
  \caption{Divisive clustering algorithm in the child layer under supervision}  
  \label{alg:KmeansChildLayer}  
  \begin{algorithmic}[1]  
    \Require  
      $\bm{u}^m$: snapshots;  \quad
    \Statex \qquad
      $\chi_{k_1}^{m}$: characteristic function from the parent layer
    \Ensure  
      for each cluster $\mathcal{C}_{k_1}$ in parent layer
    \Statex \qquad
     $\chi_{k_1, k_2}^{m}$: the characteristic function; \quad $k_2^m$: cluster indexes of snapshots $\bm{u}^m$; \quad 
    \Statex \qquad
     $\bm{c}_{k_1, k_2}$: optimal centroids; \quad
    \For{$k_1 \gets 1$ to $K_1$}     
        \State locate the snapshots $\bm{u}^m$ in the cluster $\mathcal{C}_{k_1}$ by the characteristic function $\chi_{k_1}^{m}$,  and record the snapshot index $m_1$ of the resulting $n_{k_1}$ snapshots;
        \State extract the snapshots $\bm{u}^{m_1}$ and renumber them sequentially with $m_2 = 1, \ldots, M_2$;
        \State apply $k$-means$++$ algorithm with $K_2 = 10$ clusters (as default) to the renumbered snapshots $\bm{u}^{m_2}$ and save the characteristic function $\chi_{k_1, k_2}^{m_2}$ and the centroids $\bm{c}_{k_1, k_2}$: 
        \Statex $\bm{c}_{k_1, k_2} = \sum\limits_{m_2=1}^{M_2}  \> \chi_{k_1, k_2}^{m_2} \bm{u}^{m_2}/ \sum\limits_{m_2=1}^{M_2}  \> \chi_{k_1, k_2}^{m_2} $.
    \EndFor
  \end{algorithmic}  
\end{algorithm} 
The sub-division of the clusters in the parent layer leads to a more detailed network model of the local structures.
According to the spectral content of the dynamics, multiple layers are introduced to extract the coherent dynamics. 
The naming method introduced in \S~\ref{SubSubSec:HiClustering} can clearly locate all clusters in the hierarchy.
When dealing with chaotic flow regime, there is no clear frequency boundary. We stop modelling the incoherent components $\bm{u}^{\prime}$ with a simple network model which contains all the chaotic dynamics.
Snapshots of velocity field can be highly compressed by a lossless POD to accelerate the clustering algorithm, as detailed in appendix~\ref{Sec:POD}.

\subsection{Hierarchical network model at $\Rey=80$}
\label{Sec:HNM80}

At $\Rey=80>\Rey_{2}$, the system has already undergone a supercritical Hopf bifurcation and two coincidental supercritical pitchfork bifurcations on the steady solution and the symmetric limit cycle.
As a result, three unstable steady solutions, one unstable (symmetric) limit cycle and two stable (asymmetric) mirror-conjugated limit cycles exist in the state space and organize the dynamics. 
Thus, there are six invariant sets, the two stable limit cycles being the attractors of the flow state.

The HiCNM in the first layer is based on the clustering results of the low-pass filtered data set (\S~\ref{SubSec:HNM80_L1}). The local dynamics for some typical regimes is further presented with the subclusters in the second layer (\S~\ref{SubSec:HNM80_L2}). 

\subsubsection{Hierarchical network model in Layer 1}
\label{SubSec:HNM80_L1}

\begin{figure}
\centering
$(a)$ \raisebox{-0.6\height}{\includegraphics[width=.75\linewidth]{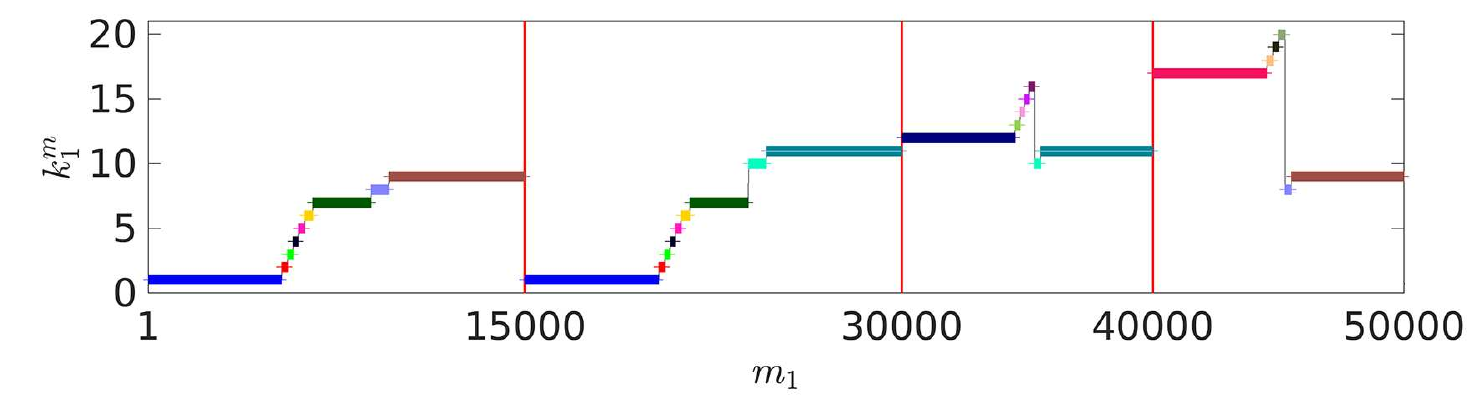}} \\
\begin{tabular}{cc}
$(b)$ \raisebox{-0.5\height}{\includegraphics[width=.35\linewidth]{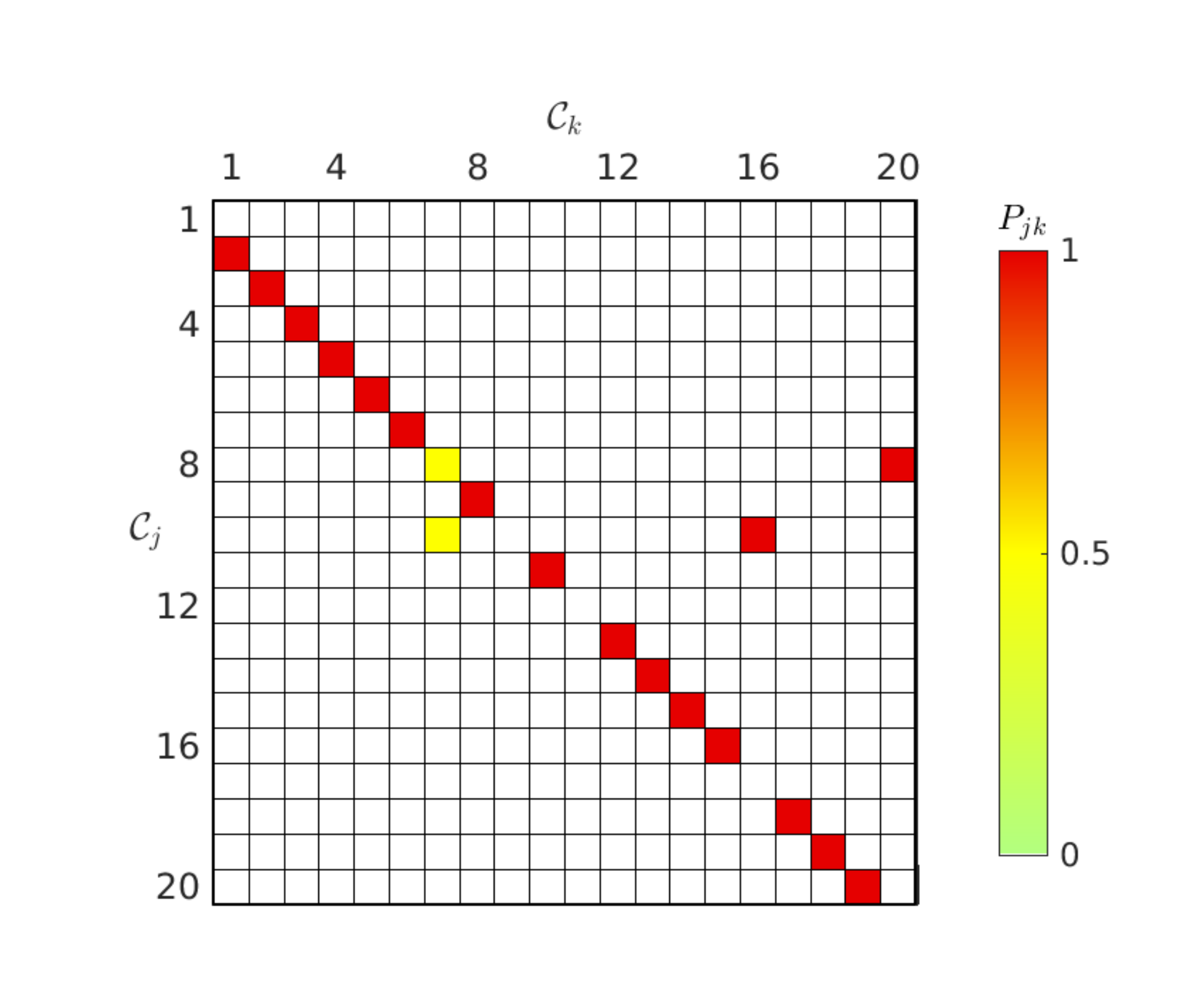}} &
$(c)$ \raisebox{-0.5\height}{\includegraphics[width=.35\linewidth]{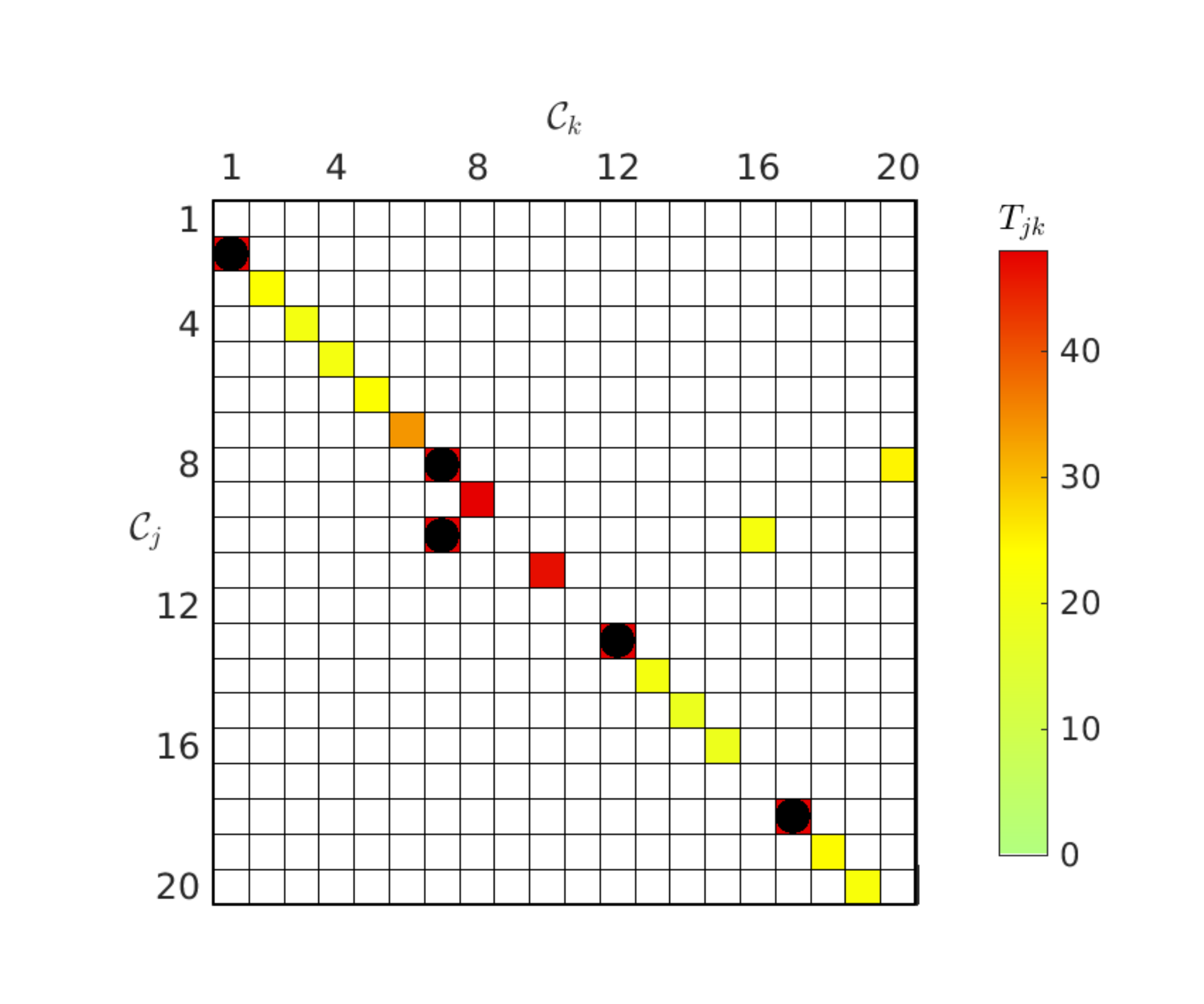}} \\
\end{tabular}
\caption{Cluster-based analysis at $\Rey= 80$ in layer 1: 
(a) transition illustrated with cluster label, 
(b) transition matrix, 
(c) residence time matrix. 
Since subclusters are not considered, the snapshot index in layer 1 $m_1$ is identical to the original index $m$.
The colorbar indicates the values of the terms. Residence time larger than 100 is marked with a solid black circle and excluded from the colorbar.
An extremely long residence time in a cluster indicates data density, and such a cluster is generally associated with an invariant set.}
\label{Fig:Matrix80_layer1}
\end{figure}

$K_1=20$ clusters are used to cluster the snapshots from the low-pass filtered data set which removes the coherent structures with frequency $f_c = 0.1074$. 
To visualize the cluster topology, we apply the classical multidimensional scaling (MDS) to represent the high-dimensional centroids in a two-dimensional subspace $[\gamma_1, \gamma_2]^{\rm T}$, while the distances between the centroids are preserved \citep{kaiser2014jfm}. 
\begin{figure}
\centering
\includegraphics[width=.65\linewidth]{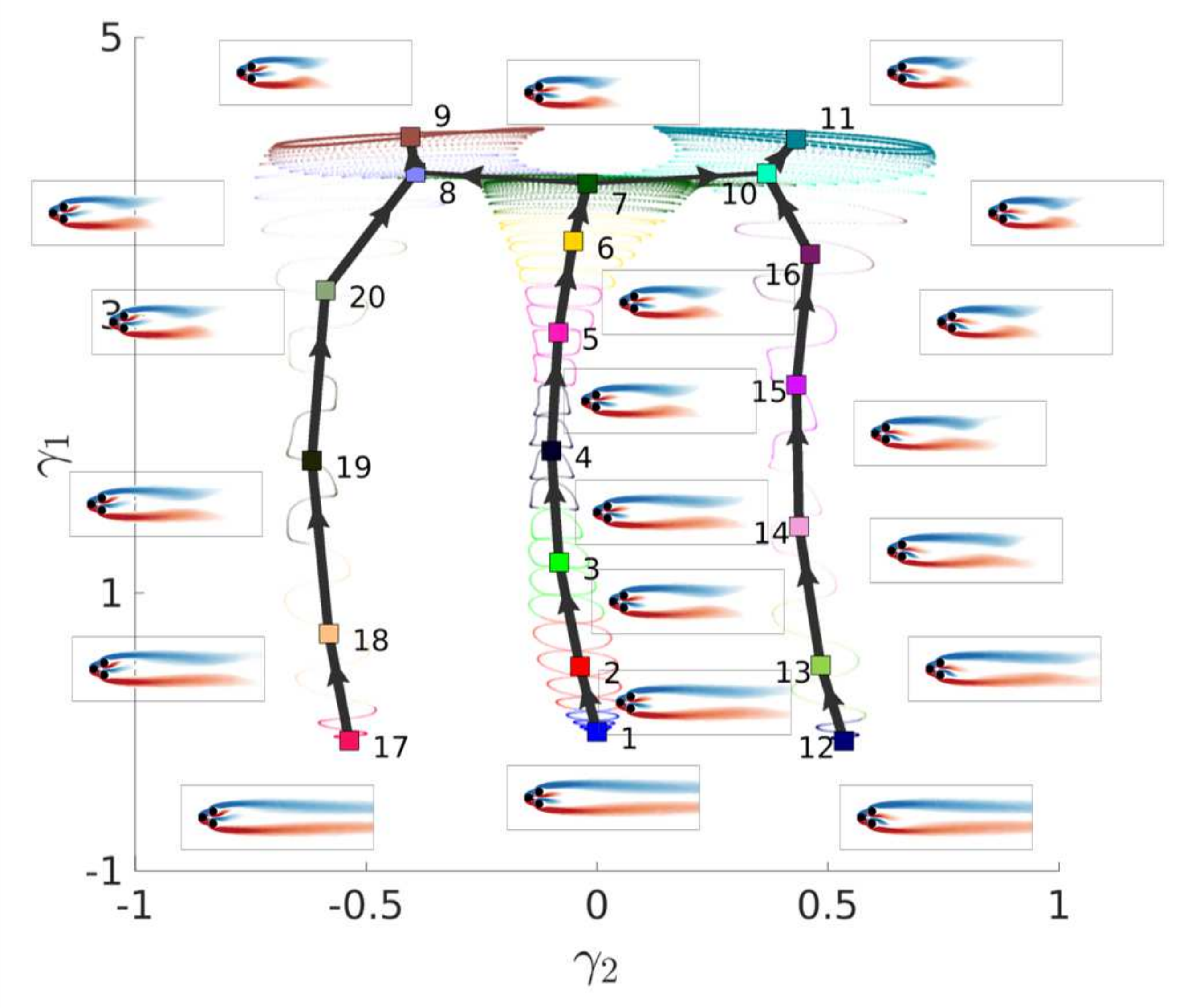} 
\caption{Graph of transitions between clusters in layer 1 at $\Rey= 80$.
Cluster centroids are marked with the colored squares, with their vorticity fields in color with $[-1.5, 1.5]$.
The snapshots belonging to them are marked as small dots with the same colors.
The transitions between clusters are shown with arrows, where the line width presents the probability of transition.}
\label{Fig:ClusterAnalysis80_layer1}
\end{figure}
As shown in figure \ref{Fig:ClusterAnalysis80_layer1}, the six exact solutions of the Navier-Stokes equations that organize the state space are well identified. 
The vorticity field of the resulting centroids can be understood as the slowly varying mean-flow field $<\bm{u}>_T$ along the transient dynamics, with $T\gg 2\pi /\omega_c$. 
The three steady solutions belong respectively to clusters $\mathcal{C}_1$ (symmetric steady solution $\bm{u}_s$), $\mathcal{C}_{12}$ (asymmetric steady solution $\bm{u}_s^-$) and $\mathcal{C}_{17}$ (asymmetric steady solution $\bm{u}_s^+$).
The three limit cycles are caught by the time-averaged flow in clusters $\mathcal{C}_7$ (symmetric mean-flow field, centroid $\bar{\bm{u}}^0$), $\mathcal{C}_9$  (asymmetric upward mean-flow field, centroid  $\bar{\bm{u}}^+$) and $\mathcal{C}_{11}$ (asymmetric downward mean-flow field, centroid $\bar{\bm{u}}^-$).
A network of four transient trajectories connects these clusters as follows:
\begin{enumerate}[${Trajectory\, }$1 :]
\item \ $\mathcal{C}_1\  (\bm{u}_s) \rightarrow \ldots \rightarrow \mathcal{C}_7\  (\bar{\bm{u}}^0)\  \rightarrow \mathcal{C}_8\ \rightarrow \mathcal{C}_9\  (\bar{\bm{u}}^+)$;
\item \ $\mathcal{C}_1\  (\bm{u}_s) \rightarrow \ldots \rightarrow \mathcal{C}_7\  (\bar{\bm{u}}^0)\  \rightarrow \mathcal{C}_{10} \rightarrow \mathcal{C}_{11}\  (\bar{\bm{u}}^-)$;
\item \ $\mathcal{C}_{12}\ (\bm{u}_s^-) \rightarrow  \ldots \rightarrow \mathcal{C}_{16} \rightarrow \mathcal{C}_{10} \rightarrow \mathcal{C}_{11}\  (\bar{\bm{u}}^-)$;
\item \ $\mathcal{C}_{17}\ (\bm{u}_s^+) \rightarrow  \ldots \rightarrow \mathcal{C}_{20} \rightarrow \mathcal{C}_8\ \rightarrow \mathcal{C}_9\  (\bar{\bm{u}}^+)$.
\end{enumerate}
We notice that most of the transitions between clusters are with $100\%$ probability, except for the bifurcating cluster $\mathcal{C}_{7}$ with half probability to transition to clusters $\mathcal{C}_{8}$ or $\mathcal{C}_{10}$.
The transition matrix in figure~\ref{Fig:Matrix80_layer1}(b) illustrates the probability of all the transitions. The probability is $1$ for the surely directed transitions. By contrast, $P_{7\, 8}=0.5$ and $P_{7\, 10}=0.5$ for the bifurcating cluster.
After entering into $\mathcal{C}_{8}$ and $\mathcal{C}_{10}$, the flow will surely enters the two clusters $\mathcal{C}_{9}$ and $\mathcal{C}_{11}$ respectively. 
The two clusters $\mathcal{C}_{9}$ and $\mathcal{C}_{11}$ catch the permanent regimes, from which the dynamics cannot escape, imposing all the terms in the ninth and eleventh columns $P_{j\, 9} = P_{j\, 11} = 0,\ \forall j$.

The time ordering of the cluster transitions is illustrated in figure~\ref{Fig:Matrix80_layer1}(a). The red vertical line separates the four trajectories. All the trajectories are irreversible transition from one steady solution to one of the two stable periodic solutions.
In figure~\ref{Fig:Matrix80_layer1}(c), the filled black circles emphasize the transitions starting from $\mathcal{C}_{1}$, $\mathcal{C}_{12}$ , $\mathcal{C}_{17}$ and $\mathcal{C}_{7}$. 
The two attracting clusters $\mathcal{C}_{9}$ and $\mathcal{C}_{11}$ have no transition to any other clusters. Hence, all the terms in the ninth and eleventh columns $T_{j\, 9}, T_{j\, 11}$ are 0, $\forall j$. The residence time associated with clusters $\mathcal{C}_{9}$ and $\mathcal{C}_{11}$ is infinite.  

According to the above discussion, the network model in this layer has successfully identified the six invariant sets of the dynamics, four being unstable, the two others being the attractors of the system.

\subsubsection{Hierarchical network model in layer 2}
\label{SubSec:HNM80_L2}
We apply algorithm~\ref{alg:KmeansChildLayer} based on the clustering result for the layer $\mathcal{L}_{1}$ in \S~\ref{SubSec:HNM80_L1}.
In the second layer $\mathcal{L}_{2}$, we will isolate and analyze the clusters $\mathcal{C}_{k_1},\ k_1=9, 7, 1$ associated with three invariant sets of the dynamics.

\subsubsection*{The permanent regime in cluster $\mathcal{C}_{9}$}

Cluster $\mathcal{C}_{9}$ is associated with one of the two asymmetric limit cycles. The $k$-means$++$ algorithm is directly applied to the intra-cluster snapshots $\bm{u}^m \in \mathcal{C}_{9}$. 
\begin{figure}
\begin{tabular}{c c}
\begin{minipage}{0.48\textwidth}
\begin{tabular}{c}
\includegraphics[width=.9\linewidth]{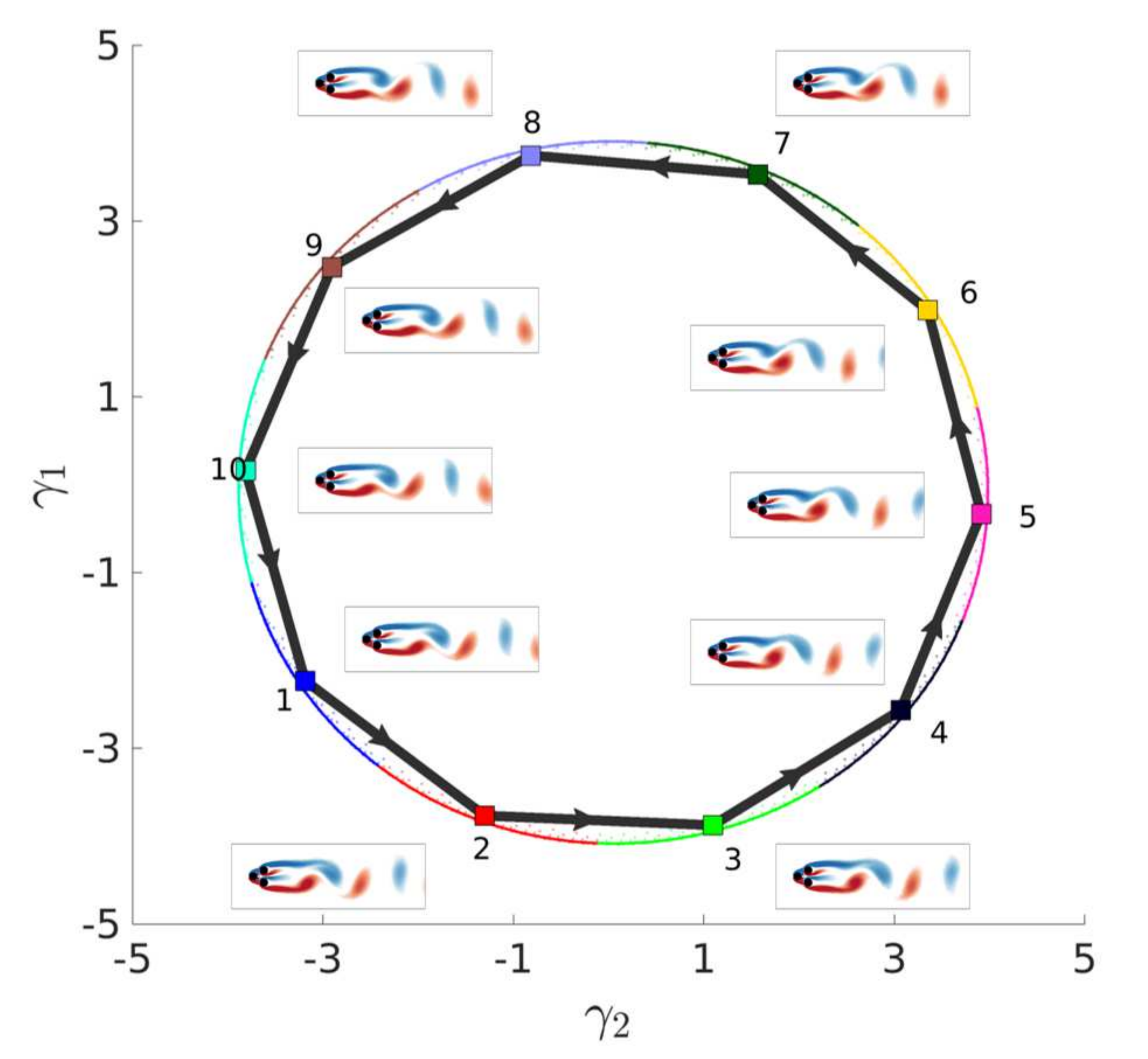}\\
$(a)$ 
\end{tabular} 
\end{minipage} &
\begin{minipage}{0.48\textwidth}
\begin{tabular}{c}
\includegraphics[width=1\linewidth]{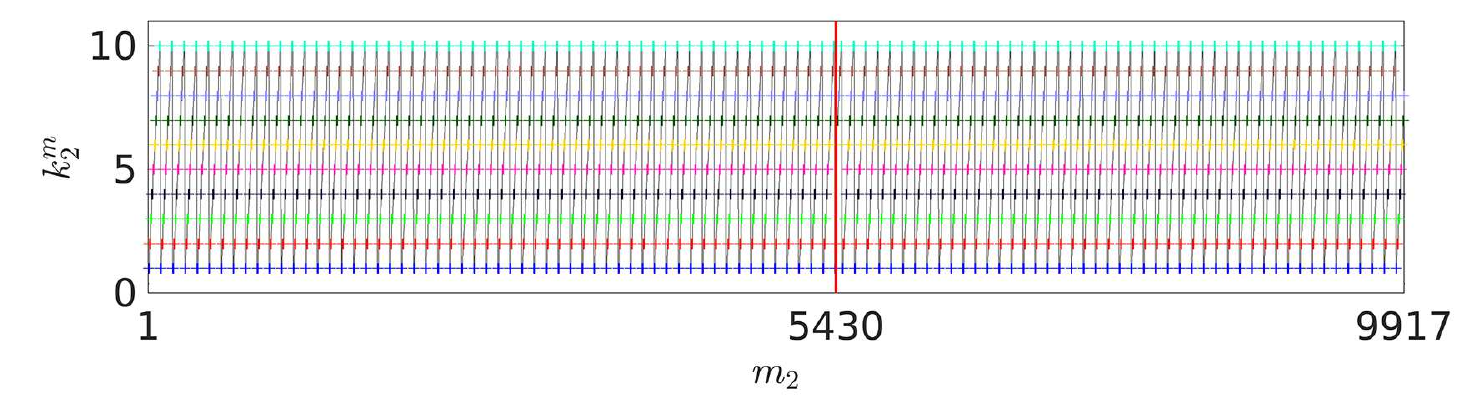} \\
$(b)$ \\
\end{tabular} 
\begin{tabular}{cc}
\includegraphics[width=.49\linewidth]{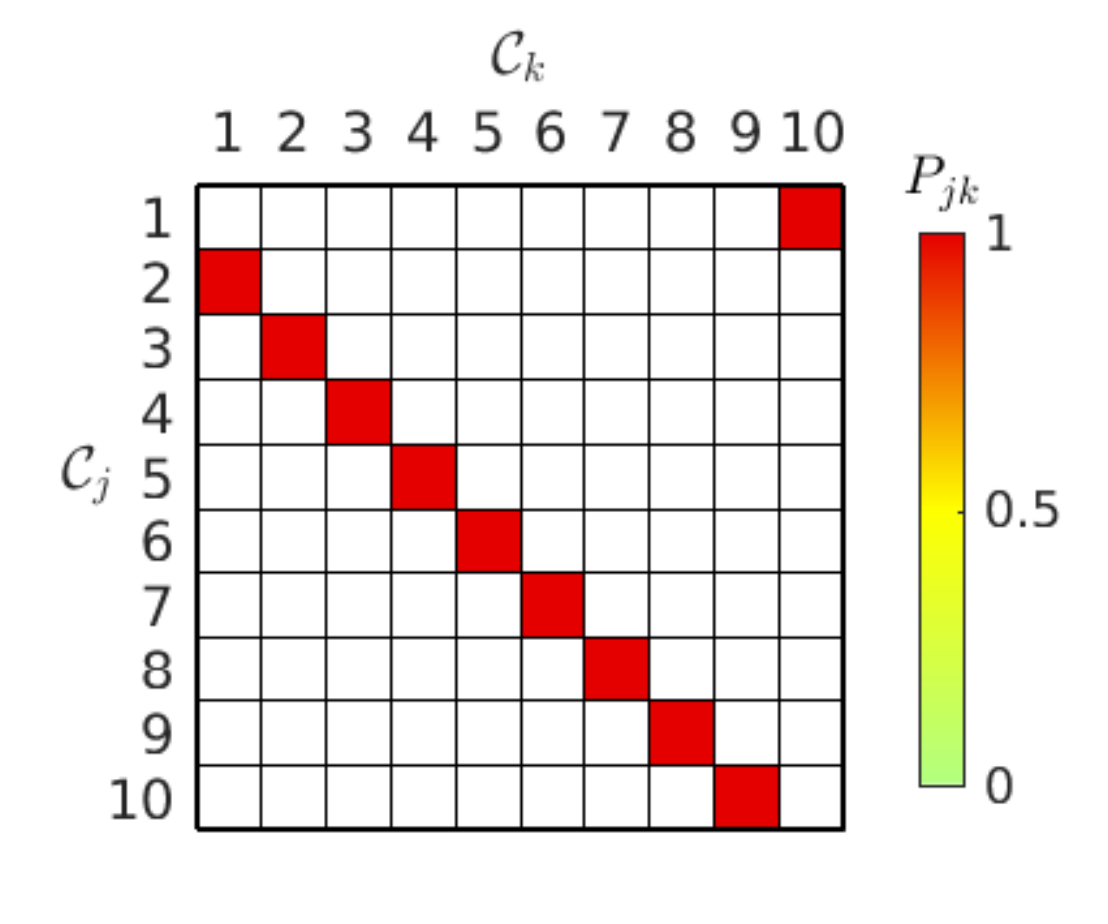} & 
\includegraphics[width=.49\linewidth]{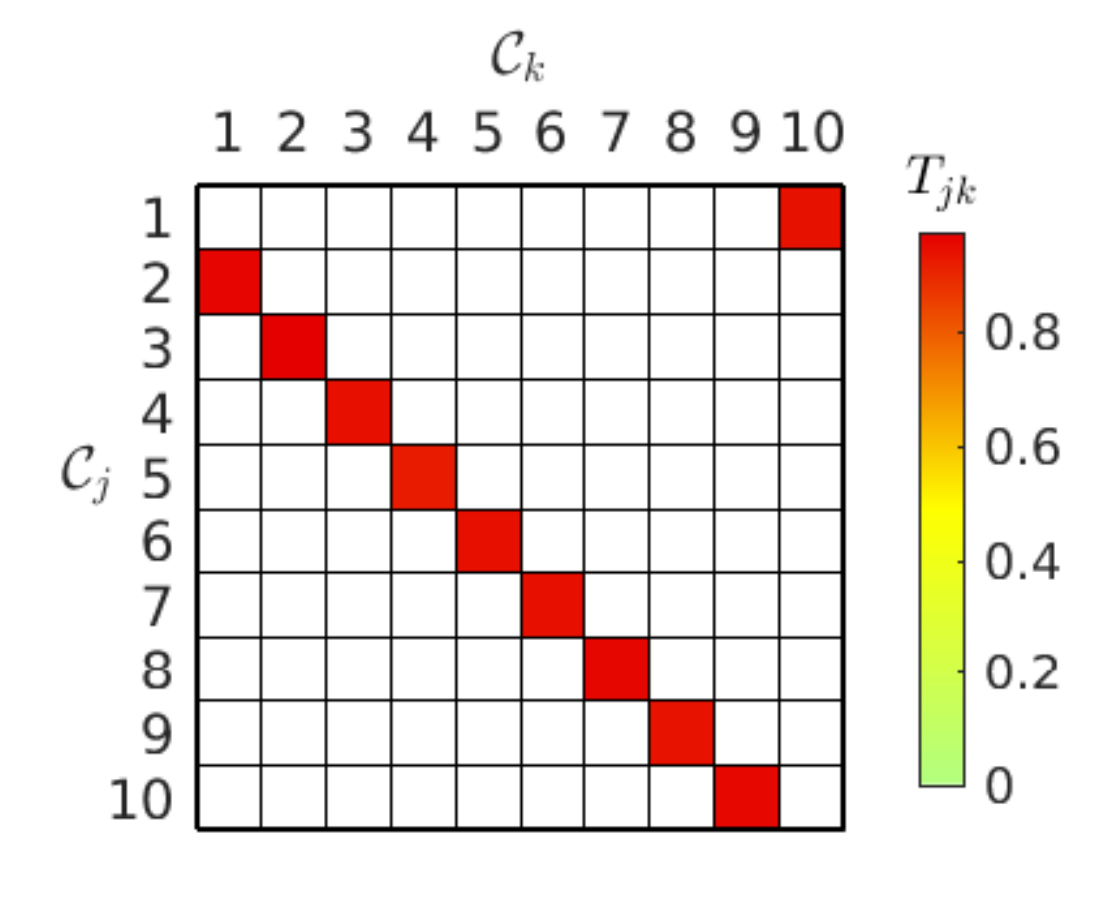} \\
$(c)$ & $(d)$ 
\end{tabular}
\end{minipage}
\end{tabular}
\caption{Cluster-based analysis in layer 2 at $\Rey= 80$ for $\mathcal{C}_{9}$: 
(a) graph of non-trivial transitions between clusters, as in figure~\ref{Fig:ClusterAnalysis80_layer1},
(b) transition illustrated with cluster label, 
(c) transition matrix, 
(d) residence time matrix, as in figure~\ref{Fig:Matrix80_layer1}.
Two trajectories pass through $\mathcal{C}_{9}$ in the parent layer, one with $m_2=1, \, \ldots , \, 5430$ and another with $m_2=5431, \, \ldots , \, 9917$. 
}
\label{Fig:ClusterAnalysis80_layer2_C9}
\end{figure}
The limit cycle in figure~\ref{Fig:ClusterAnalysis80_layer2_C9}(a) has been divided into $K_2=10$ subclusters according to algorithm~\ref{alg:KmeansChildLayer}. The centroids are distributed on the limit cycle at equal distances. The arrows between the centroids form a closed loop, which results from the periodic nature of the oscillating dynamics. The resulting centroids are phase-averaged flow fields along the complete period of the vortex shedding. The inner jet-flow of the centroids in figure~\ref{Fig:ClusterAnalysis80_layer2_C9}(a) are all deflected upwards, as expected for the attractor that belongs to cluster $\mathcal{C}_{9}$.

Figure~\ref{Fig:ClusterAnalysis80_layer2_C9}(b) associates the cluster labels to the dynamics. 
Two different transient trajectories reach $\mathcal{C}_{9}$ with entering snapshots $m = 9571$ and $45514$, one issued from $\mathcal{C}_{8}$, the other from $\mathcal{C}_{16}$, according to the network model in layer $\mathcal{L}_{1}$.
According to the entering time, the original snapshot index $m = m_2 + 9571 - 1 = 9571, \, \ldots , \, 15000$ and $m = m_2 + 45514 - 5431 = 45514, \, \ldots , \, 50000$.
The flow periodically travels along the subclusters $\mathcal{C}_{9,\, k_2},\ k_2 = 1, \ldots, 10$ as $\mathcal{C}_{9,\,1} \rightarrow \ldots \rightarrow \mathcal{C}_{9,\,10} \rightarrow \mathcal{C}_{9,\,1}$.
The limit cycle has a clear and stable transition matrix, as each cluster only has one possible destination, see figure~\ref{Fig:ClusterAnalysis80_layer2_C9}(c). 
The residence times in each clusters are uniform as shown in figure~\ref{Fig:ClusterAnalysis80_layer2_C9}(d). The sum of all residence times is $9.51$, which is close to the real time period $9.50$ of the vortex shedding in the permanent regime computed from the DNS. 

\subsubsection*{The bifurcating state in cluster $\mathcal{C}_{7}$}

Cluster $\mathcal{C}_{7}$ is associated with the symmetric limit cycle. $K_2=10$ clusters are used to classify the snapshots in this cluster.
\begin{figure}
\begin{tabular}{c c}
\begin{minipage}{0.48\textwidth}
\begin{tabular}{c}
\includegraphics[width=.9\linewidth]{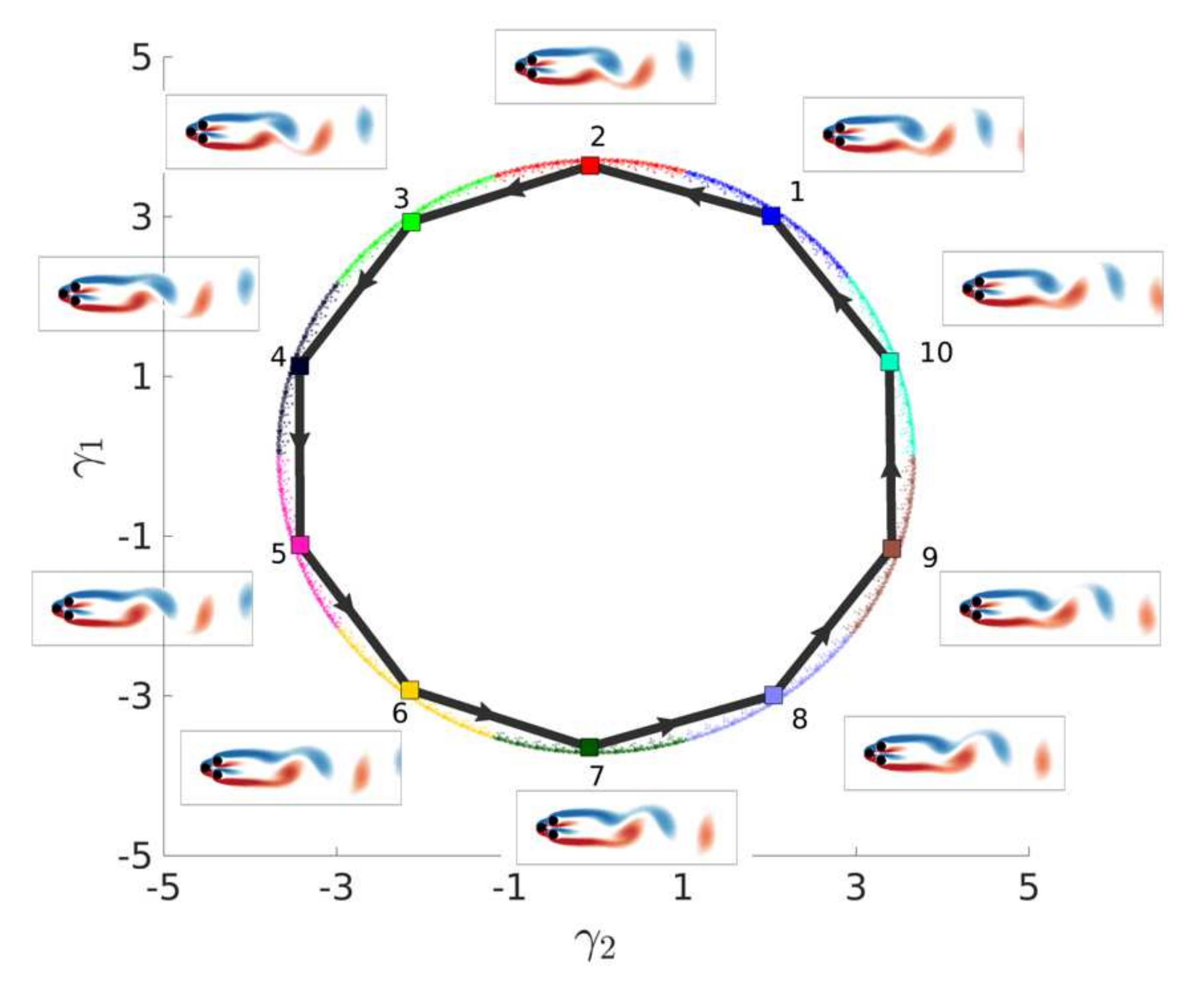}\\
$(a)$ 
\end{tabular} 
\end{minipage} &
\begin{minipage}{0.48\textwidth}
\begin{tabular}{c}
\includegraphics[width=1\linewidth]{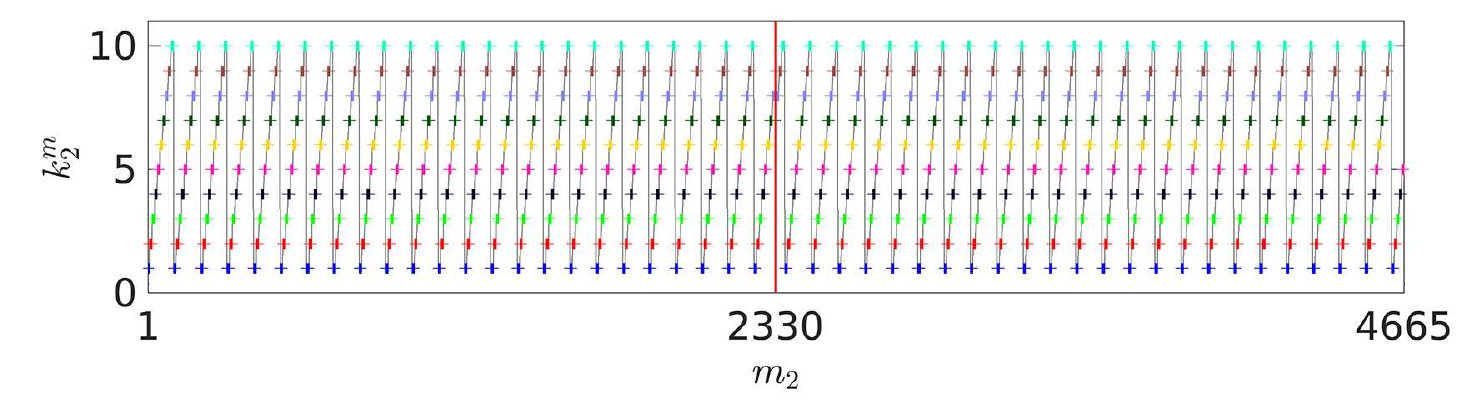} \\
$(b)$ \\
\end{tabular} 
\begin{tabular}{cc}
\includegraphics[width=.49\linewidth]{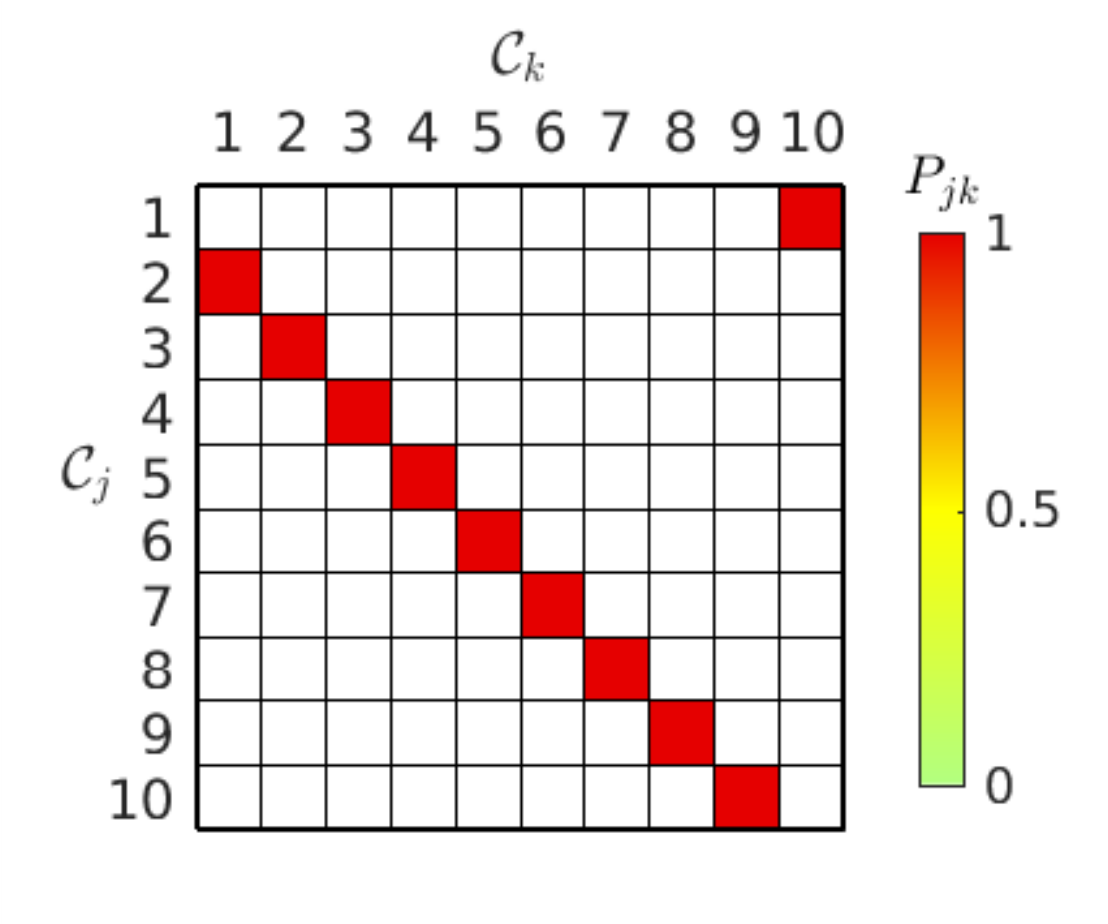} & 
\includegraphics[width=.49\linewidth]{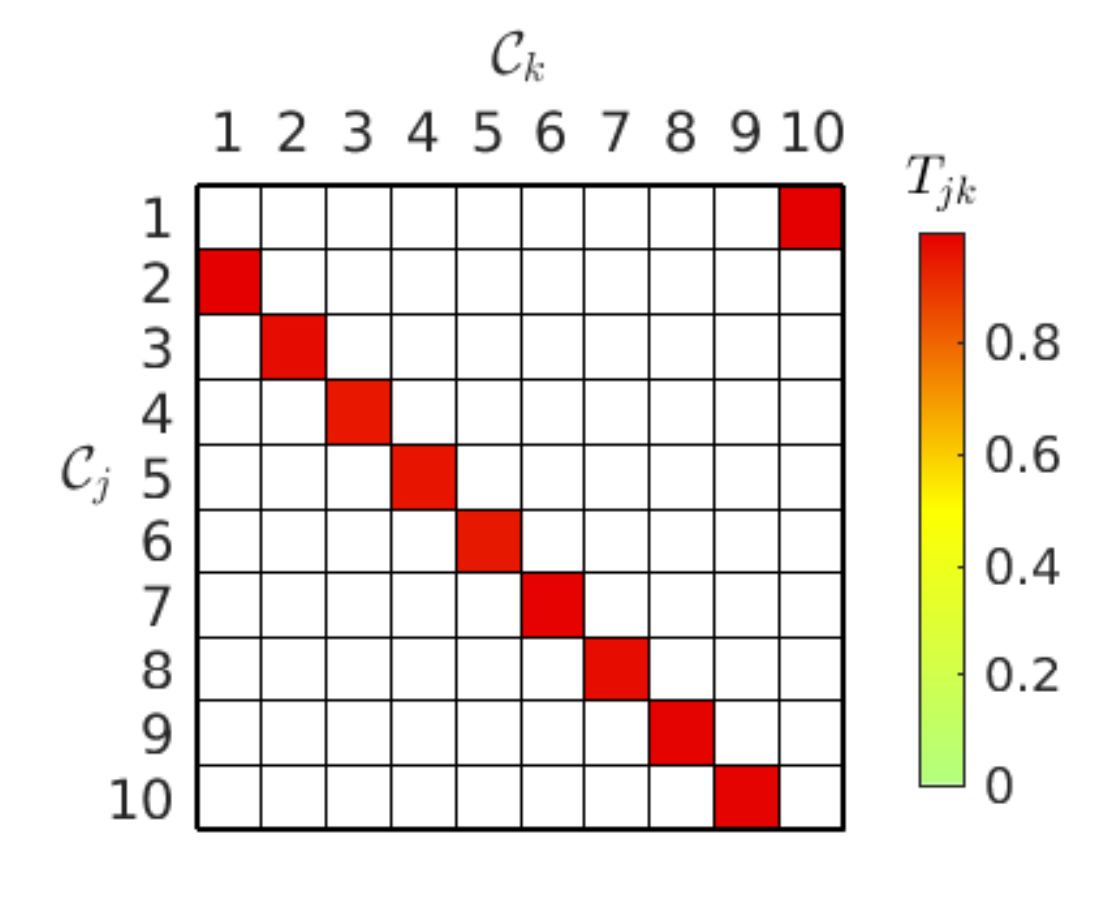} \\
$(c)$ & $(d)$ 
\end{tabular}
\end{minipage}
\end{tabular}
\caption{Cluster-based analysis in layer 2 at $\Rey= 80$ for $\mathcal{C}_{7}$: 
(a) graph of non-trivial transitions between clusters, as in figure~\ref{Fig:ClusterAnalysis80_layer1},
(b) transition illustrated with cluster label, 
(c) transition matrix, 
(d) residence time matrix, as in figure~\ref{Fig:Matrix80_layer1}.
Two trajectories pass through $\mathcal{C}_{7}$ in the parent layer, one with $m_2=1, \, \ldots , \, 2330$ and another with $m_2=2331, \, \ldots , \,  4665$.
}
\label{Fig:ClusterAnalysis80_layer2_C7}
\end{figure}

The resulting limit cycle of figure~\ref{Fig:ClusterAnalysis80_layer2_C7}(a) differs from  the limit cycle of figure~\ref{Fig:ClusterAnalysis80_layer2_C9} by its centroids. 
As expected with the symmetric limit cycle, the inner-jet is not deflected in cluster $\mathcal{C}_{7}$, while it is deflected in cluster $\mathcal{C}_{9}$.
In figure~\ref{Fig:ClusterAnalysis80_layer2_C7}(b), two different transient trajectories pass through $\mathcal{C}_{7}$ with entering snapshots $m = 6541$ and $21552$, the original snapshot index is $m = 6541, \, \ldots , \, 8870$ and $m = 21552, \, \ldots , \, 23886$.
The limit cycle has a stable transition matrix, as shown in figure~\ref{Fig:ClusterAnalysis80_layer2_C7}(c). 
The sum of the residence times of figure~\ref{Fig:ClusterAnalysis80_layer2_C7}(d) is $9.79$, again very close to the period $9.80$ of the symmetric transient vortex shedding computed from the DNS.
The bifurcating dynamics can not be detected with these subclusters but can be captured in the parent layer.

\subsubsection*{The destabilizing regime in cluster $\mathcal{C}_{1}$}

\begin{figure}
\begin{tabular}{c c}
\begin{minipage}{0.48\textwidth}
\begin{tabular}{c}
\includegraphics[width=.9\linewidth]{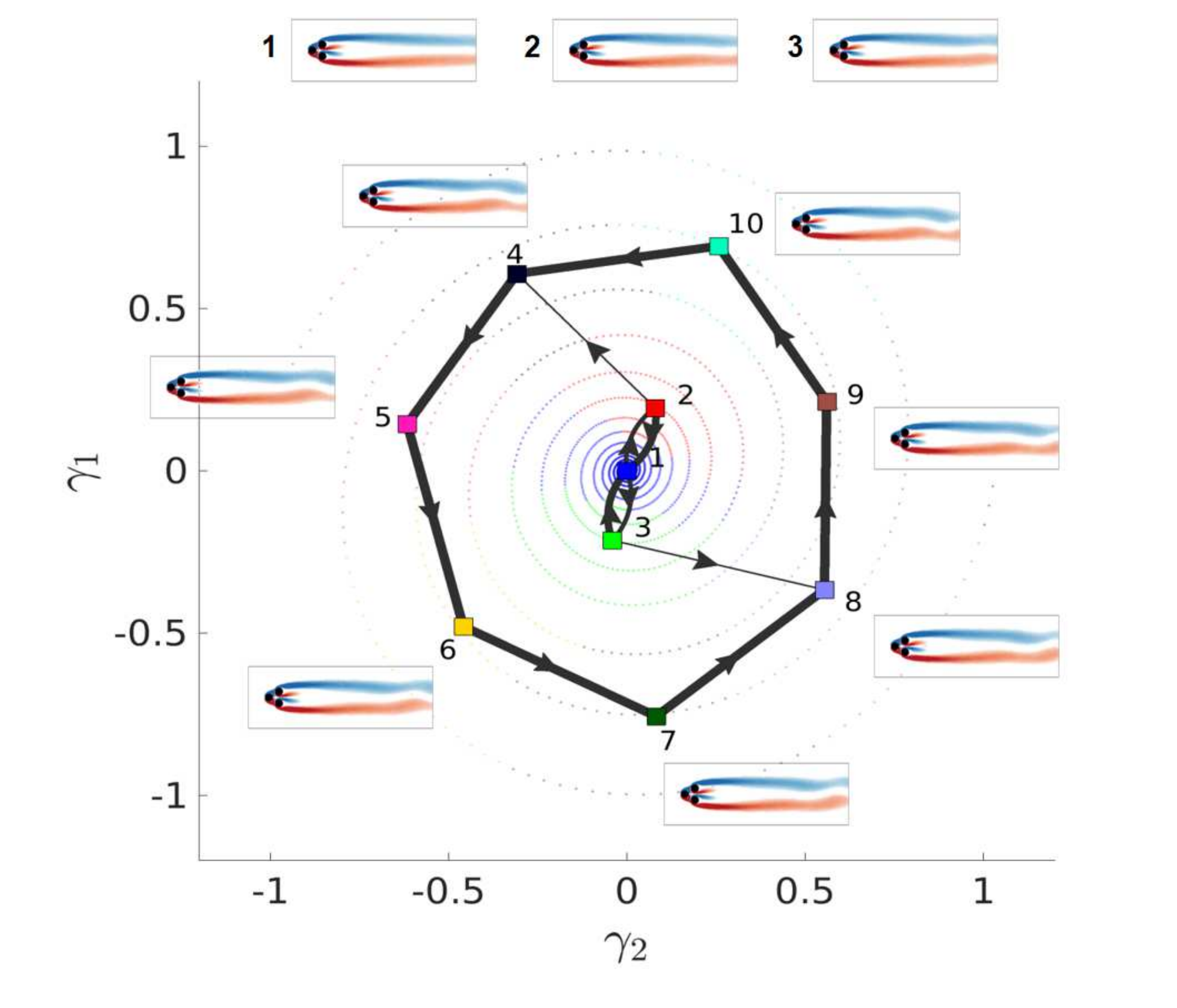}\\
$(a)$ 
\end{tabular} 
\end{minipage} &
\begin{minipage}{0.48\textwidth}
\begin{tabular}{c}
\includegraphics[width=1\linewidth]{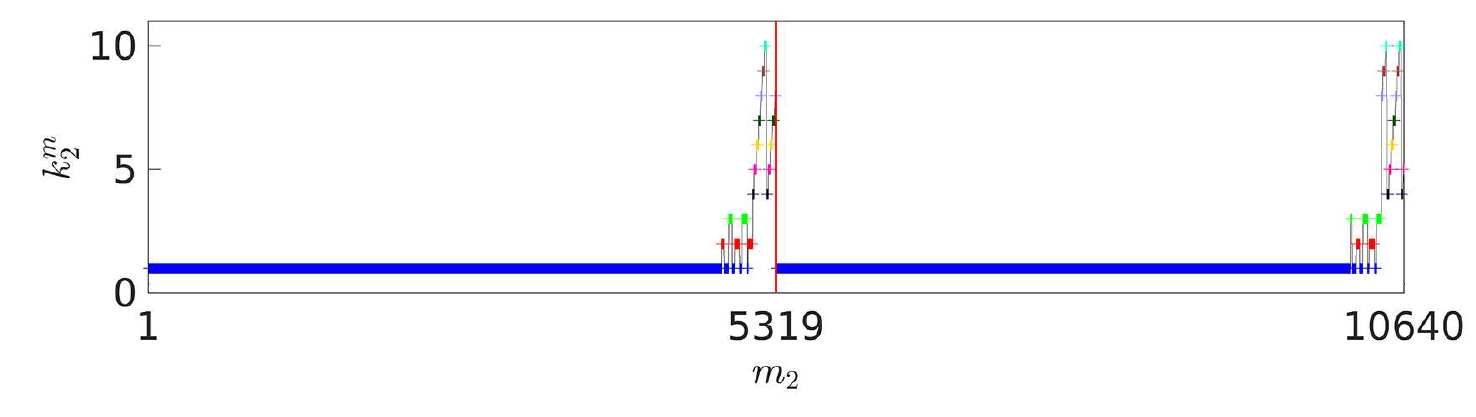} \\
$(b)$ \\
\end{tabular} 
\begin{tabular}{cc}
\includegraphics[width=.49\linewidth]{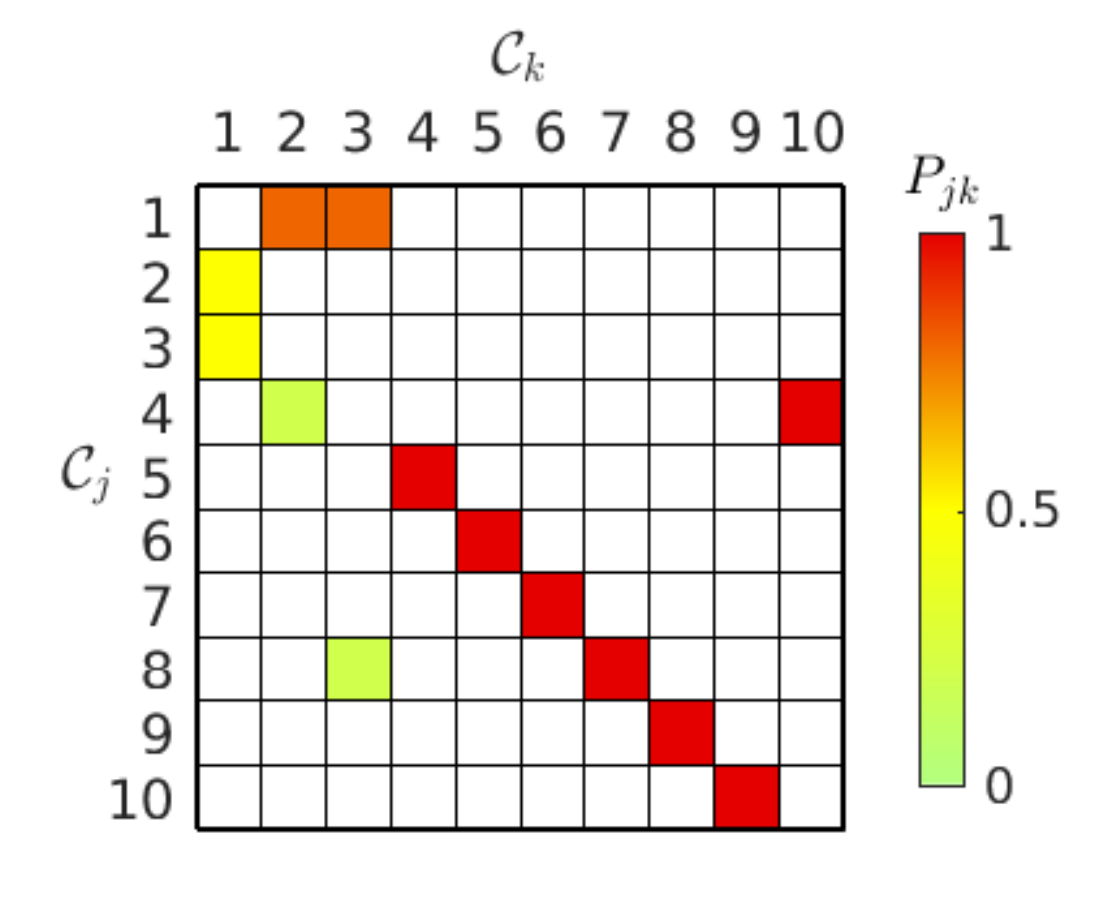} & 
\includegraphics[width=.49\linewidth]{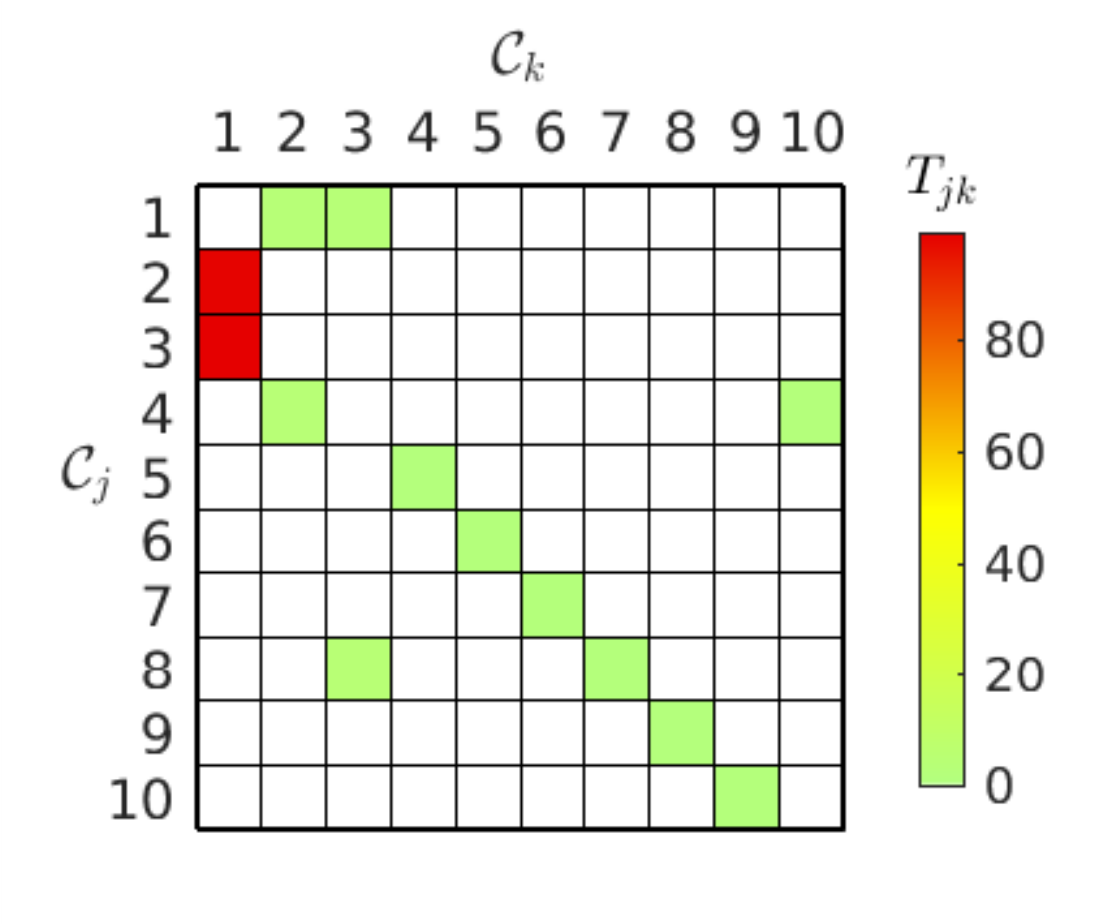} \\
$(c)$ & $(d)$ 
\end{tabular}
\end{minipage}
\end{tabular}
\caption{Cluster-based analysis in layer 2 at $\Rey= 80$ for $\mathcal{C}_{1}$: 
(a) graph of non-trivial transitions between clusters, as in figure~\ref{Fig:ClusterAnalysis80_layer1},
(b) transition illustrated with cluster label, 
(c) transition matrix, 
(d) residence time matrix, as in figure~\ref{Fig:Matrix80_layer1}.
Two trajectories pass through $\mathcal{C}_{1}$ in the parent layer, one with $m_2=1, \, \ldots , \, 5319$ and another with $m_2=5320, \, \ldots , \, 10640$.
}
\label{Fig:ClusterAnalysis80_layer2_C1}
\end{figure}

Cluster $\mathcal{C}_{1}$ is associated with two mirror-conjugated trajectories spiraling out of the symmetric steady solution. $K_2=10$ subclusters are used in the child layer $\mathcal{L}_{2}$. In the $[\gamma_1, \gamma_2]^{\rm T}$ representation of figure~\ref{Fig:ClusterAnalysis80_layer2_C1}(a), the centroids are distributed along two diverging trajectories spiraling out of the fixed point $[0,0]^{\rm T}$.  

The first three subclusters $\mathcal{C}_{1,\,1}$, $\mathcal{C}_{1,\,2}$ and $\mathcal{C}_{1,\,3}$ belong to the inner zone of the spirals, where the distribution of snapshots is dense in the $[\gamma_1, \gamma_2]^{\rm T}$ proximity map of figure~\ref{Fig:ClusterAnalysis80_layer2_C1}. As a result, three nonphysical closed-loop cycles are formed between these three clusters, namely 
$\mathcal{C}_{1,\,1} \rightarrow \mathcal{C}_{1,\,2} \rightarrow \mathcal{C}_{1,\,1}$, 
$\mathcal{C}_{1,\,1} \rightarrow \mathcal{C}_{1,\,3} \rightarrow \mathcal{C}_{1,\,1}$, and
$\mathcal{C}_{1,\,3} \rightarrow \mathcal{C}_{1,\,1} \rightarrow \mathcal{C}_{1,\,2} \rightarrow \mathcal{C}_{1,\,1} \rightarrow \mathcal{C}_{1,\,3}$.
The remaining clusters belong to the outer arms of the spirals, with a relatively sparse distribution of the snapshots. The transitions between them form a closed-loop trajectory $\mathcal{C}_{1,\,4} \rightarrow \ldots \rightarrow \mathcal{C}_{1,\,10} \rightarrow \mathcal{C}_{1,\,4}$. 
The transitions $\mathcal{C}_{1,\,2} \rightarrow \mathcal{C}_{1,\,4}$ and $\mathcal{C}_{1,\,3} \rightarrow \mathcal{C}_{1,\,8}$ correspond to the departing dynamics out of the inner zone, due to the growth of the instability. 
The varying density of distribution comes from the exponential growth of the instability. 
The flow perturbations are small in the beginning of the instability while the flow distortions evolve faster in the later stages, leading to a multiscale problem in the transient and post-transient flow dynamics. In this case, the later stages with larger distortions are obviously easier to divide into different clusters.

From figure~\ref{Fig:ClusterAnalysis80_layer2_C1}(b), two transient trajectories pass through $\mathcal{C}_{1}$ with entering snapshots $m = 1$ and $15001$. The original snapshot index is $m = 1, \, \ldots , \, 5319$ and $m = 15001, \, \ldots , \, 20321$, corresponding to the initial stage of destabilization from the symmetric steady solution.
In the transition matrix of figure~\ref{Fig:ClusterAnalysis80_layer2_C1}(c), the inner and outer portions of the spiral are also apparent. 
The inner zone is the oscillating dynamics between $\mathcal{C}_{1,\,1}$ and $\mathcal{C}_{1,\,2}$, or $\mathcal{C}_{1,\,1}$ and $\mathcal{C}_{1,\,3}$. 
The outer zone has a more obvious periodic dynamics through the remaining clusters from $\mathcal{C}_{1,\,4}$ to $\mathcal{C}_{1,\,10}$.
In figure~\ref{Fig:ClusterAnalysis80_layer2_C1}(d), the residence time in each cluster is very short compared to the residence time in $\mathcal {C}_{1,\,1}$, which the vicinity of the steady solution belong to.

\subsubsection{Dynamics reconstruction of the hierarchical network model at $\Rey=80$}
\label{SubSec:HNM80_Validation}

\begin{figure}
\centerline{
\begin{tabular}{cccc}
(a) & \raisebox{-0.5\height}{\includegraphics[width=.45\linewidth]{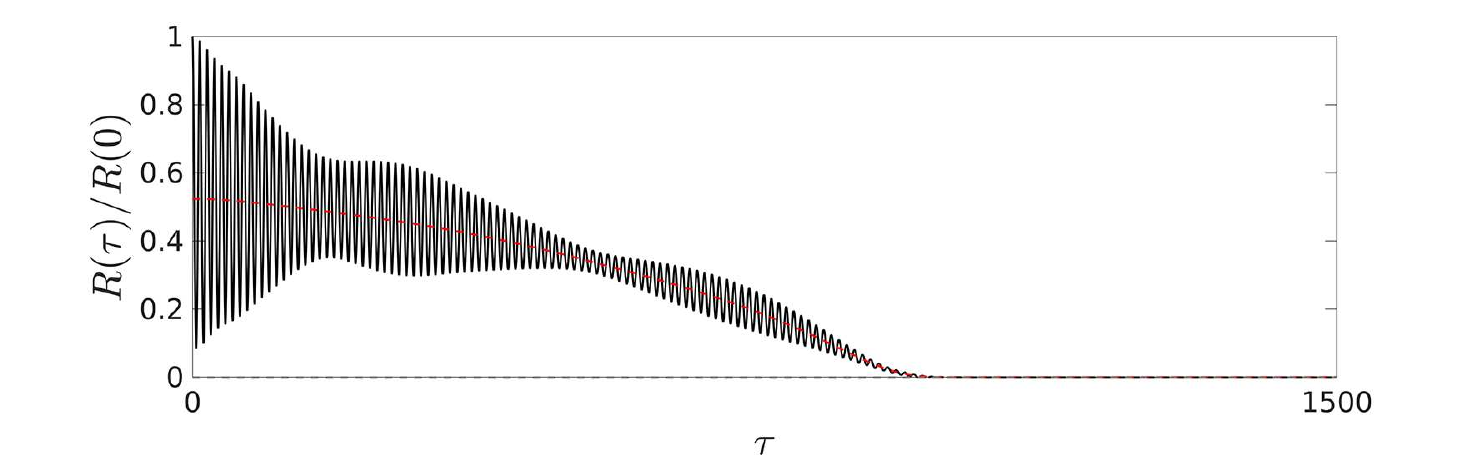}} &
(b) & \raisebox{-0.5\height}{\includegraphics[width=.45\linewidth]{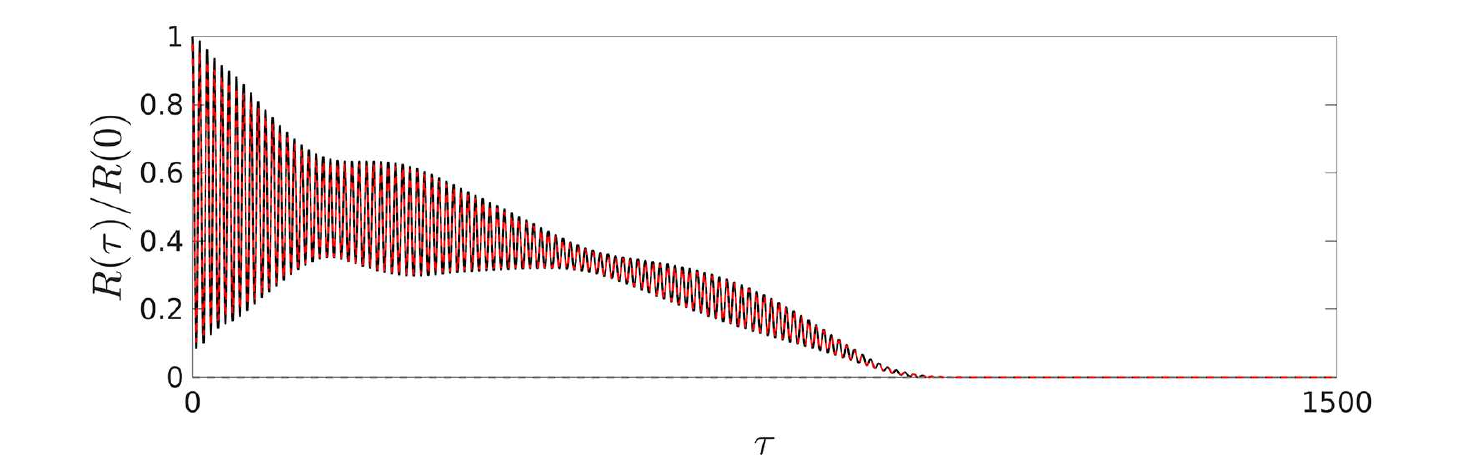}} 
\end{tabular}
}
\caption{Autocorrelation function for  $\tau \in [0, 1500)$ from DNS (black solid line) and the hierarchical network model (red dashed line) in the two layers: (a) $\mathcal{L}_{1}$ and (b) $\mathcal{L}_{2}$, at $\Rey=80$}
\label{Fig:Validation80}
\end{figure}

Figure~\ref{Fig:Validation80} shows the autocorrelation function of the DNS and the HiCNM in different layers.
As the autocorrelation function has been normalized by $R(0)$, the unit one presents the level of  kinetic energy of the whole transition.   
For the transient and post-transient dynamics, the autocorrelation function vanishes with increasingtime shift, as shown in figure~\ref{Fig:Validation80}(b).
The autocorrelation function of the DNS identifies the dominant frequency.
In layer $\mathcal{L}_{1}$, no oscillation can be identified, due to the centroids by averaging the snapshots within clusters in the state space. The RMSE of the autocorrelation function is $R_{\rm rms}^1= 17.46$.
In layer $\mathcal{L}_{2}$, the autocorrelation function of the model matches perfectly over the entire range, with $R_{\rm rms}^2=1.18$, which quantifies the accuracy of the cluster-beasd model.

\subsection{Advantages of HiCNM as compared to CNM}
\label{Sec:CNM80andHiCNM80}

In this sub-section, we compare the results of the HiCNM in its second layer to the standard CNM with the same number of clusters. The hierarchical structure can systematically present the global trend and local dynamics, which improve the graphic interpretation of transient and post-transient, multi-frequency, multi-attractor behaviours.
\begin{figure}
\centering
$(a)$ \raisebox{-0.6\height}{\includegraphics[width=.7\linewidth]{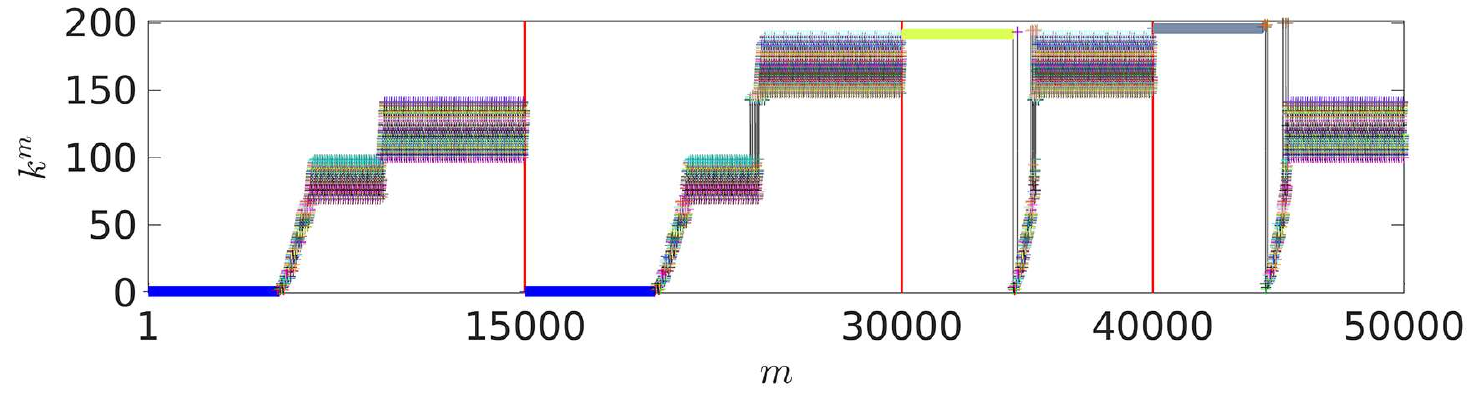}} \\
$(b)$ \raisebox{-0.6\height}{\includegraphics[width=.7\linewidth]{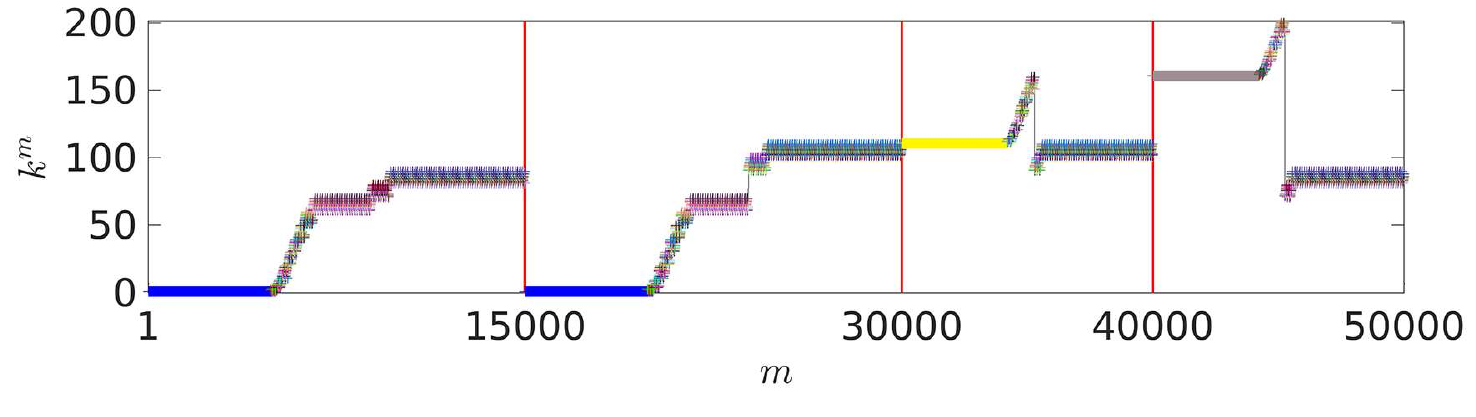}} \\
\begin{tabular}{cc}
\raisebox{-0.5\height}{\includegraphics[width=.49\linewidth]{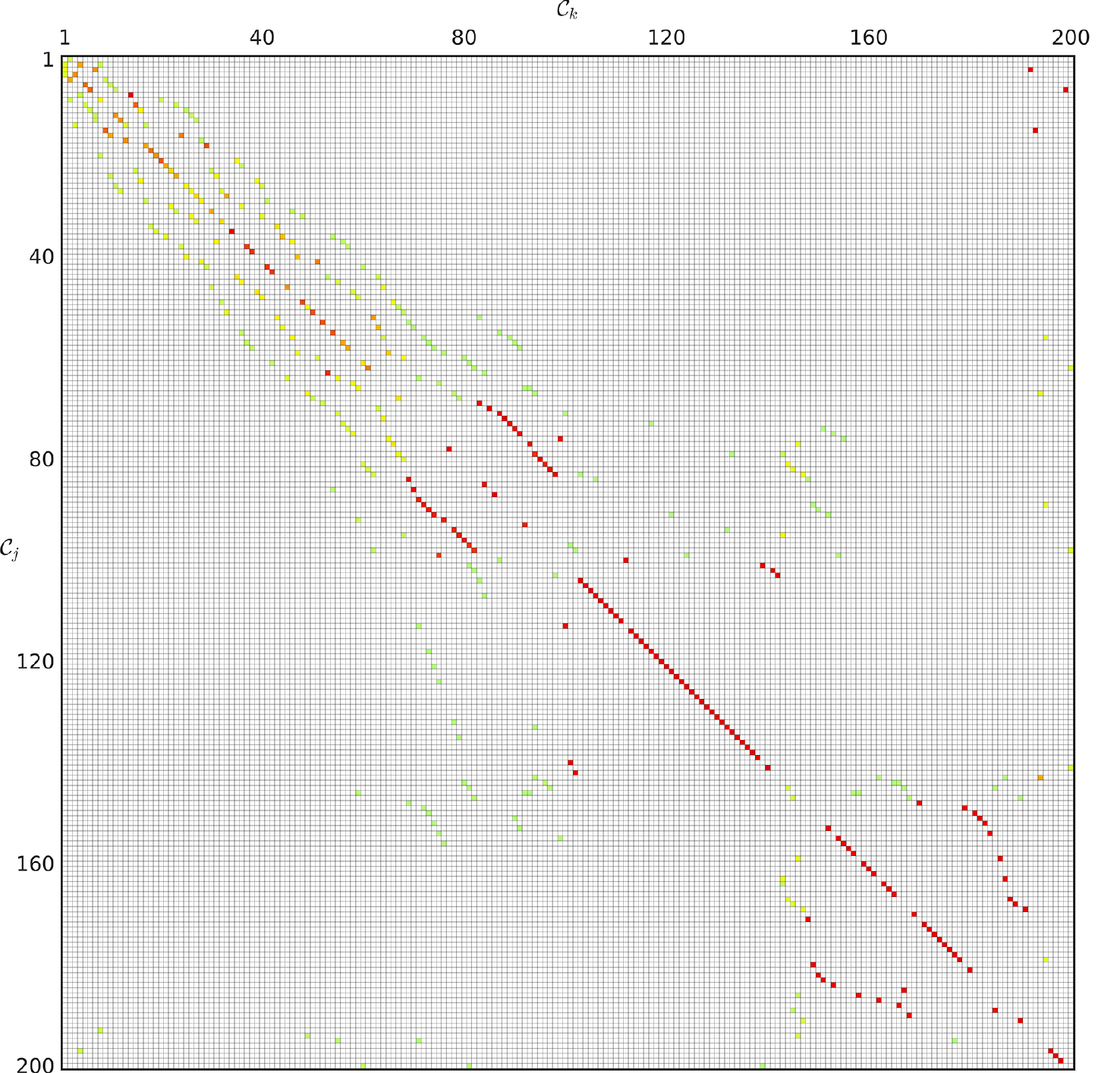}} &
\raisebox{-0.5\height}{\includegraphics[width=.49\linewidth]{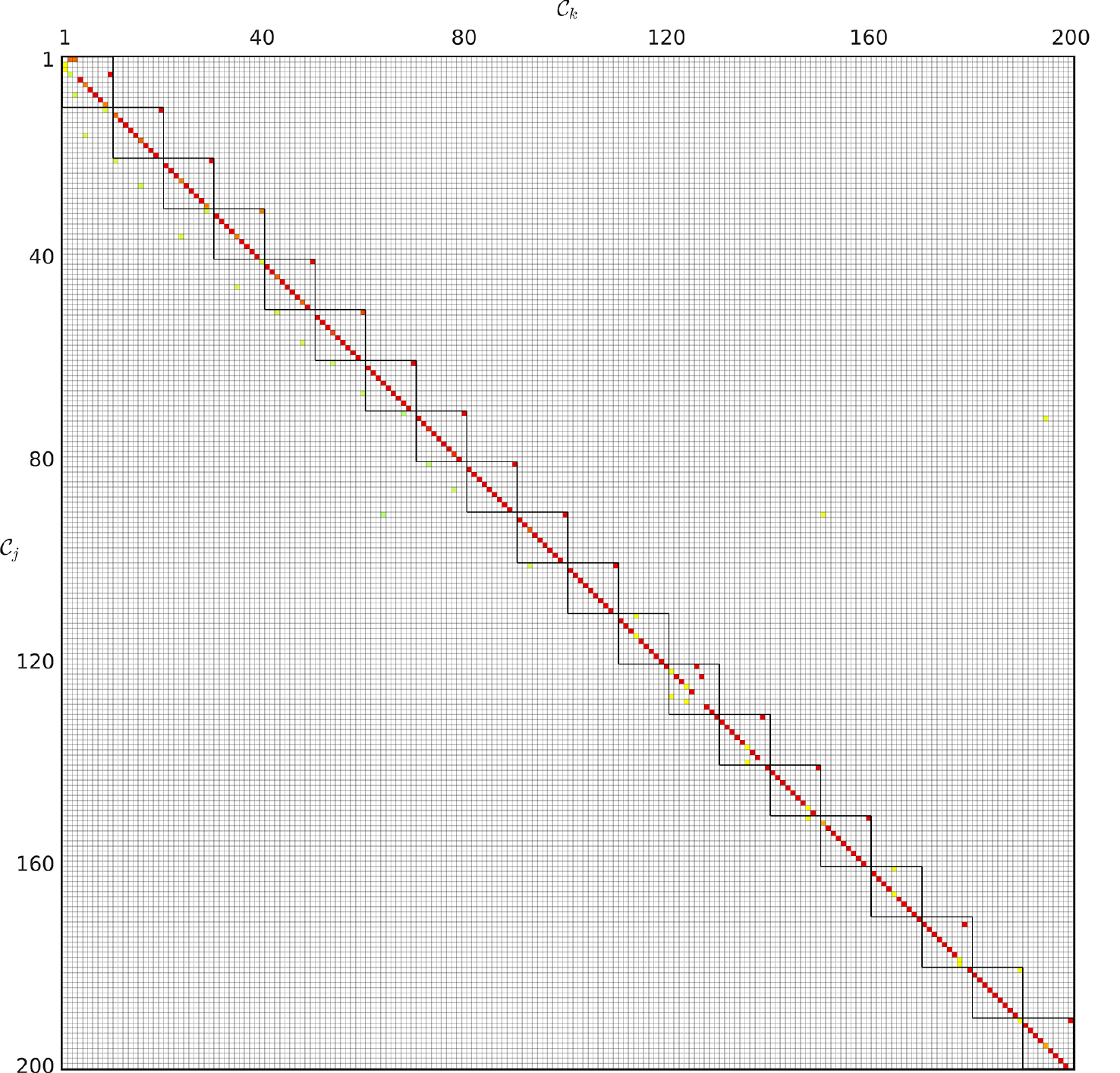}} \\
$(c)$ & $(d)$ \\
\end{tabular}
\caption{Cluster-based analysis at $\Rey= 80$ for the fluctuating flow: transition illustrated with cluster label and transition matrix of the standard network model (a, c) and the hierarchical network model in the second layer (b, d), as in figure~\ref{Fig:Matrix80_layer1}.}
\label{Fig:Matrix80_CNM}
\end{figure}

\subsubsection{Standard CNM at $\Rey= 80$}
\label{Sec:CNM80}
We show a standard network model treating the transient dynamics at $\Rey=80$ with $K=200$ clusters.
The directed graph is not shown here as too many clusters overlapped in the two-dimensional subspace $[\gamma_1, \gamma_2]^{\rm T}$, losing the interpretation of the dynamics.
In figure~\ref{Fig:Matrix80_CNM}(a), the six exact solutions are well classified. The three steady solutions are divided into $\mathcal{C}_1$, $\mathcal{C}_{192}$ and $\mathcal{C}_{196}$ separately.
The three limit cycles are identified with three blocks of oscillating labels, from $\mathcal{C}_{69}$ to $\mathcal{C}_{99}$, from $\mathcal{C}_{100}$ to $\mathcal{C}_{142}$, and from $\mathcal{C}_{148}$ to $\mathcal{C}_{191}$.
However, the transient trajectories starting with the asymmetric steady solutions are misidentified, as both travelling through the same clusters from $\mathcal{C}_{2}$ to $\mathcal{C}_{68}$ and even the block of the symmetric limit cycle.
It indicates that the clustering algorithm failed to distinguish the symmetry-breaking in the transient dynamics starting with the different steady solutions.
Too many clusters also make the transition matrix hard to read as illustrated in figure~\ref{Fig:Matrix80_CNM}(c), and also for the residence time matrix not shown here.
Ignoring the transitions with low probability in the transition matrix, the transitions in the above-mentioned three blocks are almost definite, indicating the correct identification of the cycles. 
However, the transient dynamics from $\mathcal{C}_{2}$ to $\mathcal{C}_{68}$ is random. We can hardly find any relevant feature for the building of vortex shedding or symmetric breaking. The cluster distribution directly affects the analysis of the dynamics.  

In summary, a large number of clusters tends to increase the resolution of the identified network model, but will misidentify the dynamics with the random transitions between clusters during the transient state. 
The standard network model fails to describe the transient dynamics between different invariant sets.
This kind of problem comes from the poor distribution of clusters and can be solved by the hierarchical clustering strategy, as shown in the following sub-section.

\subsubsection{HiCNM at $\Rey= 80$ in Layer 2}
\label{Sec:HiCNM80_Regroup}

Based on the hierarchical network model at $\Rey=80$ recorded in \S~\ref{Sec:HNM80}, we build a network model ensembling all the clusters in the layer $\mathcal{L}_2$, which also contains $K=200$ clusters.
For ease of illustration, the cluster indexes of two layers $\mathcal{C}_{k_1, k_2}$ are denoted with a single index $\mathcal{C}_{k}$, with $k=1,\,\ldots,\,200$.

From figure~\ref{Fig:Matrix80_CNM}(b),
we found that the cluster distribution is very uniform during the transient states and the post-transient states.
The six exact solutions are well classified, and much fewer clusters are used for the limit cycles.
The three steady solutions being divided into $\mathcal{C}_1$, $\mathcal{C}_{111}$ and $\mathcal{C}_{161}$ separately. The three limit cycles are identified with three blocks of oscillating labels, from $\mathcal{C}_{61}$ to $\mathcal{C}_{70}$, from $\mathcal{C}_{81}$ to $\mathcal{C}_{90}$, and from $\mathcal{C}_{101}$ to $\mathcal{C}_{110}$.
Even the transient states between the limit cycles can also be identified, with two blocks from $\mathcal{C}_{71}$ to $\mathcal{C}_{80}$, and from $\mathcal{C}_{91}$ to $\mathcal{C}_{100}$.
The transient trajectories starting with three different steady solutions are also well separated.
The transition matrix in figure~\ref{Fig:Matrix80_CNM}(d) are almost full of definite transitions, which shows clearer transient dynamics than the figure~\ref{Fig:Matrix80_CNM}(c).
The global matrix keeps the local dynamics for each cluster $\mathcal{C}_{k_1}$ in the first layer $\mathcal{L}_{1}$ , as shown in each small block of 10 clusters, and ensembles them together.
The transitions from block to block have a much lower probability, comparing with the cycling transitions within the block, which makes them hard to see in the figure.
However, this kind of transition between the blocks should be also definite, as shown in \S~\ref{SubSec:HNM80_L1} for the hierarchical model in the first layer.
Analogously, the multiscale problem of the dynamics in different layers also exists in the global residence time matrix, which makes the small scale terms unseeable.
Therefore, we suggest analyzing the slow-varying mean flow and the local dynamics separately in different layers, as in \S~\ref{Sec:HNM80}, to avoid the influence of different scales.

\subsection{Hierarchical network model for the quasi-periodic dynamics at \Rey=105}
\label{Sec:HNM105}

At $\Rey=105$, the flow dynamics is quasi-periodic \citep{deng2020jfm}. 
The inner jet oscillations are modulated at a non-commensurate low frequency. The flow dynamics considered in the first cluster layer is low-pass filtered (\S~\ref{SubSec:HNM105_L1}). 
The basic limit cycle associated with the dominant frequency $\omega_c$ of the vortex shedding is clustered and analysed in layer 2 (\S~\ref{SubSec:HNM105_L2}). 
The low-frequency modulations of the vortex shedding are further described in layer 3 (\S~\ref{SubSec:HNM105_L3}).

\subsubsection{Hierarchical network model in Layer 1}
\label{SubSec:HNM105_L1}
\begin{figure}
\centering
\includegraphics[width=.65\linewidth]{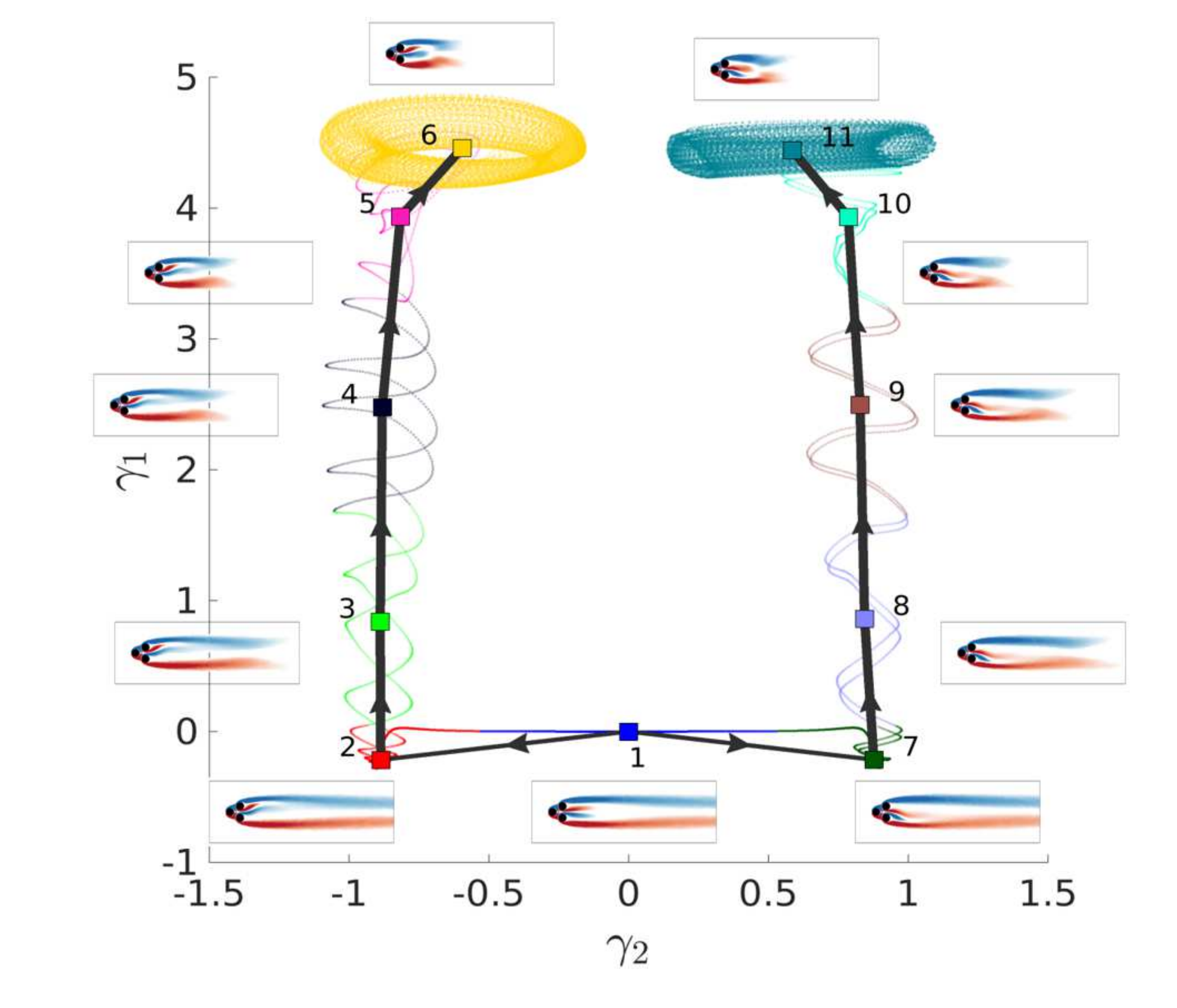} 
\caption{Graph of transitions between clusters in layer 1 at $\Rey= 105$, displayed as in figure~\ref{Fig:ClusterAnalysis80_layer1}.}
\label{Fig:ClusterAnalysis105_layer1}
\end{figure}

The frequency of the coherent component $\tilde{\bm{u}}$ is $f_c = 0.1172$.  
The $K_1=11$ clusters are used on the low-pass filtered data set, following algorithm~\ref{alg:FilteredKmeans}.
The non-trivial transitions are shown in the two-dimensional subspace $[\gamma_1, \gamma_2]^{\rm T}$ of figure~\ref{Fig:ClusterAnalysis105_layer1}.

The two attractors belong to clusters $\mathcal{C}_6$ and $\mathcal{C}_{11}$. 
The symmetric and asymmetric steady solutions belong respectively to clusters $\mathcal{C}_1$, $\mathcal{C}_2$ and $\mathcal{C}_{7}$.
The transient dynamics observed at $\Rey=105$ are different from those identified in figure~\ref{Fig:ClusterAnalysis80_layer1} for $\Rey=80$.
Figure~\ref{Fig:ClusterAnalysis105_layer1} shows four trajectories, initiated from mirror-conjugated initial conditions close to the symmetric and asymmetric steady solutions: \begin{enumerate}[${Trajectory\, }$1 :]
\item \ $\mathcal{C}_1\ (\bm{u}_s) \rightarrow \mathcal{C}_2 (\bm{u}_s^+) \rightarrow \ldots \rightarrow \mathcal{C}_6 (\bar{\bm{u}}^+)$;
\item \ $\mathcal{C}_1\ (\bm{u}_s) \rightarrow \mathcal{C}_7 (\bm{u}_s^-) \rightarrow \ldots\rightarrow \mathcal{C}_{11} (\bar{\bm{u}}^-)$;
\item \ $\mathcal{C}_2 (\bm{u}_s^+) \rightarrow \ldots \rightarrow \mathcal{C}_6 (\bar{\bm{u}}^+)$;
\item \ $\mathcal{C}_7 (\bm{u}_s^-) \rightarrow \ldots\rightarrow \mathcal{C}_{11} (\bar{\bm{u}}^-)$.
\end{enumerate}
The state trajectories start from the symmetric steady solution to one of the two asymmetric steady solutions, then converge to the corresponding attracting torus.
The two trajectories on the left side have a significant phase delay, while the two on the right side are almost in the same phase. 
This phase difference is a random function that depends on the initial condition.
The two tori also look different because the feature vectors associated with $[\gamma_1, \gamma_2]^{\rm T}$ are asymmetrical from the multidimensional scaling.

\begin{figure}
\centering
$(a)$ \raisebox{-0.6\height}{\includegraphics[width=.65\linewidth]{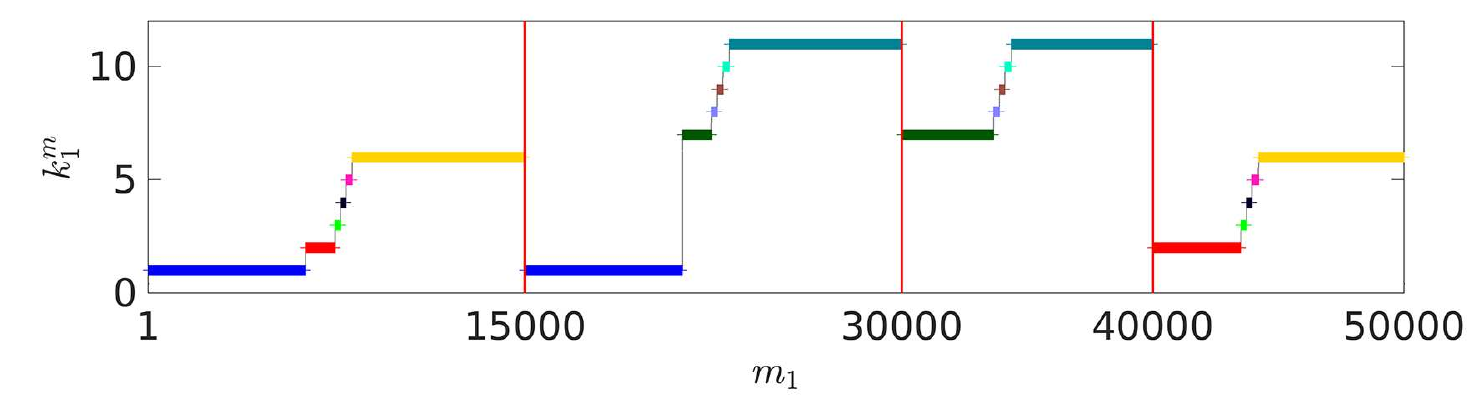}} \\
\begin{tabular}{cc}
$(b)$ \raisebox{-0.5\height}{\includegraphics[width=.31\linewidth]{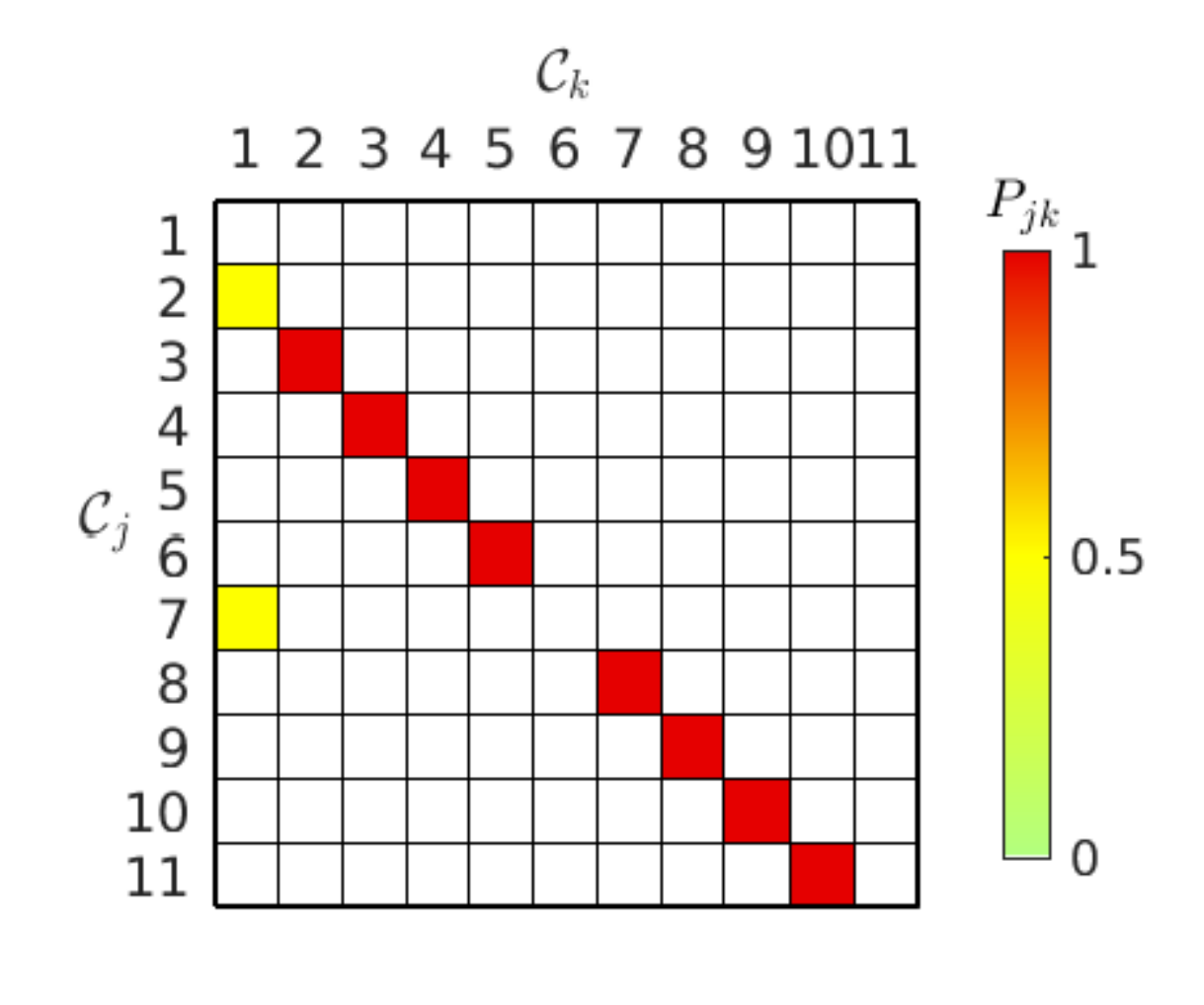}} &
$(c)$ \raisebox{-0.5\height}{\includegraphics[width=.31\linewidth]{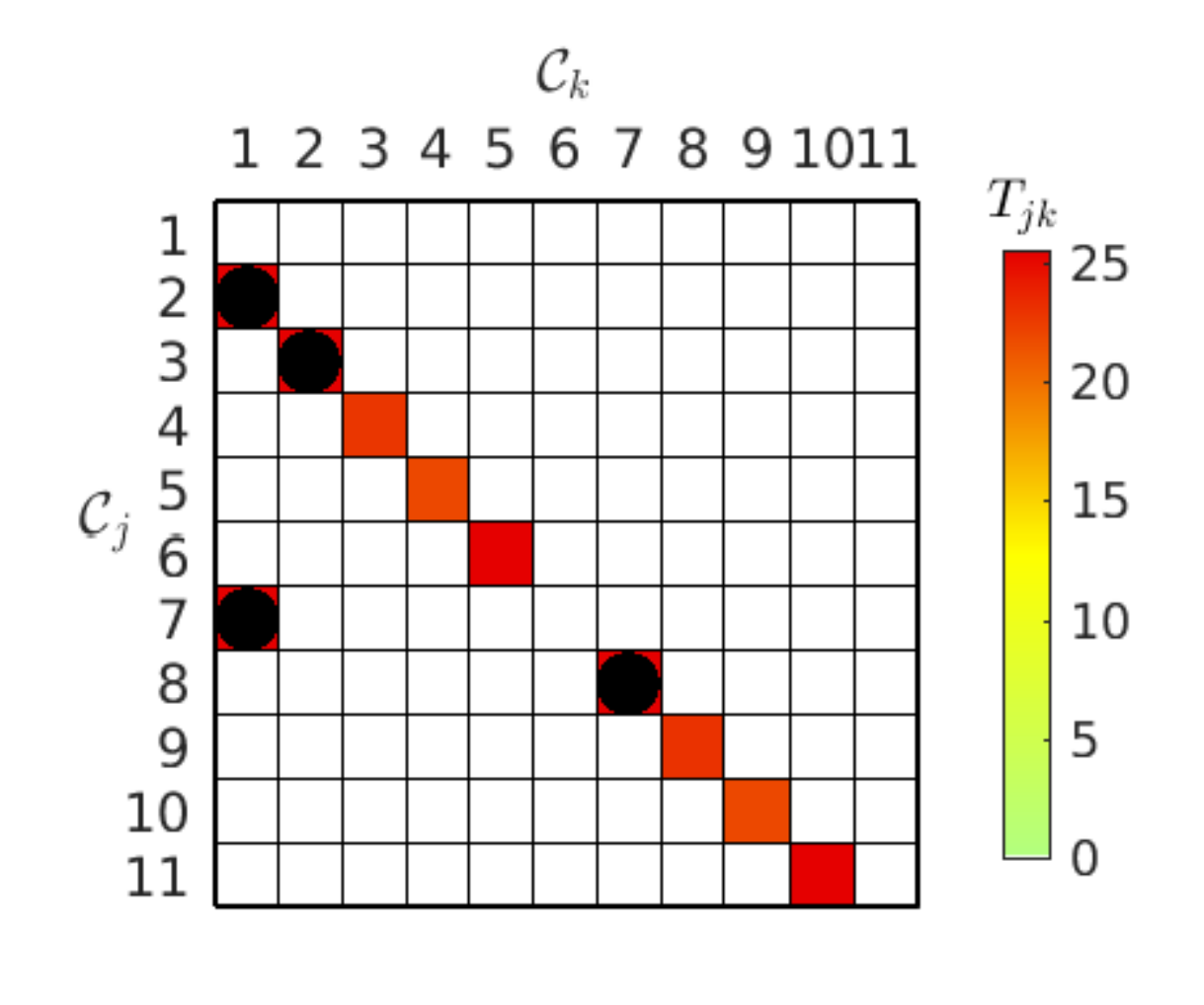}} \\
\end{tabular}
\caption{Cluster-based analysis at $\Rey= 105$ in layer 1:
 (a) transition illustrated with cluster label, 
(b) transition matrix, 
(c) residence time matrix, displayed as in figure~\ref{Fig:Matrix80_layer1}.
}
\label{Fig:Matrix105_layer1}
\end{figure}

In figure~\ref{Fig:Matrix105_layer1}(a), the evolution of the cluster label of the snapshots illustrates four irreversible transient dynamics. The transition matrix in figure~\ref{Fig:Matrix105_layer1}(b) exhibits two red diagonals associated with trajectories to the final state, and two yellow elements associated with the initial state starting close to the symmetric steady solutions in cluster $\mathcal{C}_1$.
The transition from  $\mathcal{C}_1$ to clusters $\mathcal{C}_{2}$ and $\mathcal{C}_{7}$ each have 1/2 probability.
The two attracting clusters $\mathcal{C}_{6}$ and $\mathcal{C}_{11}$, associated with the attractors of the system, have no transition to any other clusters, as all the terms in the sixth and eleventh columns $P_{j\, 6} = P_{j\, 11} = 0,\ \forall j$. 
In the residence time matrix of figure~\ref{Fig:Matrix105_layer1}(c), the filled black circles indicates three typical clusters $\mathcal{C}_{1}$, $\mathcal{C}_{2}$ and $\mathcal{C}_{7}$, which correspond to the vicinity of the three steady solutions. All the terms in the sixth and eleventh columns $T_{j\, 6}, T_{j\, 11}$ are 0 $\forall j$, which means that the residence time is infinite, as expected for attractors.  

\subsubsection{Hierarchical network model in Layer 2}
\label{SubSec:HNM105_L2}
We focus on the permanent regime in the cluster $\mathcal{C}_6$, and apply the sub-division clustering algorithm~\ref{alg:KmeansChildLayer}, which results in $K_2=10$ subclusters in layer $\mathcal{L}_{2}$.
\begin{figure}
\begin{tabular}{c c}
\begin{minipage}{0.48\textwidth}
\begin{tabular}{c}
\includegraphics[width=.9\linewidth]{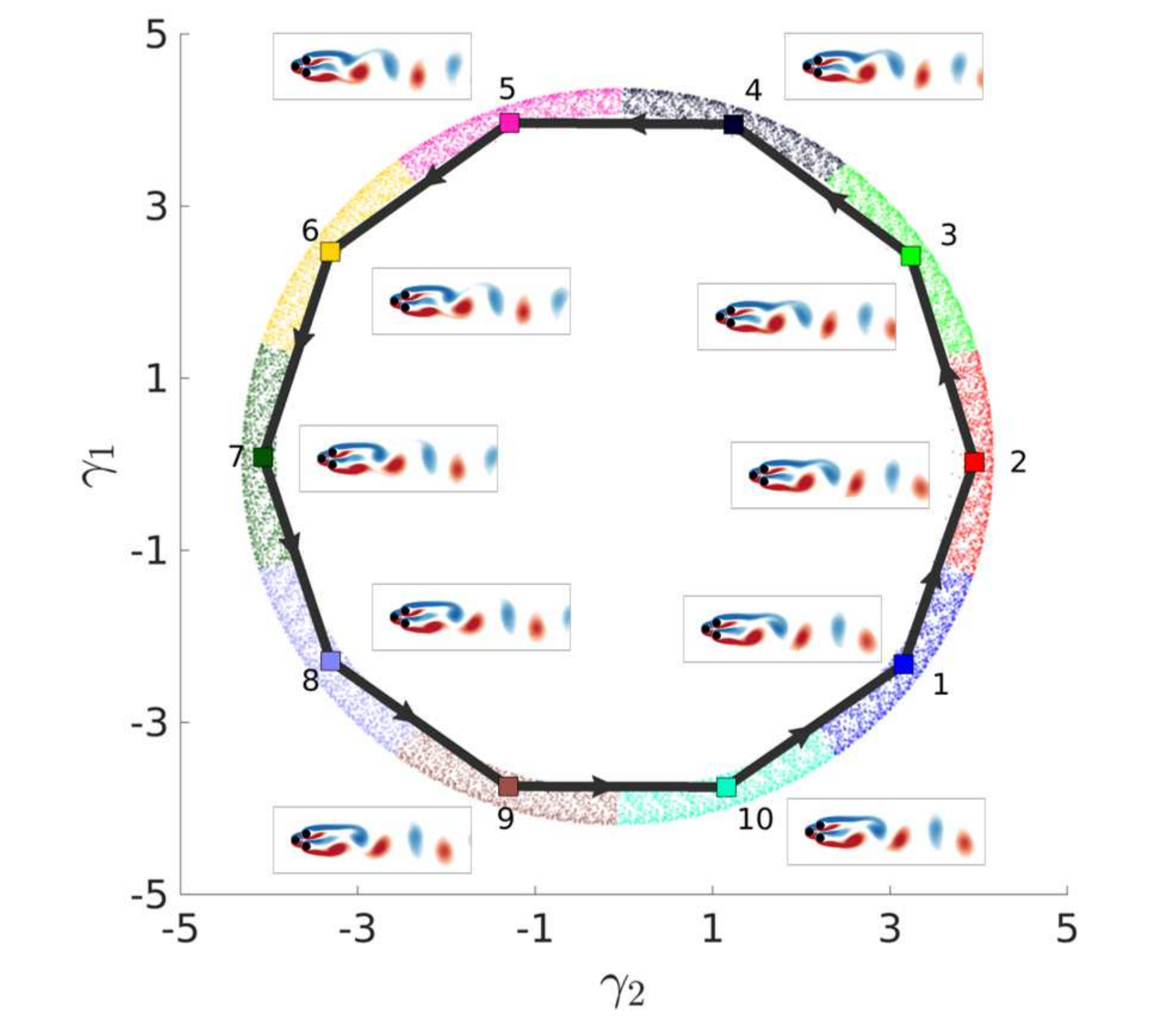}\\
$(a)$ 
\end{tabular} 
\end{minipage} &
\begin{minipage}{0.48\textwidth}
\begin{tabular}{c}
\includegraphics[width=1\linewidth]{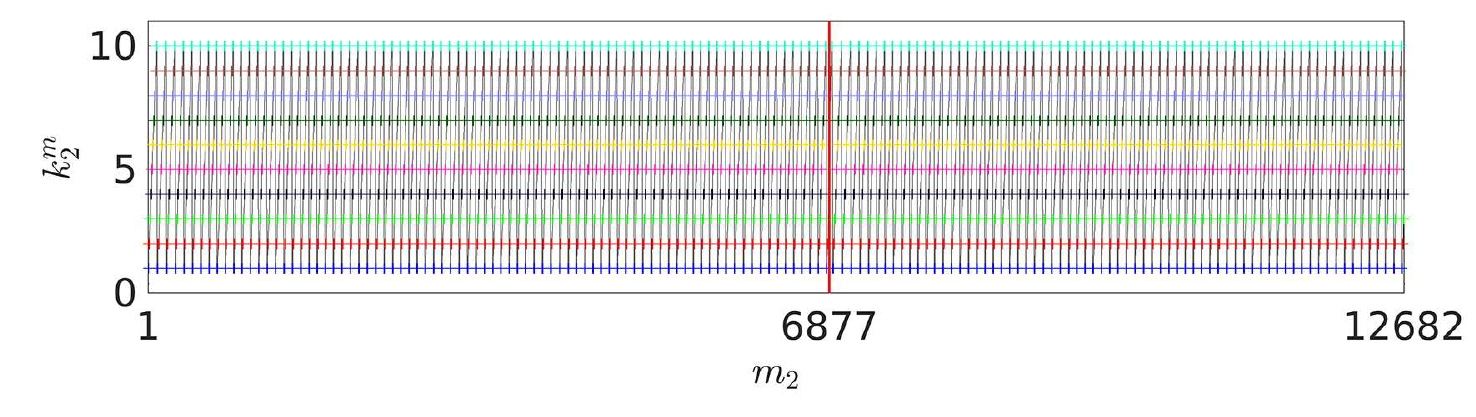} \\
$(b)$ \\
\end{tabular} 
\begin{tabular}{cc}
\includegraphics[width=.49\linewidth]{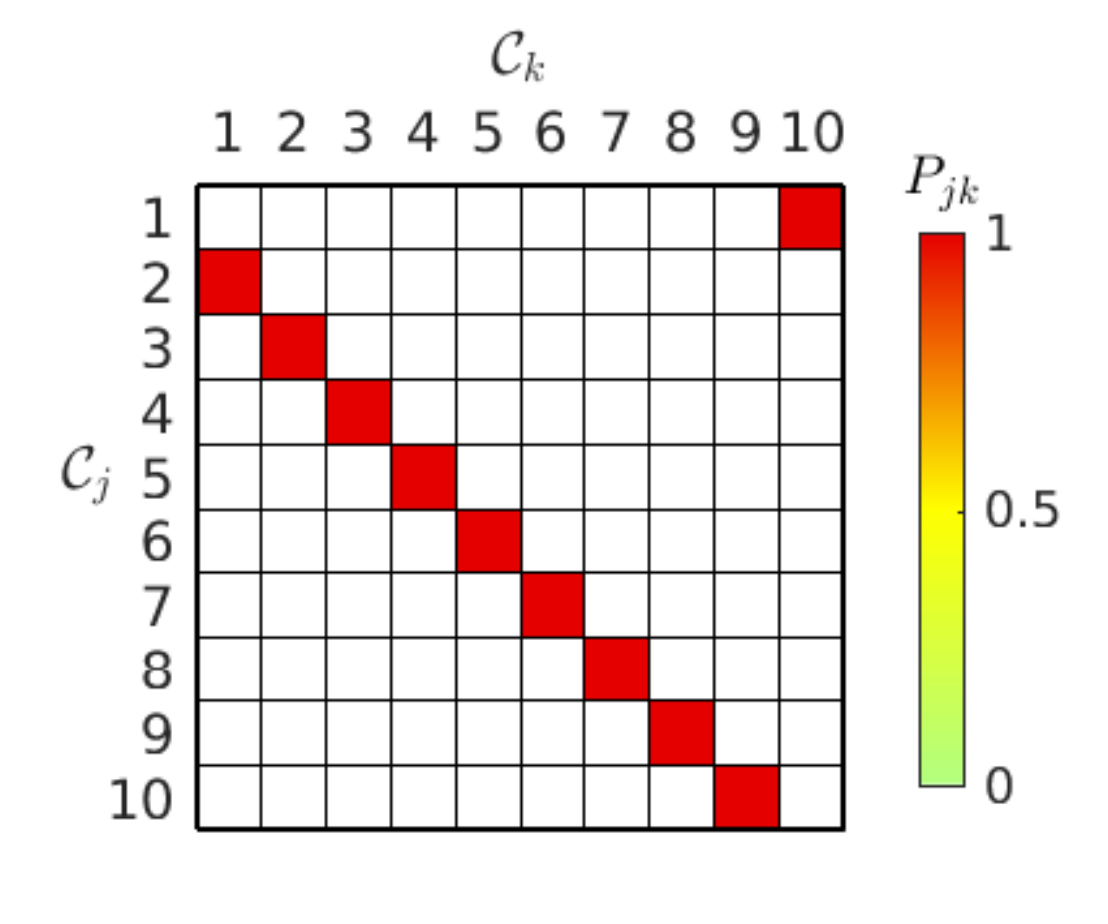} & 
\includegraphics[width=.49\linewidth]{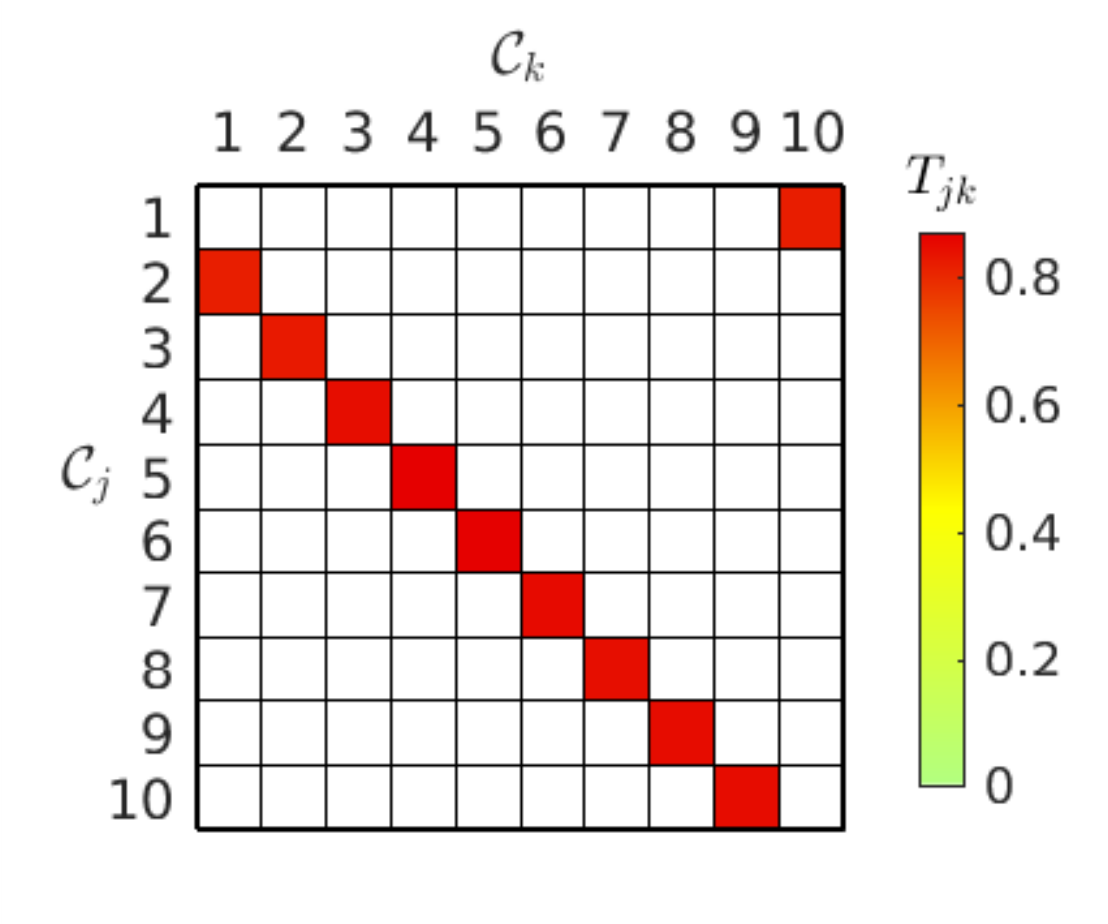} \\
$(c)$ & $(d)$ 
\end{tabular}
\end{minipage}
\end{tabular}
\caption{Cluster-based analysis in layer 2 at $\Rey= 105$ for $\mathcal{C}_{6}$: 
(a) graph of non-trivial transitions between clusters, as in figure~\ref{Fig:ClusterAnalysis80_layer1},
(b) transition illustrated with cluster label, 
(c) transition matrix, 
(d) residence time matrix, as in figure~\ref{Fig:Matrix80_layer1}.
Two trajectories pass through $\mathcal{C}_{6}$ in the parent layer, one with $m_2=1, \, \ldots , \, 6877$ and another with $m_2=6878, \, \ldots , \, 12682$.
}
\label{Fig:ClusterAnalysis105_layer2_C6}
\end{figure}

As illustrated in figure~\ref{Fig:ClusterAnalysis105_layer2_C6}(a), a closed orbit between the 10 clusters is found. 
Figure~\ref{Fig:ClusterAnalysis105_layer2_C6}(b) shows two transient trajectories pass through $\mathcal{C}_{6}$, the original snapshot index is $m = 8124, \, \ldots , \, 15000$ and $m = 44196, \, \ldots , \, 50000$, involving the asymptotic regime of an attractor.
The clusters in this closed orbit have a clear transition rule, as is evidenced by the transition matrix of figure~\ref{Fig:ClusterAnalysis105_layer2_C6}(c).
The sum of the residence times is $8.45$ for the cycle of  $\mathcal{C}_3 \rightarrow \mathcal{C}_4 \rightarrow \ldots\rightarrow \mathcal{C}_{10} \rightarrow \mathcal{C}_{3}$ from the residence time matrix in figure~\ref{Fig:ClusterAnalysis105_layer2_C6}(d).

\subsubsection{Hierarchical network model in Layer 3}
\label{SubSec:HNM105_L3}

Based on the detected cycle found in layer $\mathcal{L}_{2}$, the entering snapshots of one subcluster can be used to sample the quasi-periodic regime. 
The snapshots recurrently enter into each cluster of layer $\mathcal{L}_{2}$. 
The entering states, defined by $T_{o,\,k_1,\,k_2}^{m}$ in Eq.\eqref{Eqn:CharacteristicFunction2}, are the entry in the clusters.
The entering snapshots of a given cluster can be considered as hits in a ``Poincar\'e section'' \cite{guckenheimer2013}. 

The clustering algorithm in the layer $\mathcal{L}_{3}$ is applied to the snapshots of all the Poincar\'e sections. We still apply the sub-division clustering algorithm~\ref{alg:KmeansChildLayer}, with $T_{o,\,k_1,\,k_2}^{m}$ as characteristic function.

We consider the entry in cluster $\mathcal{C}_{6,\,10}$ for illustration, and build a network model in the third layer $\mathcal{L}_{3}$.
In the following, the cluster symbol in the first two layers $\mathcal{C}_{k_1=6,\,k_2 = 10, k_3}$ is omitted. 
\begin{figure}
\begin{tabular}{c c}
\begin{minipage}{0.48\textwidth}
\begin{tabular}{c}
\includegraphics[width=.9\linewidth]{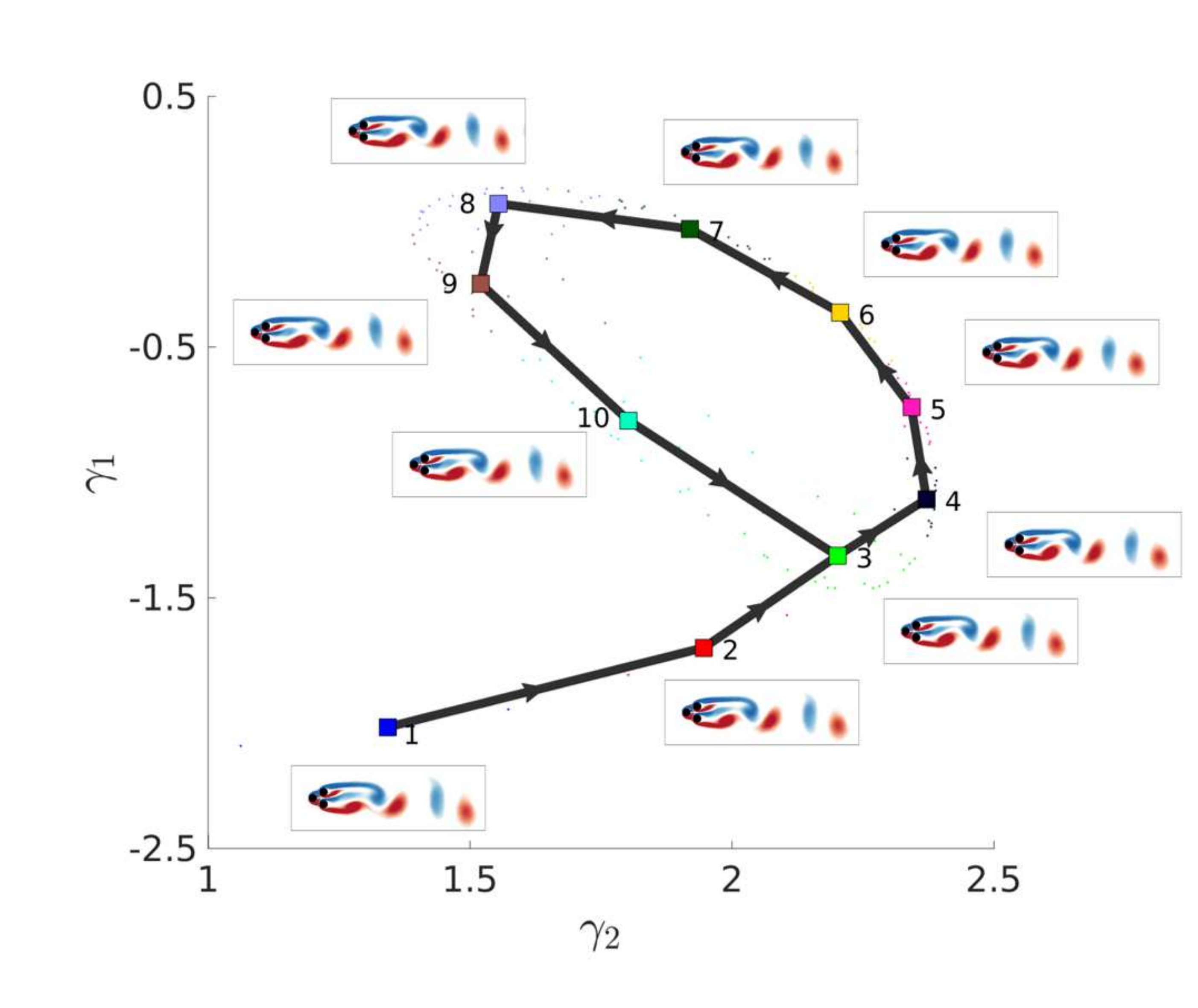}\\
$(a)$ 
\end{tabular} 
\end{minipage} &
\begin{minipage}{0.48\textwidth}
\begin{tabular}{c}
\includegraphics[width=1\linewidth]{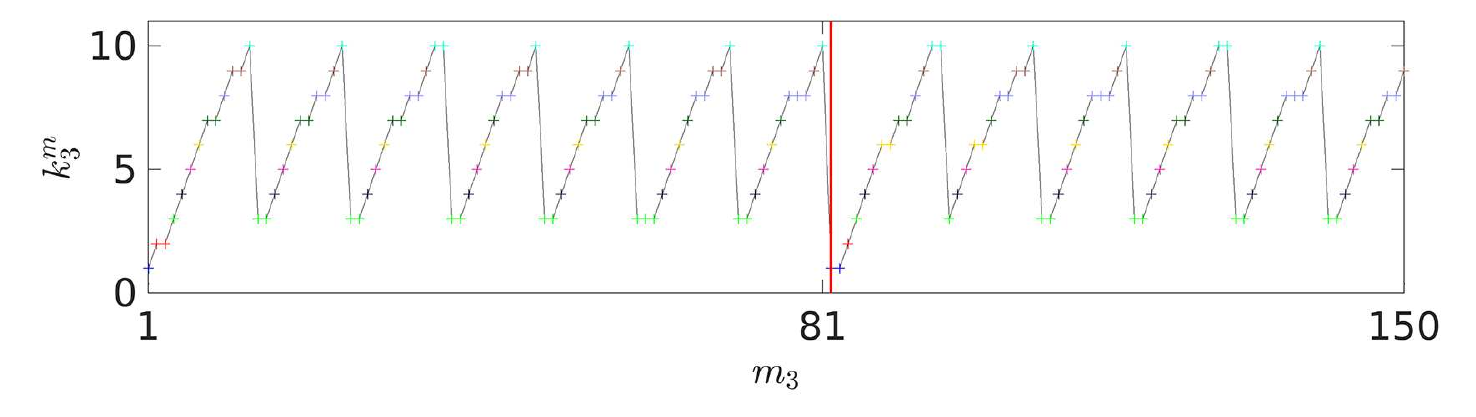} \\
$(b)$ \\
\end{tabular} 
\begin{tabular}{cc}
\includegraphics[width=.49\linewidth]{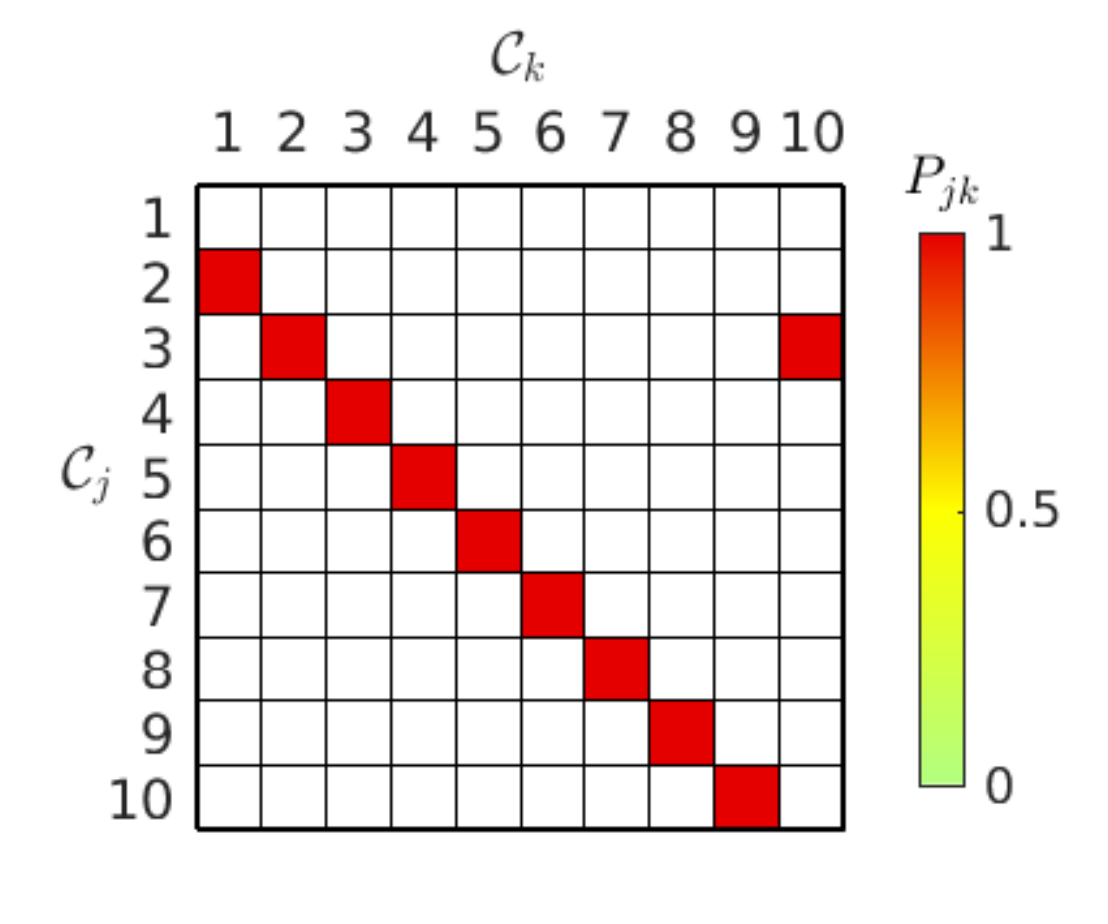} & 
\includegraphics[width=.49\linewidth]{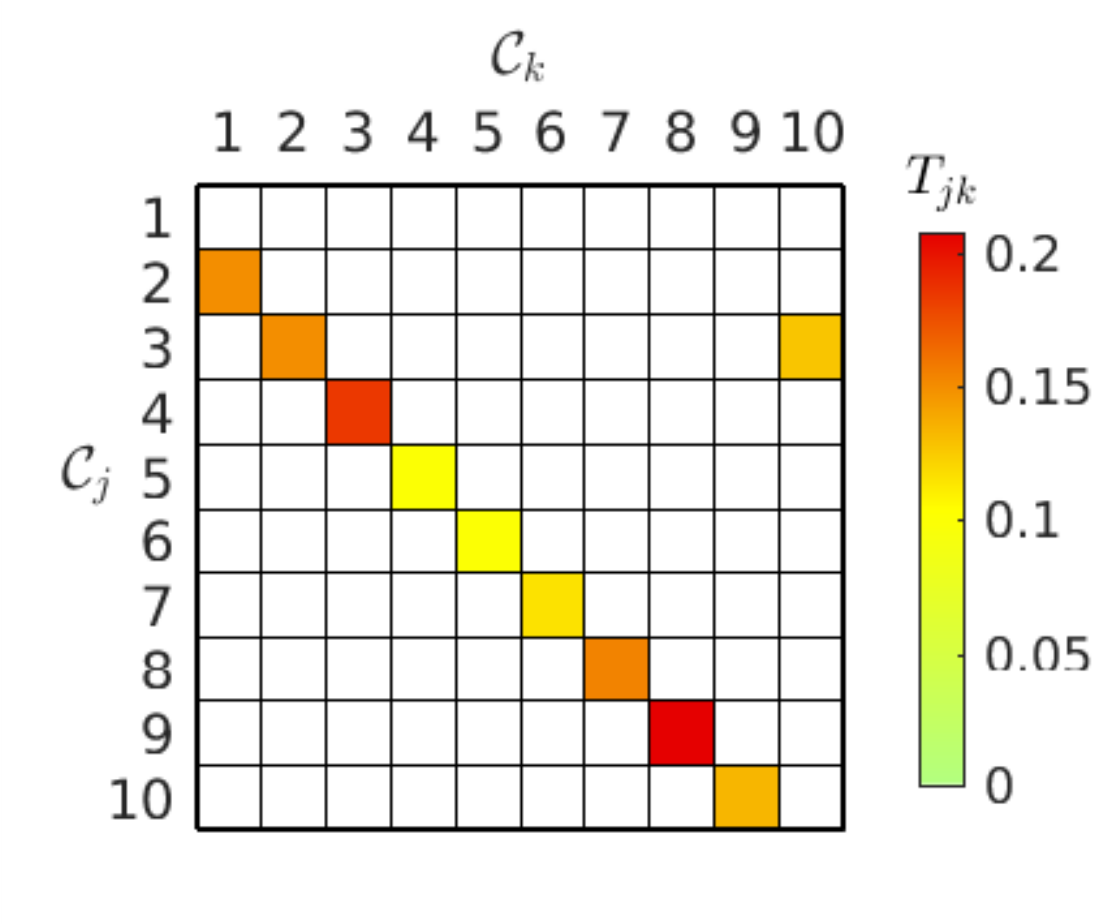} \\
$(c)$ & $(d)$ 
\end{tabular}
\end{minipage}
\end{tabular}
\caption{Cluster-based analysis in layer 3 at $\Rey= 105$ for $\mathcal{C}_{6,\,10}$: 
(a) graph of non-trivial transitions between clusters, as in figure~\ref{Fig:ClusterAnalysis80_layer1},
(b) transition illustrated with cluster label, 
(c) transition matrix, 
(d) residence time matrix, as in figure~\ref{Fig:Matrix80_layer1}.
Two trajectories are detected in the ``Poincar\'e section'' of $\mathcal{C}_{6,\,10}$, one with $m_3=1, \, \ldots , \, 81$ and another with $m_3=82, \, \ldots , \, 150$.
}
\label{Fig:ClusterAnalysis105_layer3_C10}
\end{figure}

As shown by the graph of non-trivial transitions shown in figure~\ref{Fig:ClusterAnalysis105_layer3_C10}(a), there exists a cycle $\mathcal{C}_3 \rightarrow \mathcal{C}_4 \rightarrow \ldots \rightarrow \mathcal{C}_{10} \rightarrow \mathcal{C}_{3}$. 
The periodically changing cluster label in figure~\ref{Fig:ClusterAnalysis105_layer3_C10}(b) indicates the existence of a recurrent dynamics. The original cluster index corresponds to two sets of discrete snapshots in the Poincar\'e section, with interval approximate to the periodic detected in the $\mathcal{L}_{2}$.
From the transition matrix of figure~\ref{Fig:ClusterAnalysis105_layer3_C10}(c), the two clusters $\mathcal{C}_1$ and $\mathcal{C}_2$ are not part of the cycle, but they form the transient part of the dynamics, before entering the cycle. This indicates that the low frequency modulations only start after the high-frequency oscillations have started in the second layer.
In other words, the low frequency does not exist during the building process of the vortex shedding, but appears after the vortex shedding has developed to a certain degree.

From the residence time matrix of figure~\ref{Fig:ClusterAnalysis105_layer3_C10}(d), the number of snapshots that belong to the cycle in the Poincar\'e section, determined by averaging over multiple trajectories, is $11.41$. 
The clustering analysis in the second layer $\mathcal{L}_{2}$ indicates that the trajectories periodically hit the Poincar\'e section with $8.45$, by summing up the elements in the blocks from $\mathcal{C}_3$ to $\mathcal{C}_{10}$ in figure~\ref{Fig:ClusterAnalysis105_layer3_C10}(d).
Hence, the resulting period of the cyclic process, considering both frequencies, is around $96.41$, which is very close to the real period of $97.10$ determined from the DNS.

\subsubsection{Dynamics reconstruction of the hierarchical network model at $\Rey=105$}
\label{SubSec:HNM105_Validation}

\begin{figure}
\centerline{
\begin{tabular}{cccc}
(a) & \raisebox{-0.5\height}{\includegraphics[width=.45\linewidth]{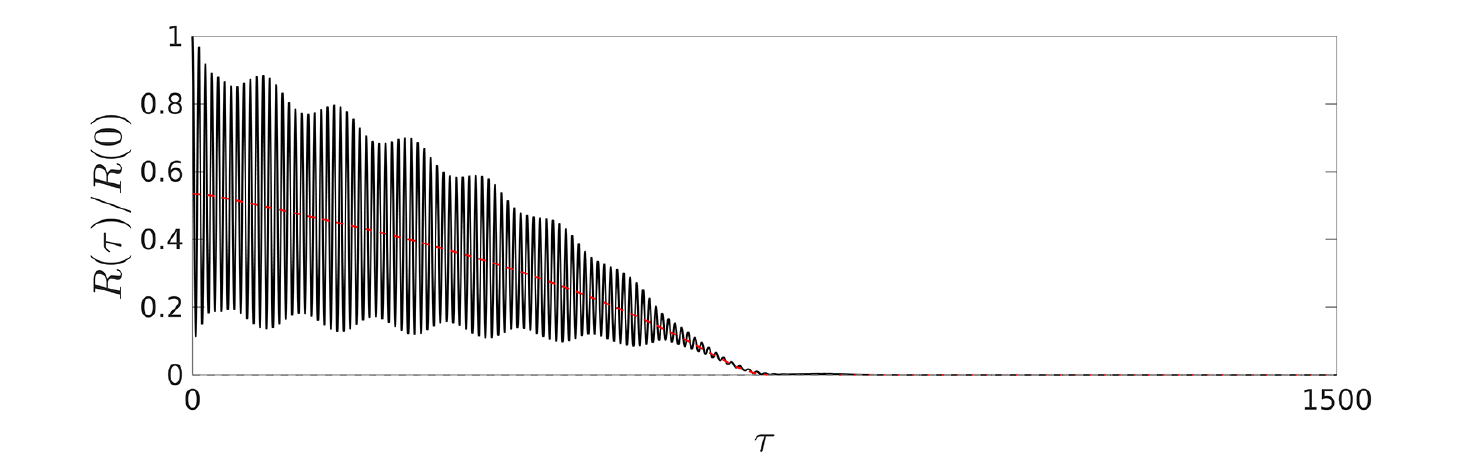}} &
(b) & \raisebox{-0.5\height}{\includegraphics[width=.45\linewidth]{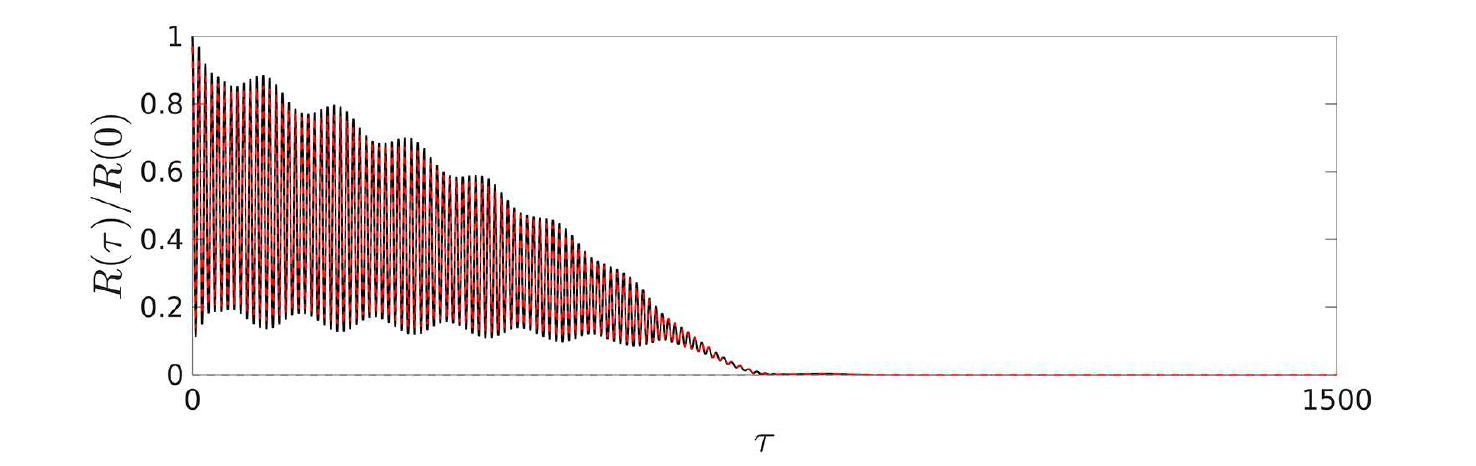}} \\
(c) & \raisebox{-0.5\height}{\includegraphics[width=.45\linewidth]{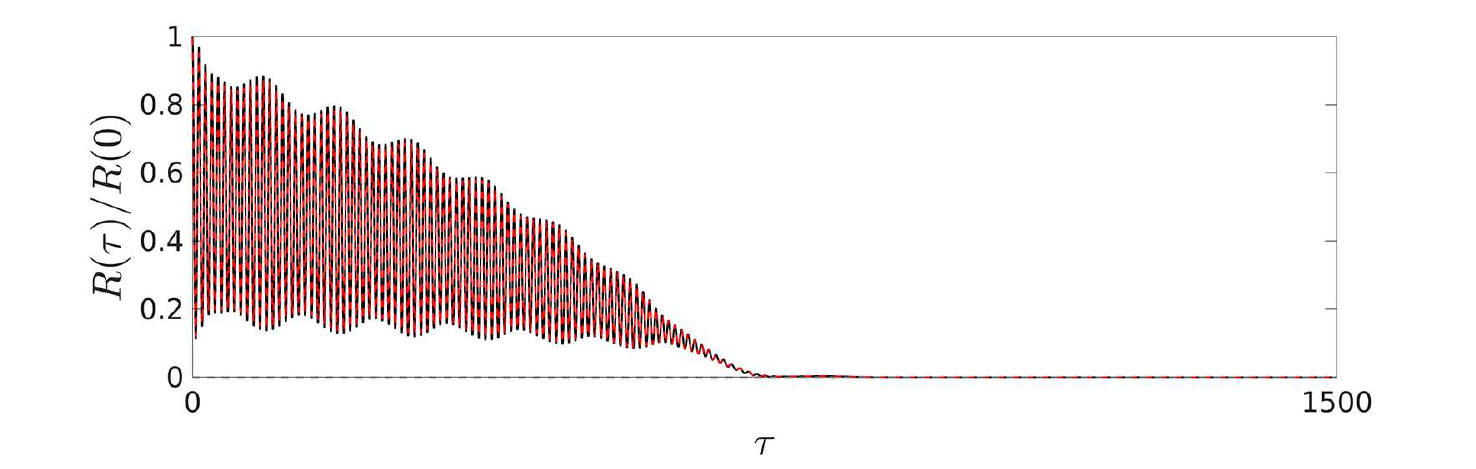}} 
\end{tabular}
}
\caption{Autocorrelation function for  $\tau \in [0, 1500)$ from DNS (black solid line) and the hierarchical network model (red dashed line) in the two layers: (a) $\mathcal{L}_{1}$, (b) $\mathcal{L}_{2}$, and (c) $\mathcal{L}_{3}$, at $\Rey=105$}
\label{Fig:Validation105}
\end{figure}

Figure~\ref{Fig:Validation105} shows the autocorrelation function of the DNS and the HiCNM in the three layers. 
The autocorrelation function of the DNS identifies the two dominant frequencies of the dynamics.
In layer $\mathcal{L}_{1}$, no oscillation can be identified, and the RMSE of the autocorrelation function is $R_{\rm rms}^1=22.20$.
In layer $\mathcal{L}_{2}$, the autocorrelation function of the model matches well with the high-frequency oscillations.
The low-frequency oscillations can be also found, but the amplitude does not fit well. The error is $R_{\rm rms}^2=1.52$, which is good enough for the accuracy.
In layer $\mathcal{L}_{3}$, the amplitude of the low-frequency oscillations can be better reproduced. The error is further reduced to $R_{\rm rms}^3=0.77$ with higher accuracy.

\subsection{Hierarchical network model at \Rey=130}
\label{Sec:HNM130}

At $\Rey=130$, the asymptotic dynamics is chaotic. We apply the clustering algorithm~\ref{alg:FilteredKmeans} first to the low-pass filtered data set (\S~\ref{SubSec:HNM130_L1}), before considering some typical flow regimes in the subclusters (\S~\ref{SubSec:HNM130_L2}).

\subsubsection{Hierarchical network model in layer 1}
\label{SubSec:HNM130_L1}
\begin{figure}
\centering
\includegraphics[width=.7\linewidth]{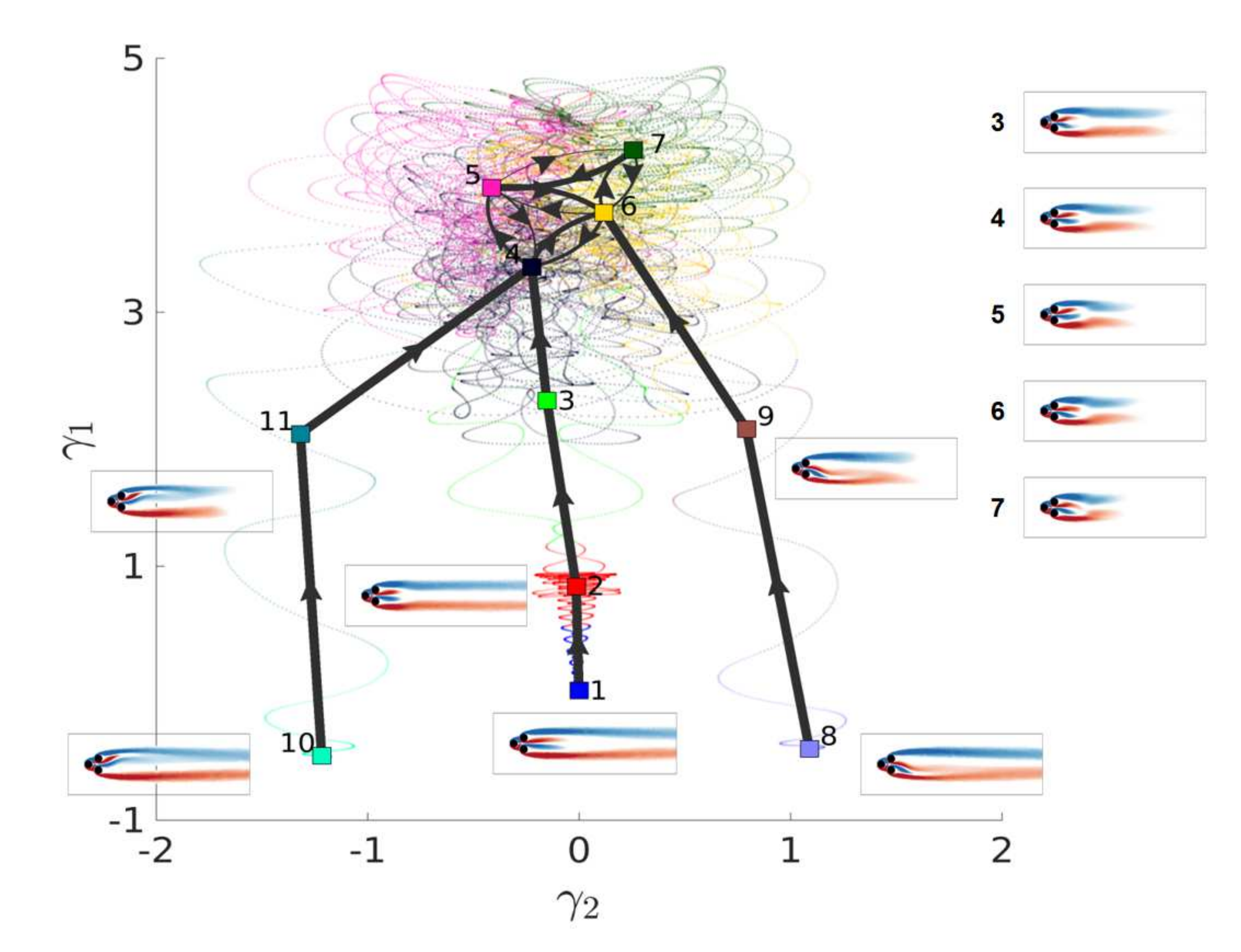} 
\caption{Graph of transitions between clusters in layer 1 at $\Rey= 130$, displayed as in figure~\ref{Fig:ClusterAnalysis80_layer1}.}
\label{Fig:ClusterAnalysis130_layer1}
\end{figure}

In the first layer, the filtered data set is used to analyze the mean-field dynamics with $K_1=11$ clusters. The detected frequency associated with the coherent component $\tilde{\bm{u}}$ is $f_c = 0.1225$.
The clustering algorithm~\ref{alg:FilteredKmeans} is applied to the data set, and the non-trivial transitions are shown in the two-dimensional subspace $[\gamma_1, \gamma_2]^{\rm T}$ of figure~\ref{Fig:ClusterAnalysis130_layer1}.
Four trajectories are found, each issued from one of the three steady solutions. The symmetric steady solution and the two asymmetric steady solutions respectively belong to clusters $\mathcal{C}_1$, $\mathcal{C}_{8}$, and $\mathcal{C}_{10}$.
These clusters evolve through $\mathcal{C}_1 \rightarrow \mathcal{C}_2 \rightarrow \mathcal{C}_3$, $\mathcal{C}_8 \rightarrow \mathcal{C}_9$ and $\mathcal{C}_{10} \rightarrow \mathcal{C}_{11}$ before entering into the same chaotic cloud, consisting of the remaining clusters. 

\begin{figure}
\centering
$(a)$ \raisebox{-0.6\height}{\includegraphics[width=.7\linewidth]{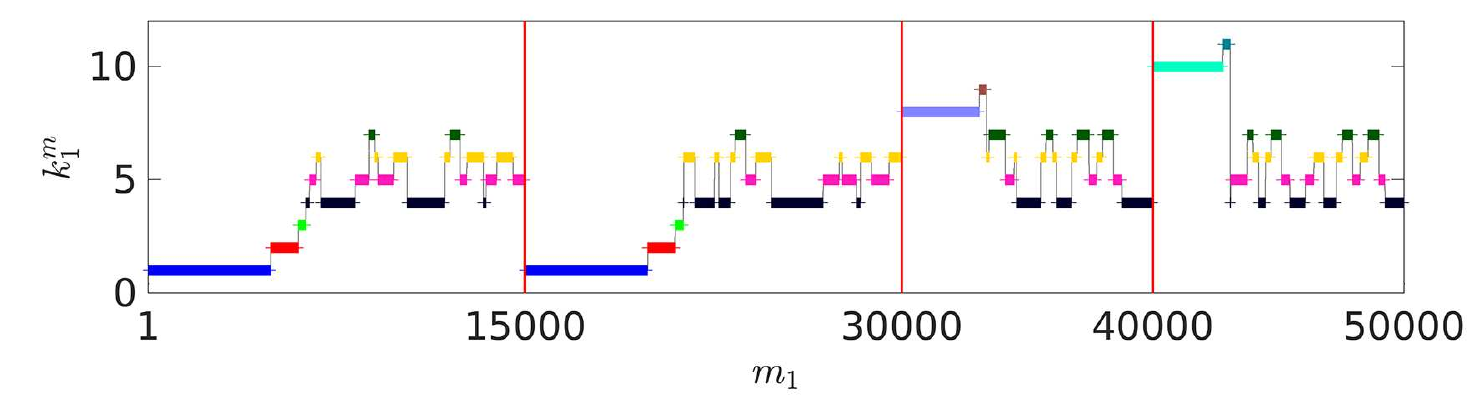}} \\
\begin{tabular}{cc}
$(b)$ \raisebox{-0.5\height}{\includegraphics[width=.31\linewidth]{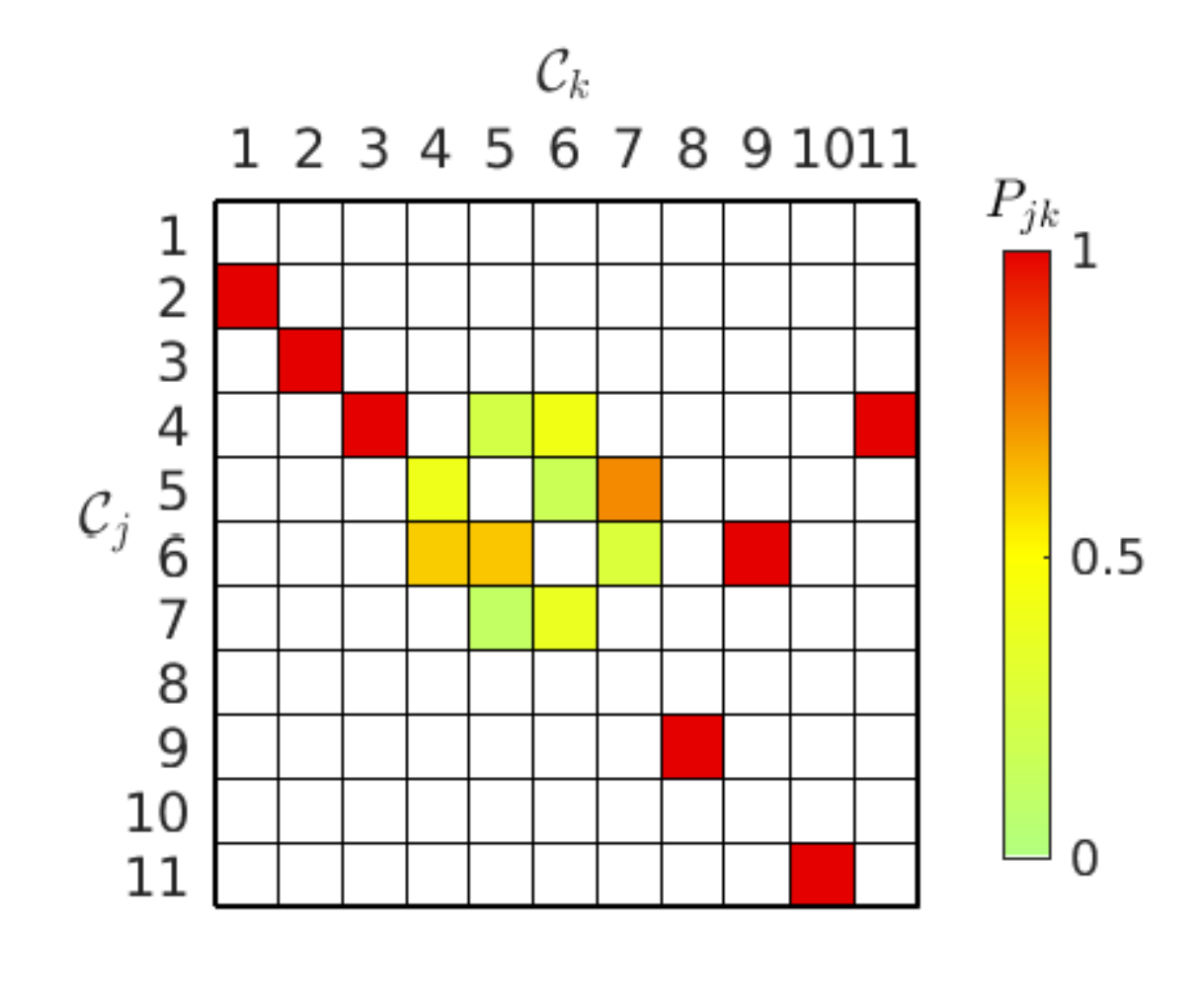}} &
$(c)$ \raisebox{-0.5\height}{\includegraphics[width=.31\linewidth]{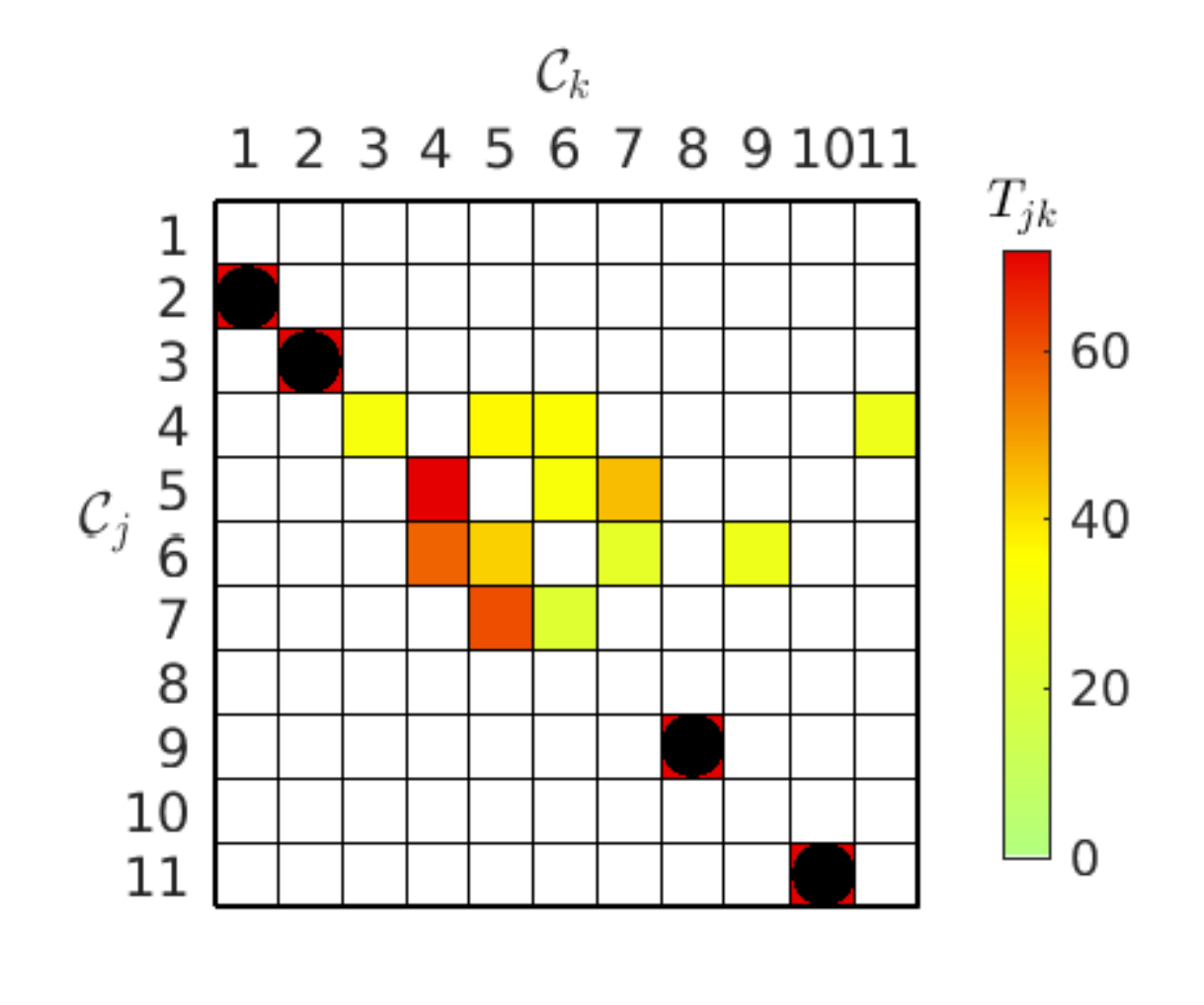}} \\
\end{tabular}
\caption{Cluster-based analysis at $\Rey= 130$ in layer 1:
(a) transition illustrated with cluster label, 
(b) transition matrix, 
(c) residence time matrix, displayed as in figure~\ref{Fig:Matrix80_layer1}.
}
\label{Fig:Matrix130_layer1}
\end{figure} 

There is no obvious periodic block of oscillating dynamics in the transient dynamics, as illustrated by the cluster labels of figure~\ref{Fig:Matrix130_layer1}(a). However, the initial destabilizing process is characterized by a very long residence time in the clusters to which the steady solutions belong.
From the transition matrix of figure~\ref{Fig:Matrix130_layer1}(b), three transitions leading to the chaotic region with $100\%$ probability, as $P_{3\,4}$, $P_{9\,6}$ and $P_{11\, 4}$.
Each cluster in the chaotic cloud has at least two possible destinations, with nearly equal probability. The hidden transition dynamics for this chaotic regime will be analized in the next layer.
The residence time matrix of figure~\ref{Fig:Matrix130_layer1}(c) shows four black filled circles associated with clusters $\mathcal{C}_1$, $\mathcal{C}_{8}$, $\mathcal{C}_{10}$ and $\mathcal{C}_2$. 
The steady solutions belong to the first three, while the symmetric limit cycle belongs to 
$\mathcal{C}_2$, as it will become clear in the next sections.

\subsubsection{Hierarchical network model in layer 2}
\label{SubSec:HNM130_L2}

In the second layer $\mathcal{L}_{2}$, we focus on the clusters associated with four typical states detected in layer $\mathcal{L}_{1}$: the destabilizing state from the symmetric steady solution in $\mathcal{C}_{1}$, the destabilizing state from the (upward) asymmetric steady solution in $\mathcal{C}_{10}$, the transient state with long residence time before chaos in $\mathcal{C}_{2}$, and the chaotic state in the group of clusters $\mathcal{C}_{4},\ldots,\mathcal{C}_{7}$.

\subsubsection*{The destabilizing state in the cluster $\mathcal{C}_{1}$}

\begin{figure}
\begin{tabular}{c c}
\begin{minipage}{0.48\textwidth}
\begin{tabular}{c}
\includegraphics[width=.9\linewidth]{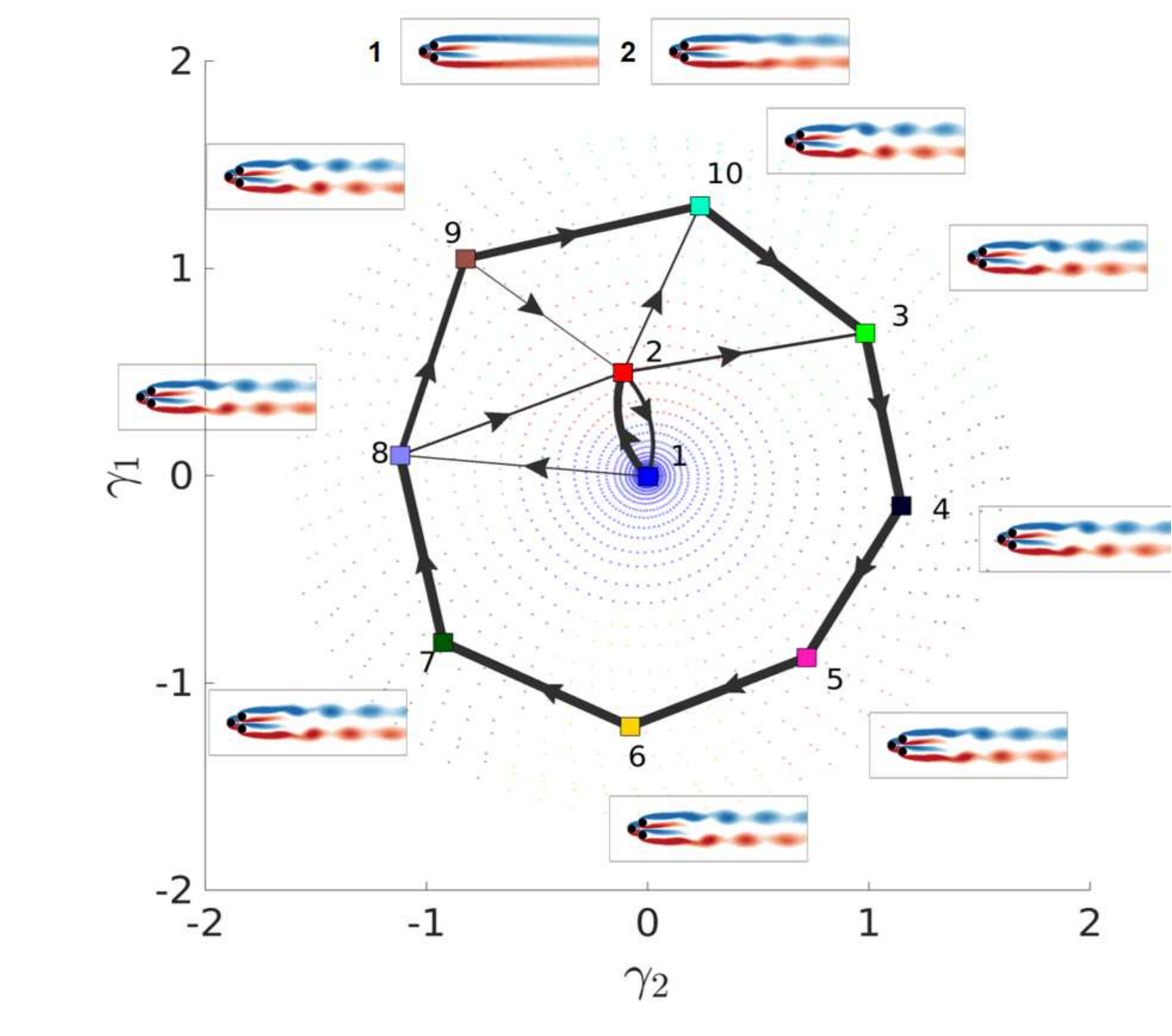}\\
$(a)$ 
\end{tabular} 
\end{minipage} &
\begin{minipage}{0.48\textwidth}
\begin{tabular}{c}
\includegraphics[width=1\linewidth]{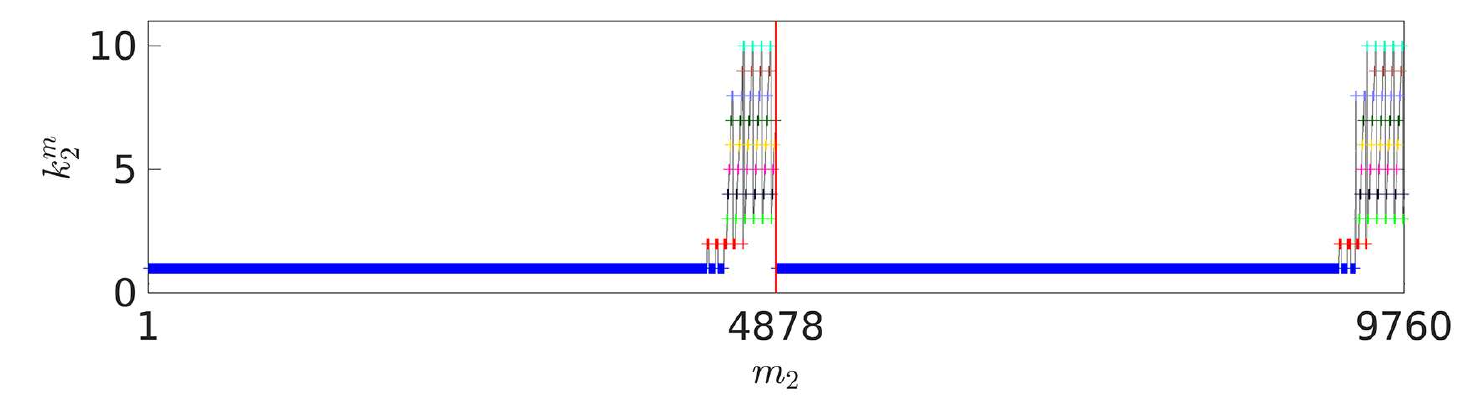} \\
$(b)$ \\
\end{tabular} 
\begin{tabular}{cc}
\includegraphics[width=.49\linewidth]{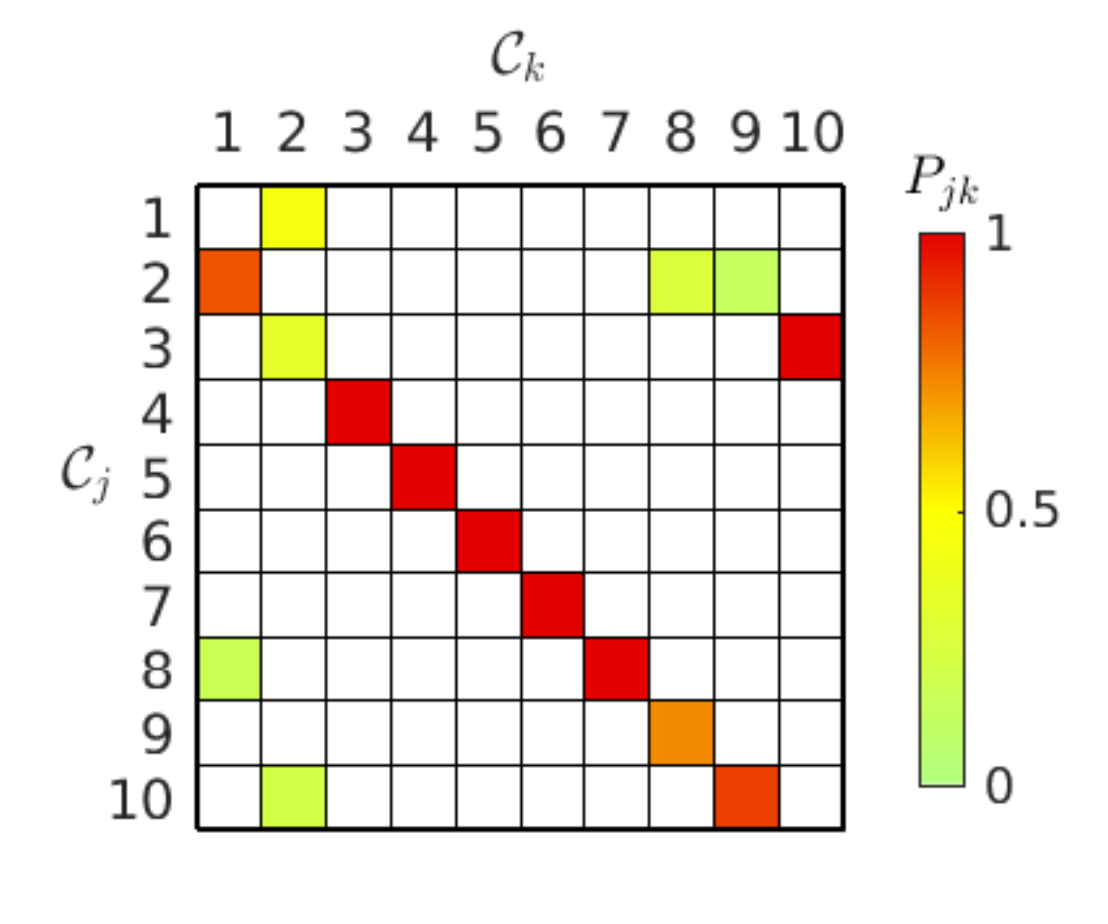} & 
\includegraphics[width=.49\linewidth]{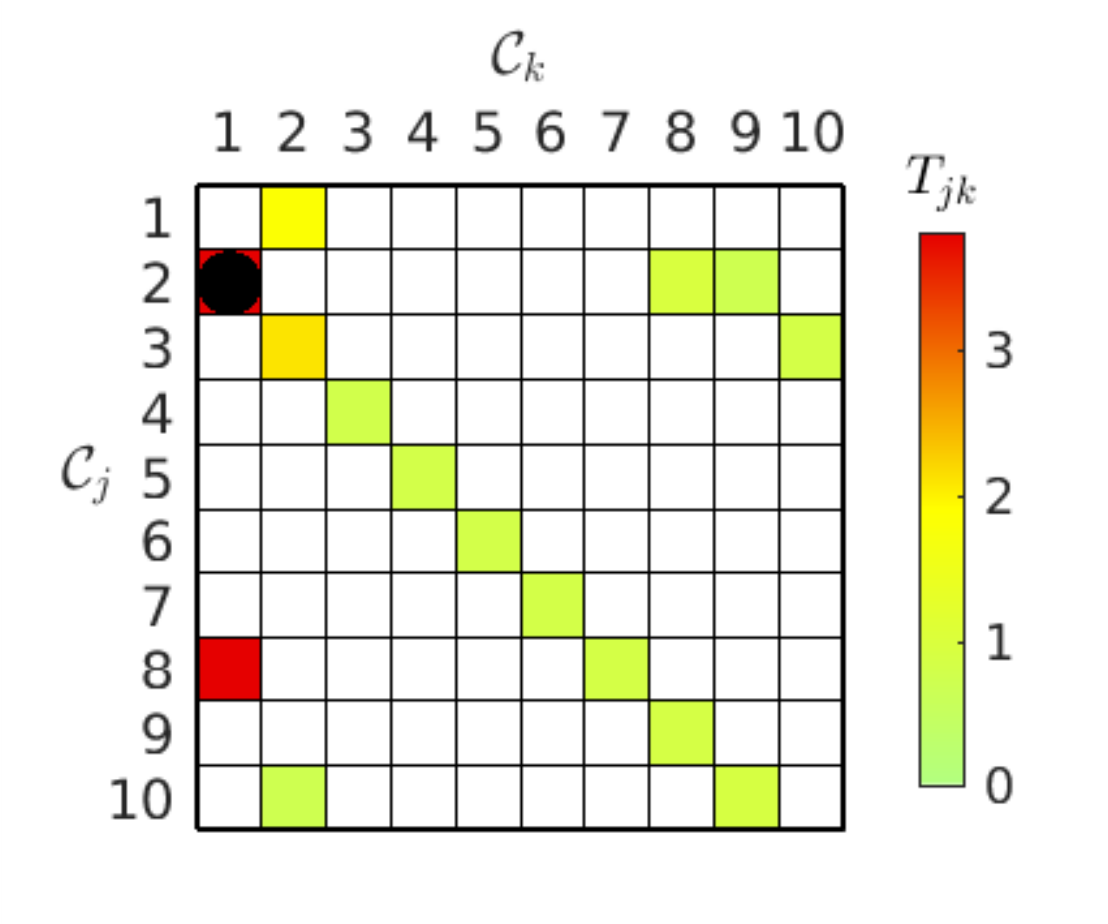} \\
$(c)$ & $(d)$ 
\end{tabular}
\end{minipage}
\end{tabular}
\caption{Cluster-based analysis in layer 2 at $\Rey= 130$ for $\mathcal{C}_{1}$: 
(a) graph of non-trivial transitions between clusters, as in figure~\ref{Fig:ClusterAnalysis80_layer1},
(b) transition illustrated with cluster label, 
(c) transition matrix, 
(d) residence time matrix, as in figure~\ref{Fig:Matrix80_layer1}.
Two trajectories pass through $\mathcal{C}_{1}$ in the parent layer, one with $m_2=1, \, \ldots , \, 4878$ and another with $m_2=4879, \, \ldots , \, 9760$.}
\label{Fig:ClusterAnalysis130_layer2_C1}
\end{figure}

Cluster $\mathcal{C}_{1}$ gathers snapshots in the initial stage of the instability starting from the symmetric steady solution. In the second layer $\mathcal{L}_{2}$, the snapshots of $\mathcal{C}_{1}$ are dispatched into $K_2 = 10$ subclusters $\mathcal{C}_{1,\,k_2}$, with $k_2 = 1,\ldots,K_2$. 
Figure~\ref{Fig:ClusterAnalysis130_layer2_C1} shows the transient trajectories with the centroids in the $[\gamma_1, \gamma_2]^{\rm T}$ plane.

Similar to figure~\ref{Fig:ClusterAnalysis80_layer2_C1}(a), the snapshots of figure~\ref{Fig:ClusterAnalysis130_layer2_C1}(a) form two diverging trajectories spiraling out of the center $[0,0]$. 
A loop is formed between the subclusters $\mathcal{C}_{1,\,1}$ and $\mathcal{C}_{1,\,2}$ in the inner zone. In the outer zone, a cycle appears with the periodic trajectory  $\mathcal{C}_{1,\,3} \rightarrow \ldots \rightarrow \mathcal{C}_{1,\,10} \rightarrow \mathcal{C}_{1,\,3}$. Figure~\ref{Fig:ClusterAnalysis130_layer2_C1}(b) shows two trajectories leaving $\mathcal{C}_{1,\,1}$. The original snapshot index is $m = 1, \, \ldots , \, 4878$ and $m = 15001, \, \ldots , \, 19882$, corresponding to the initial stage of the instability.

The transition matrix in figure~\ref{Fig:ClusterAnalysis130_layer2_C1}(c) corroborates this periodic cycle. The black filled circle in figure~\ref{Fig:ClusterAnalysis130_layer2_C1}(d) marks out the transition with a long-residence time, due to the unstable center that belong to cluster $\mathcal{C}_{1,\,1}$.

The transitions $\mathcal{C}_{1,\,1} \rightarrow \mathcal{C}_{1,\,8}$,  $\mathcal{C}_{1,\,2} \rightarrow \mathcal{C}_{1,\,3}$ and $\mathcal{C}_{1,\,2} \rightarrow \mathcal{C}_{1,\,10}$ correspond to the departing dynamics out of the inner zone, due to the development of the instability. 
We also note the returning transitions $\mathcal{C}_{1,\,8} \rightarrow \mathcal{C}_{1,\,2}$ and $ \mathcal{C}_{1,\,9} \rightarrow \mathcal{C}_{1,\,2}$. However, the latter do not mean that the flow actually returns back to the destabilizing center, as both trajectories are spiralling out of the center.
This confusing result comes from the clustering process. 
The edge between the inner and outer zones is not well defined, due to the varying density distribution of snapshots along the arms of the spirals. Cluster $\mathcal{C}_{1,\,2}$ overlaps the outer zone, to the difference of cluster $\mathcal{C}_{1,\,1}$, which fully belong to the inner zone.  
If we ignore the loop in the inner zone and merge $\mathcal{C}_{1,\,1}$ and $\mathcal{C}_{1,\,2}$, it shows a dynamical evolution from one cluster to the cycle of a group of clusters.

\subsubsection*{The destabilizing state in cluster $\mathcal{C}_{10}$}

\begin{figure}
\begin{tabular}{c c}
\begin{minipage}{0.48\textwidth}
\begin{tabular}{c}
\includegraphics[width=.9\linewidth]{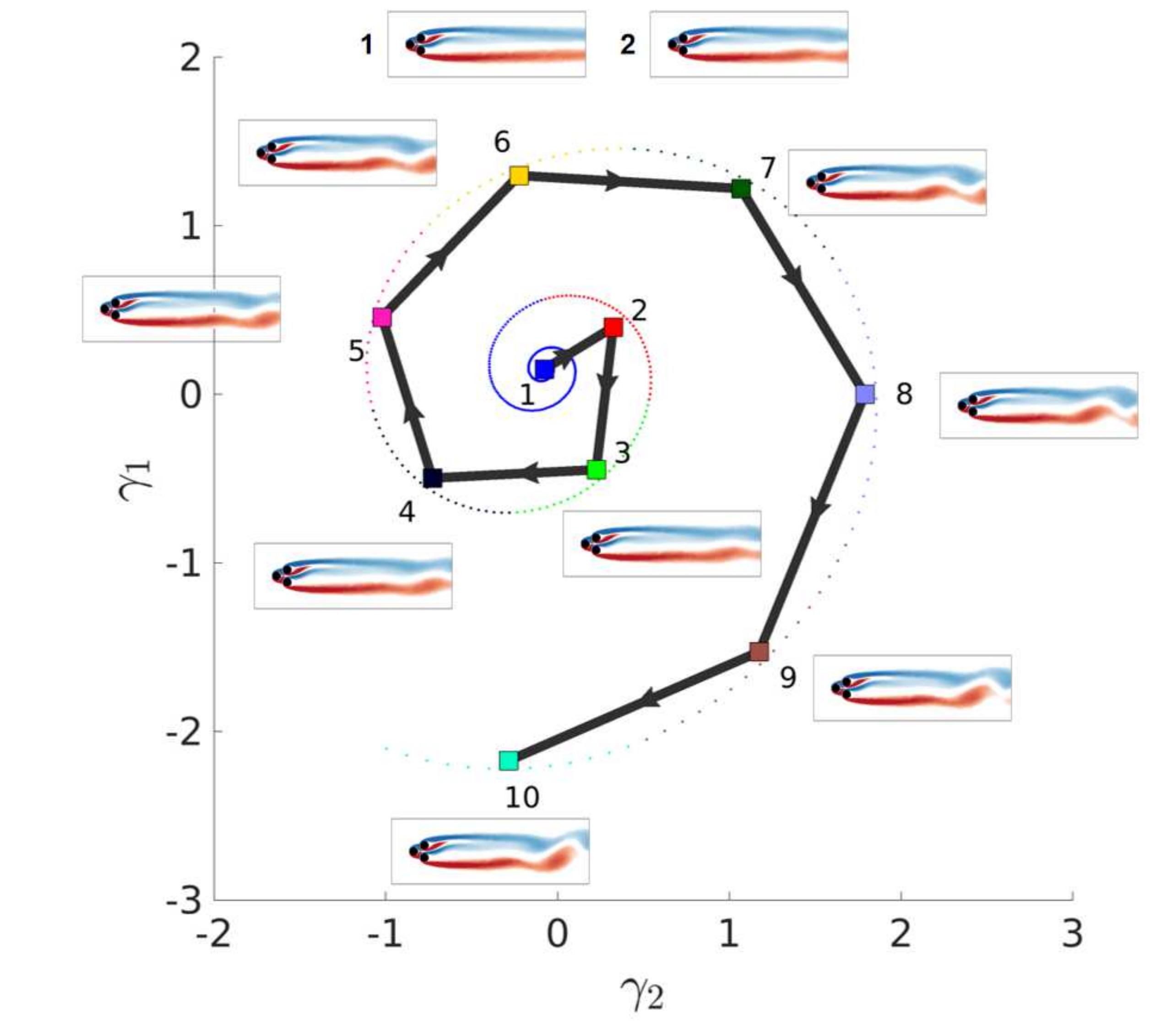}\\
$(a)$ 
\end{tabular} 
\end{minipage} &
\begin{minipage}{0.48\textwidth}
\begin{tabular}{c}
\includegraphics[width=1\linewidth]{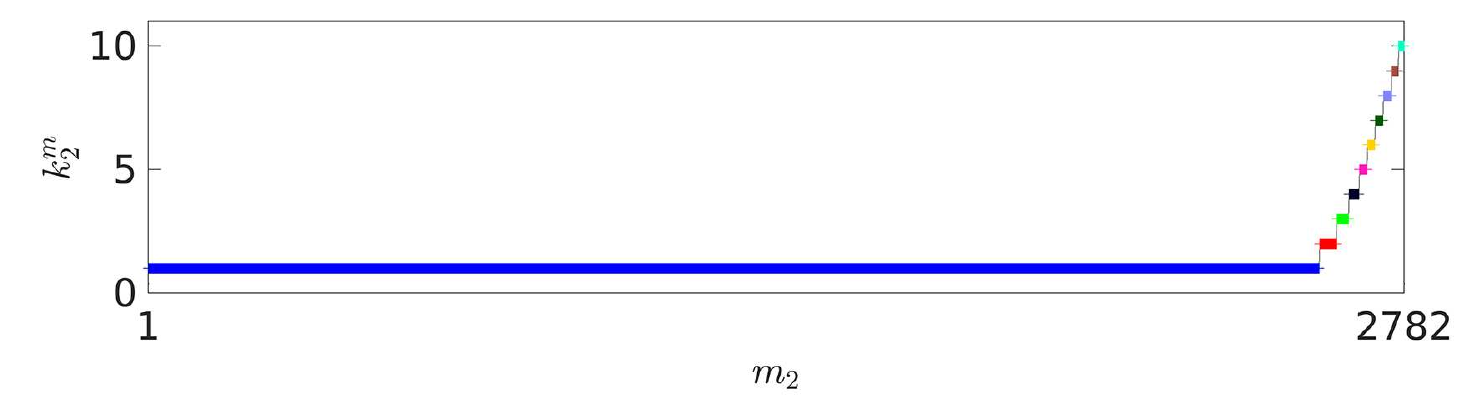} \\
$(b)$ \\
\end{tabular} 
\begin{tabular}{cc}
\includegraphics[width=.49\linewidth]{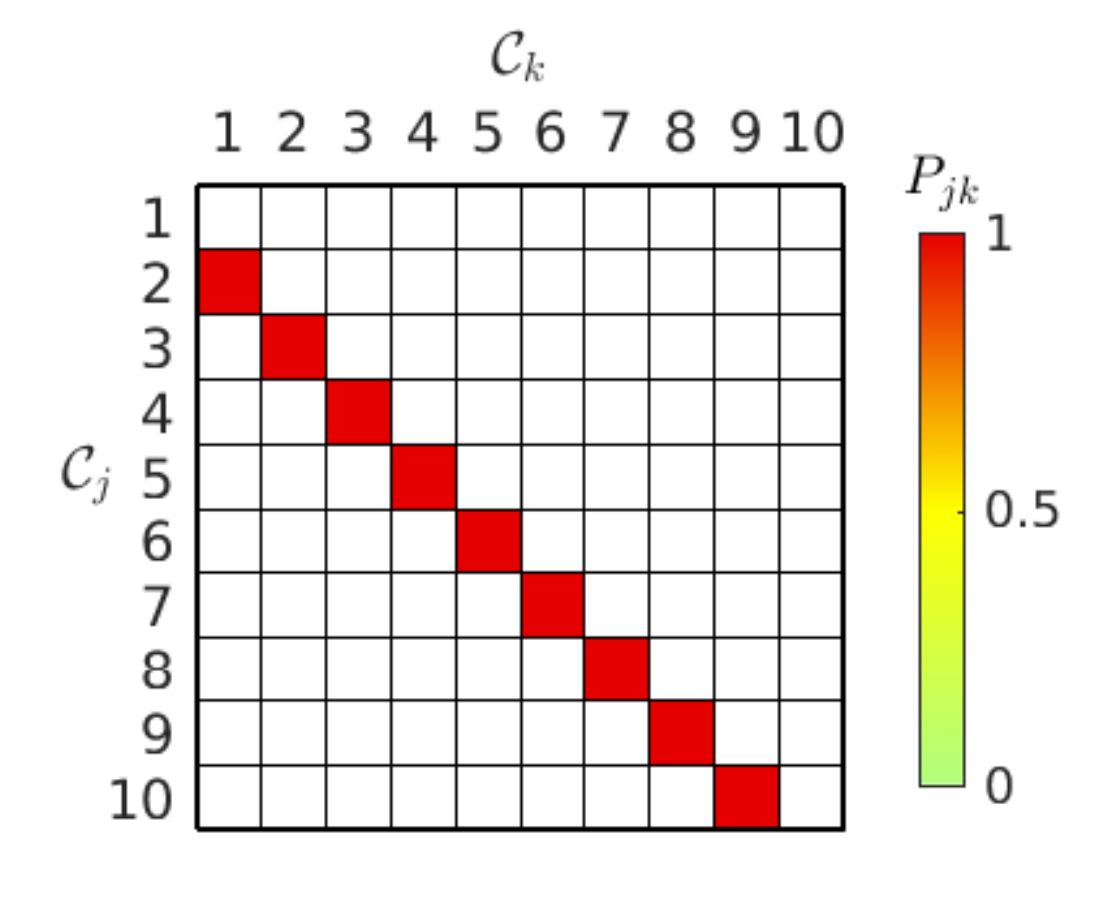} & 
\includegraphics[width=.49\linewidth]{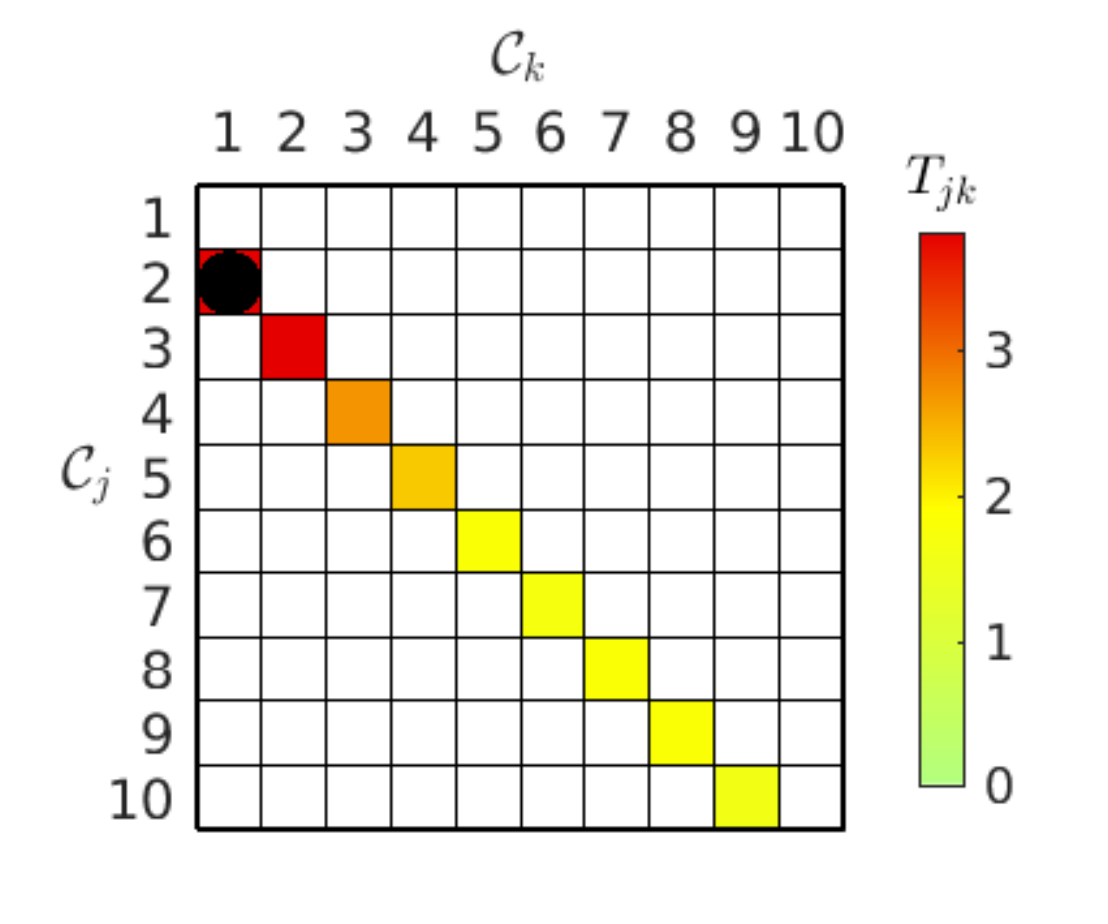} \\
$(c)$ & $(d)$ 
\end{tabular}
\end{minipage}
\end{tabular}
\caption{Cluster-based analysis in layer 2 at $\Rey= 130$ for $\mathcal{C}_{10}$: 
(a) graph of non-trivial transitions between clusters, as in figure~\ref{Fig:ClusterAnalysis80_layer1},
(b) transition illustrated with cluster label, 
(c) transition matrix, 
(d) residence time matrix, as in figure~\ref{Fig:Matrix80_layer1}.
A sole trajectorie pass through $\mathcal{C}_{10}$ in the parent layer with $m_2=1, \, \ldots , \, 2782$.}
\label{Fig:ClusterAnalysis130_layer2_C10}
\end{figure}

Cluster $\mathcal{C}_{10}$ contains the trajectory spiraling out from the upward-deflected asymmetric steady solution, as shown in figure~\ref{Fig:ClusterAnalysis130_layer2_C10}(a). The snapshots are dispatched into $K_2 = 10$ of subclusters $\mathcal{C}_{10,\,k_2}$, with $k_2 = 1,\ldots,K_2$. 
Together with figure~\ref{Fig:ClusterAnalysis130_layer2_C10}(b), it shows a one-way transition, departing from the unstable center with a sparse spiral. The original snapshot index is $m = 40001, \, \ldots , \, 42782$ for the initial stage of the destabilisation from one asymmetric steady solution.

The graph of figure~\ref{Fig:ClusterAnalysis130_layer2_C10}(a) is different from the graphs of figure~\ref{Fig:ClusterAnalysis80_layer2_C1}(a) and
figure~\ref{Fig:ClusterAnalysis130_layer2_C1}(a), the latter being associated with the symmetric steady solution.
The flow destabilisation from the asymmetric steady solution develops faster than from the symmetric steady solution, as illustrated by the linear growth rates $\sigma_{sym}=0.032$ and $\sigma_{asym}=0.106$ of the respective pairs of unstable eigenmodes. 
As a result, the distribution of snapshots is sparser in figure~\ref{Fig:ClusterAnalysis130_layer2_C10}(a) than in figure~\ref{Fig:ClusterAnalysis130_layer2_C1}(a).
As indicated by figure~\ref{Fig:ClusterAnalysis130_layer2_C10}(b), 
93.2\% of the 2782 snapshots of cluster $\mathcal{C}_{10}$ belongs to subcluster $\mathcal{C}_{10,\,1}$. 
The flow quickly travels through all the remaining clusters in only 188 snapshots.

The centroids of figure~\ref{Fig:ClusterAnalysis130_layer2_C1}(a) and figure~\ref{Fig:ClusterAnalysis130_layer2_C10}(a) have two main differences: (i) The inner jet is symmetric in the centroids of figure~\ref{Fig:ClusterAnalysis130_layer2_C1}(a) while it is deflected upwards in the controids of figure~\ref{Fig:ClusterAnalysis130_layer2_C10}(a); 
(ii) the von K\'arm\'an street of vorticies of  figure~\ref{Fig:ClusterAnalysis130_layer2_C1}(a) exhibits positive and negative vortices well apart from each other in the $y$-axis.  
In figure~\ref{Fig:ClusterAnalysis130_layer2_C10}(a), the positive and negative vortices are of a larger strength and adjacent to the $x$-axis, together with a longer shear layer.

\subsubsection*{Transient regime before chaos in cluster $\mathcal{C}_{2}$}

\begin{figure}
\begin{tabular}{c c}
\begin{minipage}{0.48\textwidth}
\begin{tabular}{c}
\includegraphics[width=.9\linewidth]{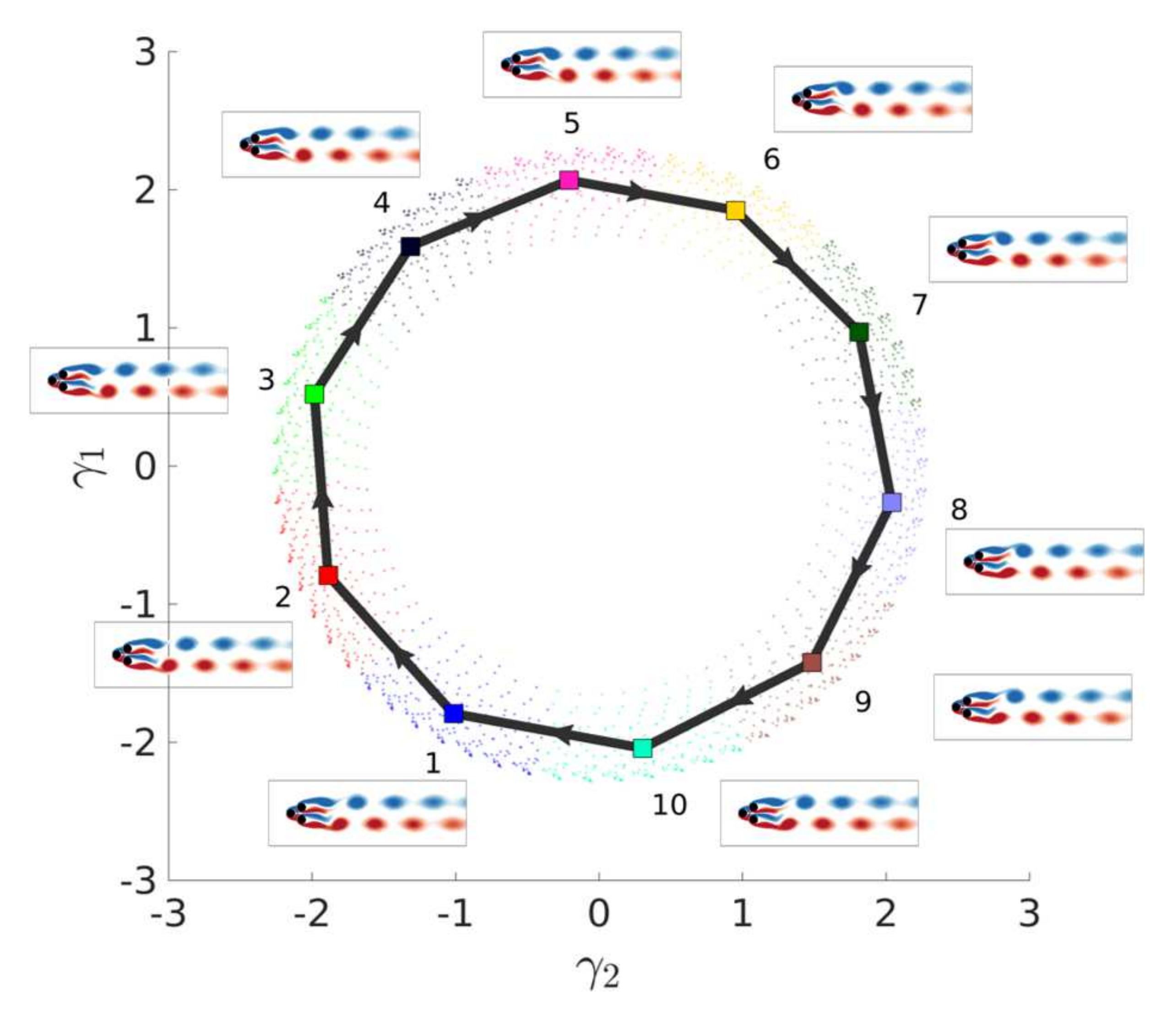}\\
$(a)$ 
\end{tabular} 
\end{minipage} &
\begin{minipage}{0.48\textwidth}
\begin{tabular}{c}
\includegraphics[width=1\linewidth]{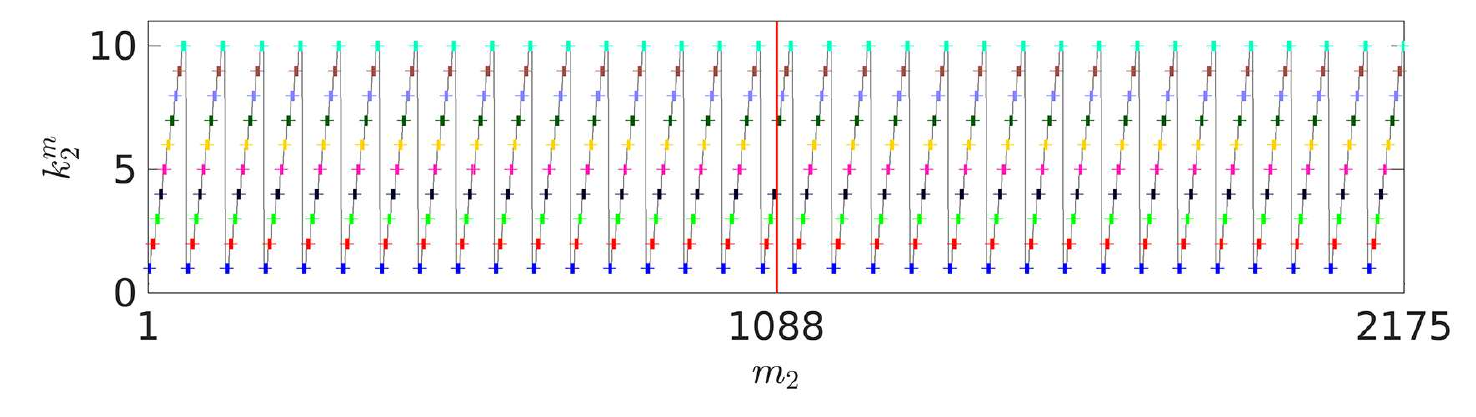} \\
$(b)$ \\
\end{tabular} 
\begin{tabular}{cc}
\includegraphics[width=.49\linewidth]{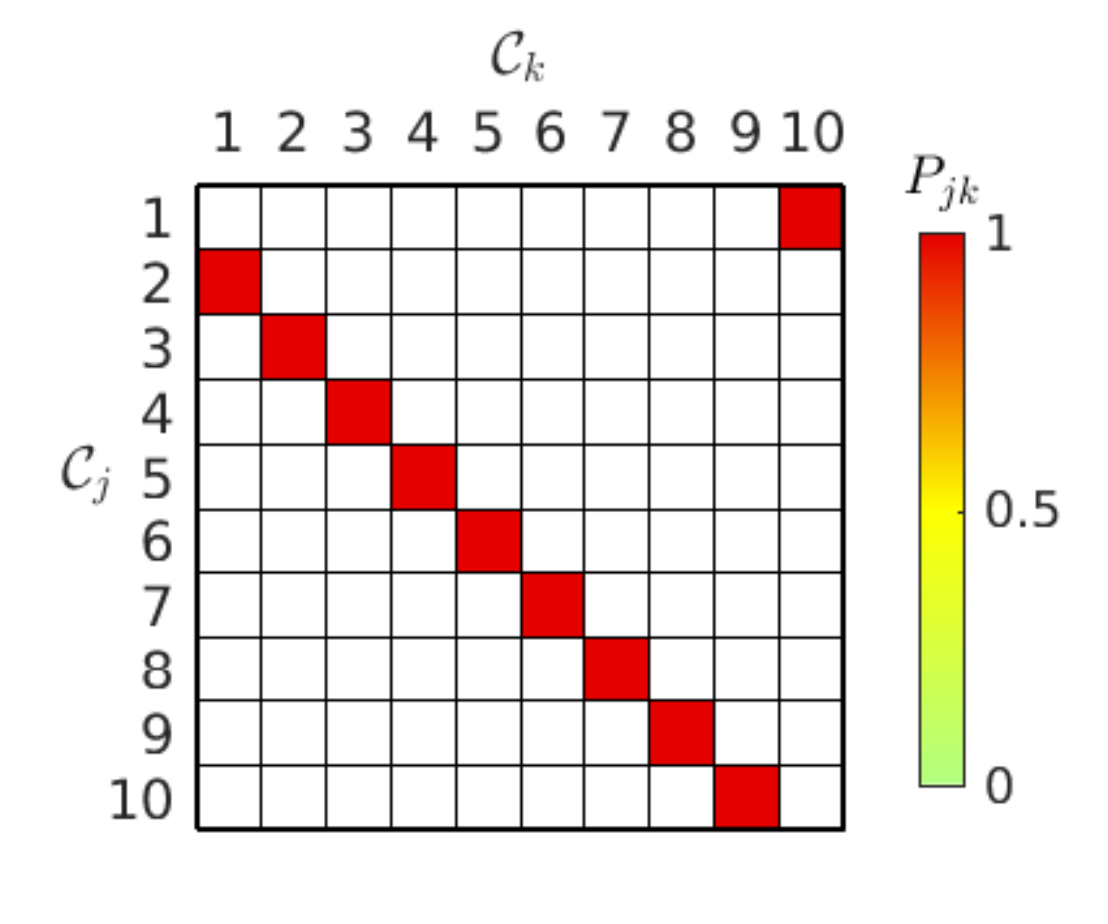} & 
\includegraphics[width=.49\linewidth]{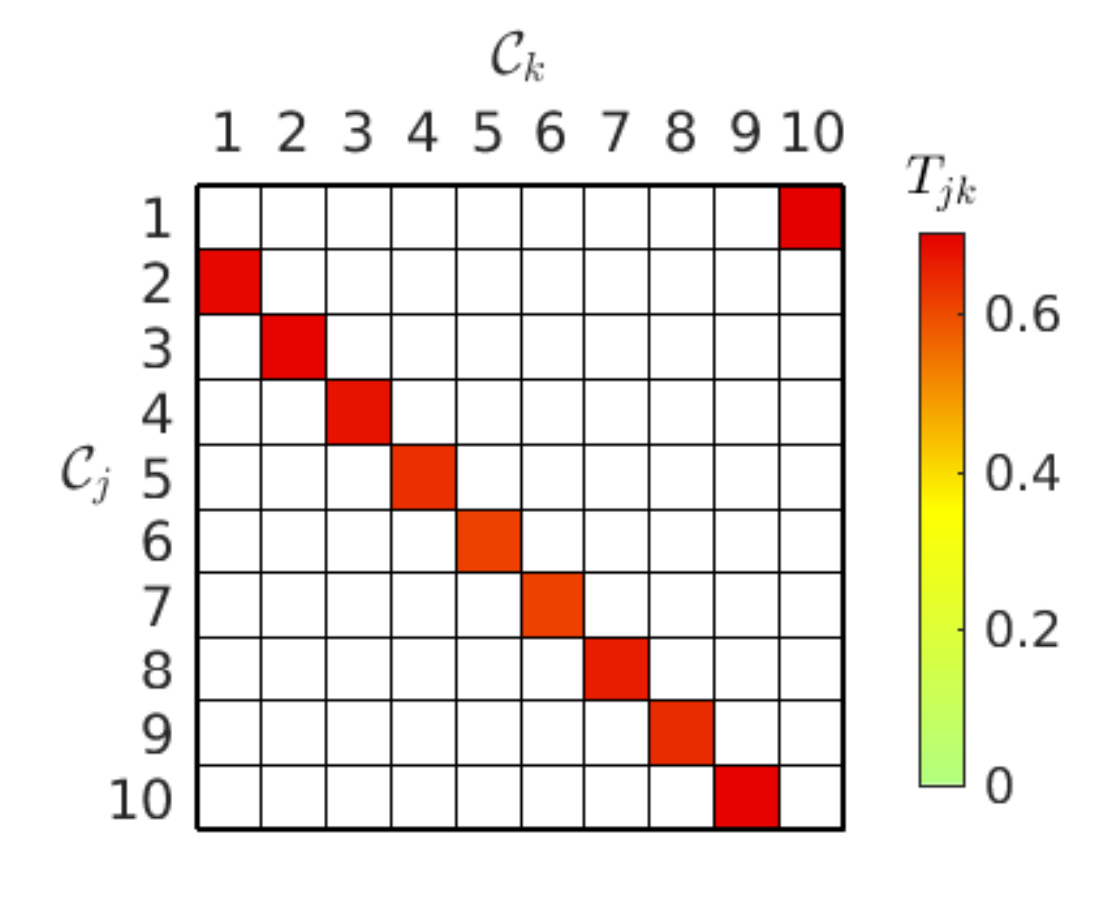} \\
$(c)$ & $(d)$ 
\end{tabular}
\end{minipage}
\end{tabular}
\caption{Cluster-based analysis in layer 2 at $\Rey= 130$ for $\mathcal{C}_{2}$: 
(a) graph of non-trivial transitions between clusters, as in figure~\ref{Fig:ClusterAnalysis80_layer1},
(b) transition illustrated with cluster label, 
(c) transition matrix, 
(d) residence time matrix, as in figure~\ref{Fig:Matrix80_layer1}.
Two trajectories pass through $\mathcal{C}_{2}$ in the parent layer, one with $m_2=1, \, \ldots , \, 1088$ and another with $m_2=1089, \, \ldots , \, 2175$.}
\label{Fig:ClusterAnalysis130_layer2_C2}
\end{figure}

Cluster $\mathcal{C}_{2}$ contains two transient trajectories which connect the symmetric steady solution in $\mathcal{C}_{1}$ to the chaotic cloud, with the original snapshot index $m = 4879, \, \ldots , \, 5966$ and $19883, \, \ldots , \, 20969$.

Figure~\ref{Fig:ClusterAnalysis130_layer2_C2}(b) shows that the flow periodically travels through the ten subclusters in the child layer $\mathcal{L}_{2}$ of cluster $\mathcal{C}_{2}$. 
As for the centroids of figure~\ref{Fig:ClusterAnalysis130_layer2_C2}, the alley of vortices do not cross the $x$-axis. The closed-transitions $\mathcal{C}_1 \rightarrow \ldots \rightarrow \mathcal{C}_{10} \rightarrow \mathcal{C}_1$ is further evidenced in the transition matrix of figure~\ref{Fig:ClusterAnalysis130_layer2_C2}(c). 
The residence times of  figure~\ref{Fig:ClusterAnalysis130_layer2_C2}(d) are rather uniform and the averaged period is 6.62.

\subsubsection*{The chaotic state in the group of clusters $\mathcal{C}_{4},\ldots,\mathcal{C}_{7}$}

When dealing with the chaotic dynamics, considering each cluster $\mathcal{C}_{4},\ldots,\mathcal{C}_{7}$ separately is doable. However, instead of building a network model for each cluster separately, we can build an overall model for these four clusters in the chaotic regime.
Therefore, for analyzing the chaotic dynamics, we consider all the clusters that belong to the chaotic cloud in layer $\mathcal{L}_{1}$ as a whole. 

\begin{figure}
\centering
\includegraphics[width=.65\linewidth]{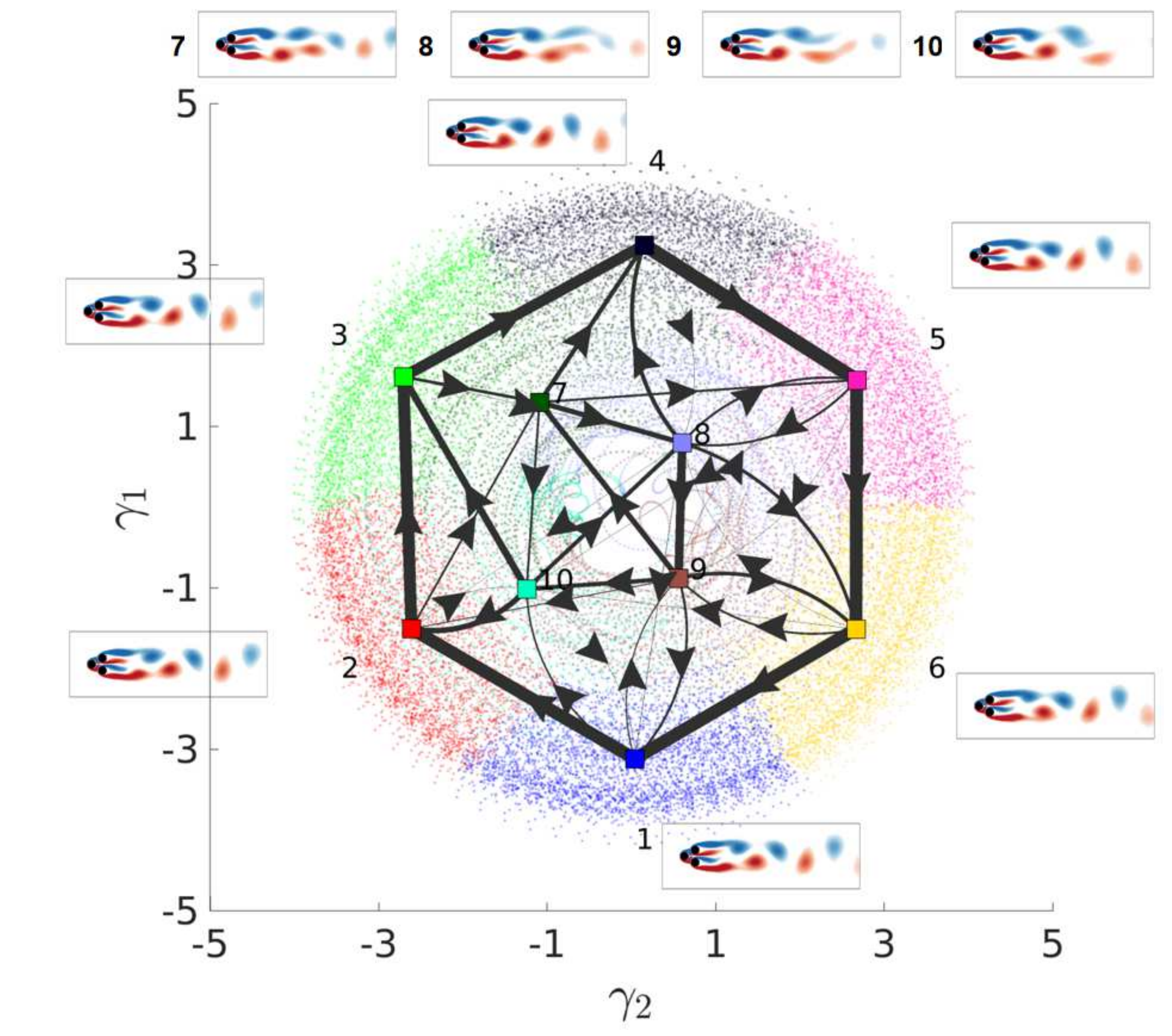} 
\caption{Graph of transitions between clusters in layer 2 at $\Rey= 130$ for the chaotic clusters of $\mathcal{L}_{1}$, displayed as in figure~\ref{Fig:ClusterAnalysis80_layer1}.}
\label{Fig:ClusterAnalysis130_layer2_CHAOS}
\end{figure}
\begin{figure}
\centering
$(a)$ \raisebox{-0.6\height}{\includegraphics[width=.6\linewidth]{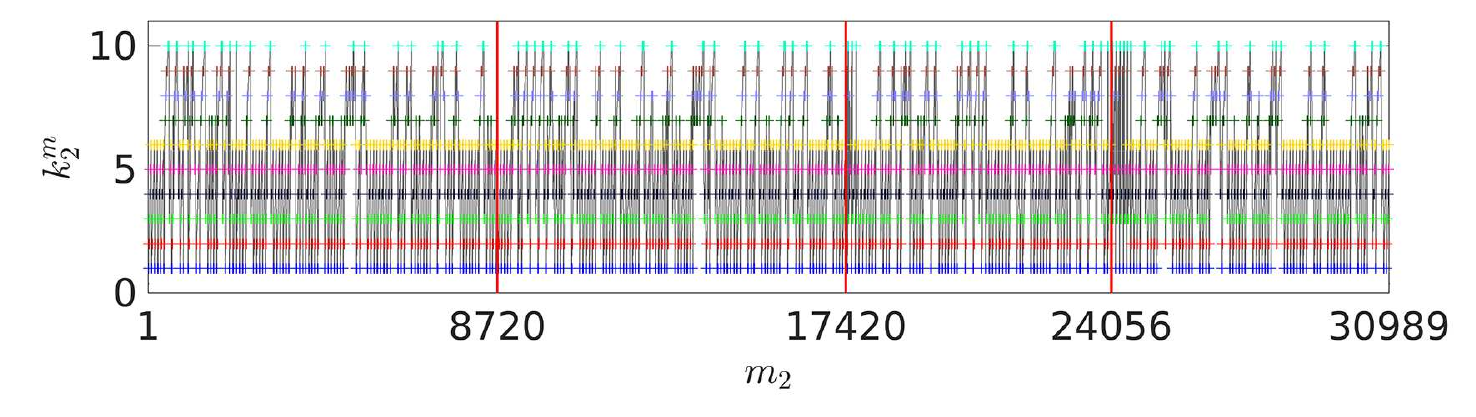}} \\
\begin{tabular}{cc}
$(b)$ \raisebox{-0.5\height}{\includegraphics[width=.25\linewidth]{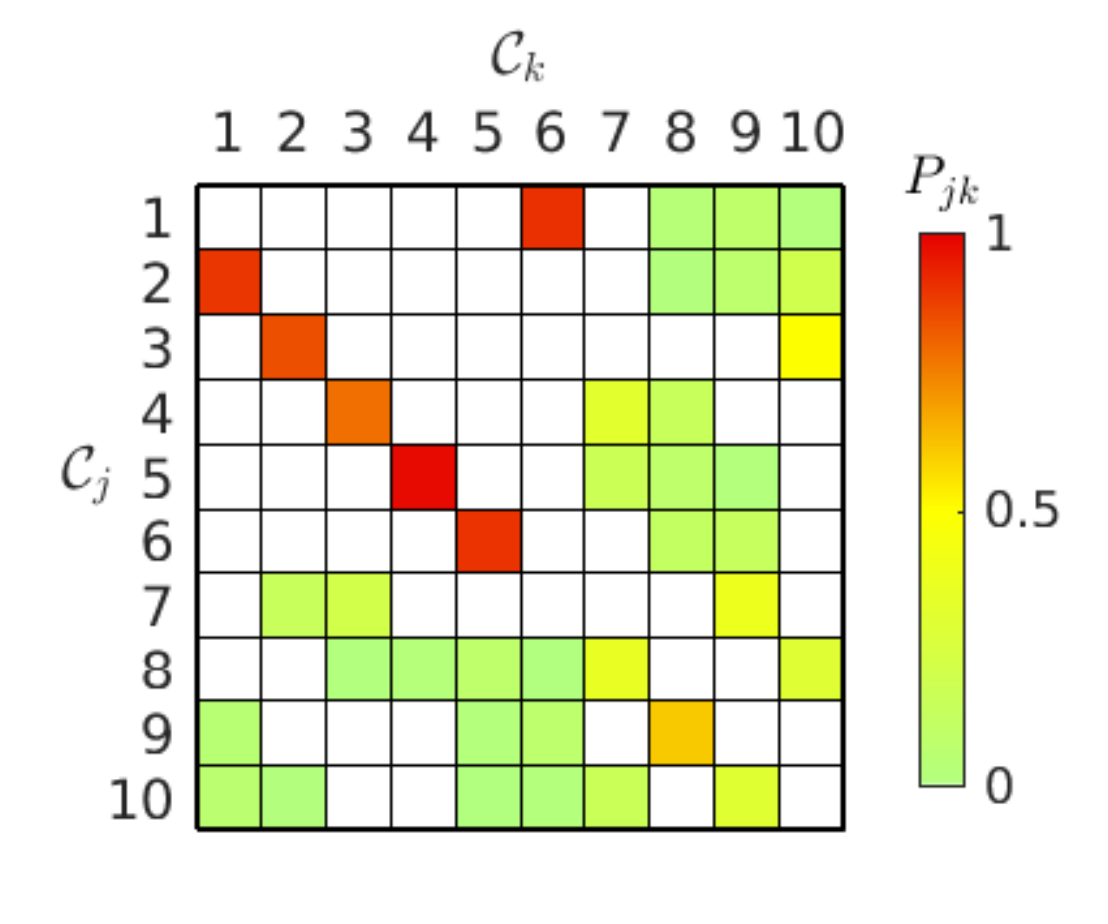}} &
$(c)$ \raisebox{-0.5\height}{\includegraphics[width=.25\linewidth]{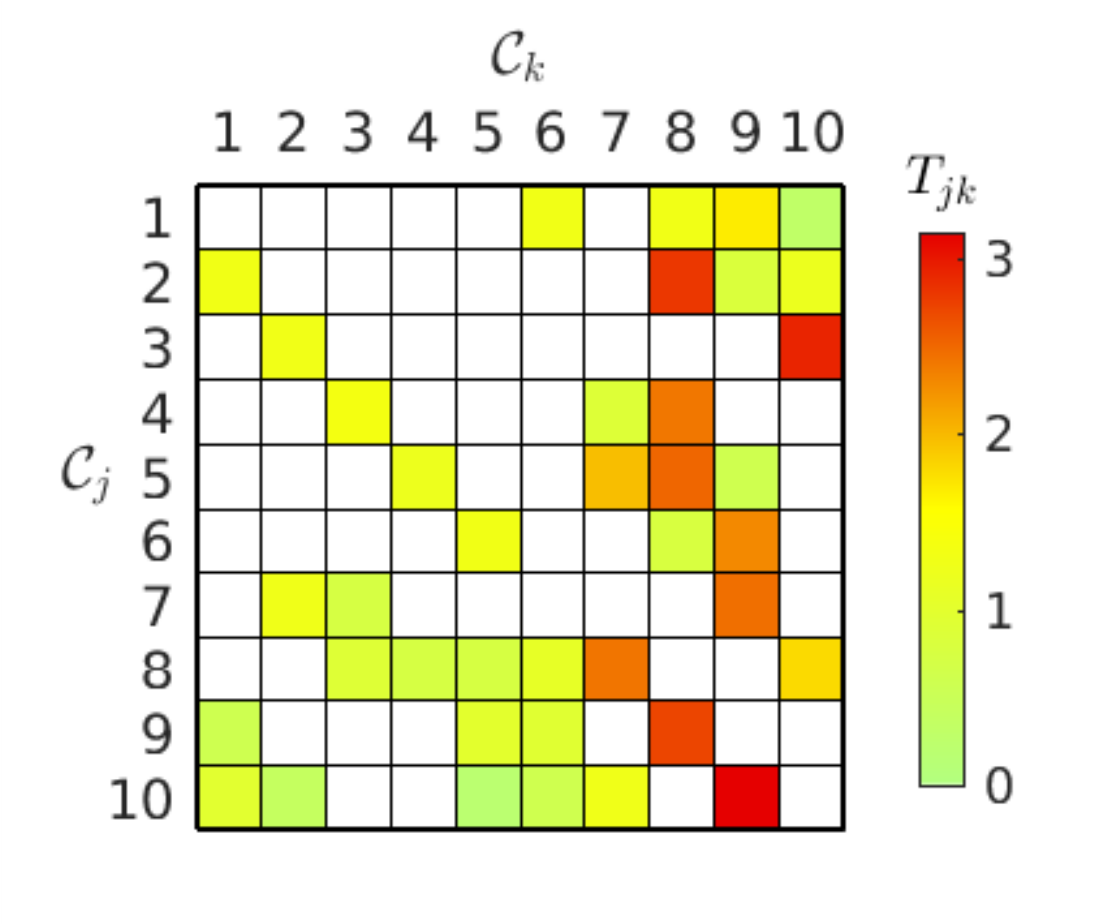}} \\
\end{tabular}
\caption{Cluster-based analysis in layer 2 at $\Rey= 130$ for the chaotic clusters of $\mathcal{L}_{1}$: 
(a) transition illustrated with cluster label, 
(b) transition matrix, 
(c) residence time matrix, as in figure~\ref{Fig:Matrix80_layer1}.
All four trajectories reach the chaotic clusters,with $m_2=1, \, \ldots , \, 8720$, $8721, \, \ldots , \, 17420$, $17421, \, \ldots , \, 24056$ and $24057, \, \ldots , \, 30989$.}
\label{Fig:Matrix130_layer2_CHAOS}
\end{figure}

From figure~\ref{Fig:Matrix130_layer1}(a), all four trajectories will reach the same chaotic attractor described by the chaotic clusters, with the original snapshot index $m=6281, \, \ldots , \, 15000$, $21301, \, \ldots , \, 30000$, $33365, \, \ldots , \, 40000$ and $43068, \, \ldots , \, 50000$.
In figure~\ref{Fig:ClusterAnalysis130_layer2_CHAOS}, the closed orbit of the clusters $\mathcal{C}_1 \rightarrow \ldots \rightarrow \mathcal{C}_{6} \rightarrow \mathcal{C}_1$ is formed, with a relatively high probability of transition between the successive clusters. 
The flow field of the centroids of the first six clusters form a complete cycle of vortex shedding. 
This is interesting, as it reminds the periodic and quasi-periodic dynamics respectively observed at $\Rey=80$ and $\Rey=105$.
The periodic block of the transition matrix of figure~\ref{Fig:Matrix130_layer2_CHAOS}(b), from $\mathcal{C}_1$ to $\mathcal{C}_6$, corroborates the existence of the periodic dynamics. 
The centroids of the clusters in the cycle present a similar structure of coherence, as can be seen in figure~\ref{Fig:ClusterAnalysis130_layer2_CHAOS}. 
There are also other possible transitions from clusters on this orbit to the remaining clusters with much smaller probability.
The residence times shown in figure~\ref{Fig:Matrix130_layer2_CHAOS}(c) for each cluster on the orbit are uniform, and the averaged period along the complete cycle is $7.78$ by summing up the elements in the block from $\mathcal{C}_1$ to $\mathcal{C}_6$.

The remaining clusters $\mathcal{C}_{7}, \ldots, \mathcal{C}_{10}$ have multiple destinations with quasi-random possibilities. Even though the probabilities of these random transitions are small, they contribute to the chaotic dynamics of the flow field, with recurrent transitions $\mathcal{C}_1 \rightarrow \mathcal{C}_{10}$, $\mathcal{C}_{9} \rightarrow \mathcal{C}_1$, and so on. 
The flow fields of the associated centroids are shown in figure~\ref{Fig:ClusterAnalysis130_layer2_CHAOS}. Their structure looks like distortions of the vortex shedding cycle formed by the first six clusters.
The network model indicates that the fully chaotic state still contains a main cycle of clusters associated with a periodic vortex shedding, together with the random jumping to the clusters associated with a stochastic disorder in the wake. In this case, the transition matrix can be used to build a stochastic model, as shown in appendix~\ref{Sec:Stochastic}.

\subsubsection{Dynamics reconstruction of the hierarchical network model at $\Rey=130$}
\label{SubSec:HNM130_Validation}

\begin{figure}
\centerline{
\begin{tabular}{cccc}
(a) & \raisebox{-0.5\height}{\includegraphics[width=.45\linewidth]{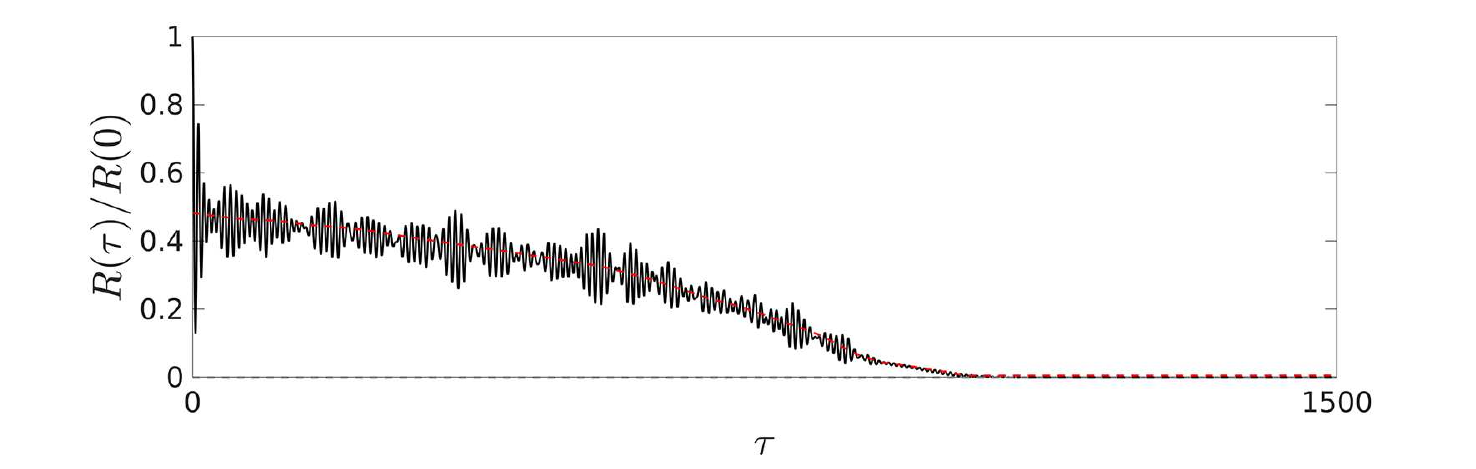}} &
(b) & \raisebox{-0.5\height}{\includegraphics[width=.45\linewidth]{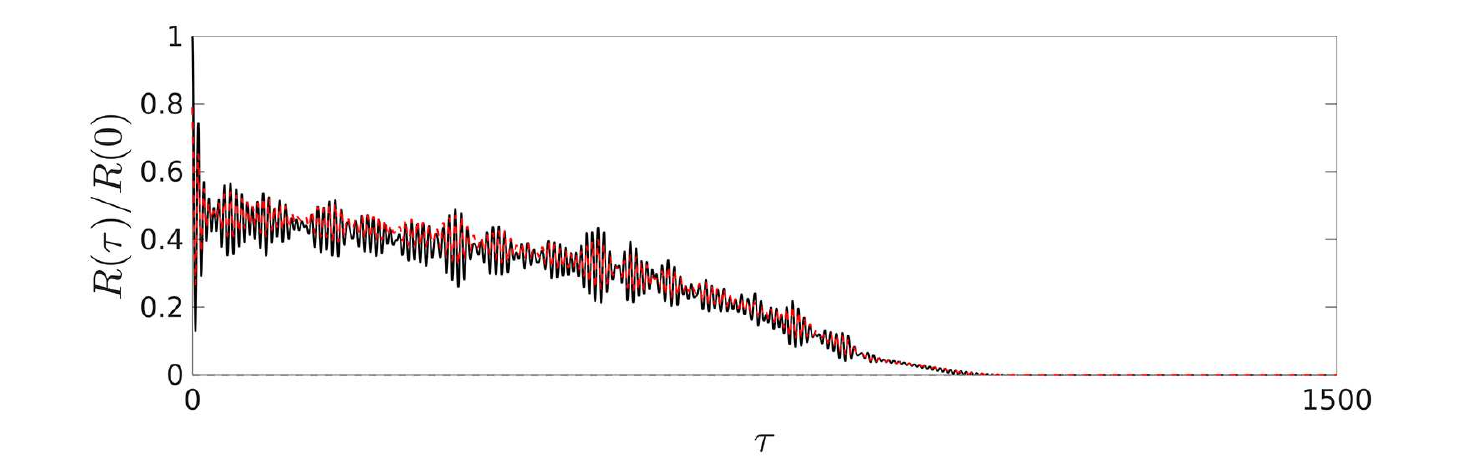}}
\end{tabular}
}
\caption{Autocorrelation function for  $\tau \in [0, 1500)$ from DNS (black solid line) and the the hierarchical network model (red dashed line)  in the two layers: (a) $\mathcal{L}_{1}$  and (b) $\mathcal{L}_{2}$, at $\Rey=130$.}
\label{Fig:Validation130}
\end{figure}

Figure~\ref{Fig:Validation130} shows the autocorrelation function of the DNS and the HiCNM in the two layers. 
We stop the hierarchical modelling in layer $\mathcal{L}_{2}$,  as both the transient and chaotic dynamics can be fairly reproduced with a limited number of clusters.
The autocorrelation function of the DNS shows chaotic oscillations with a dominant frequency.
In layer $\mathcal{L}_{1}$, no oscillation can be identified, and the RMSE of the autocorrelation function is $R_{\rm rms}^1 = 6.94$.
In layer $\mathcal{L}_{2}$, $R_{\rm rms}^2 = 3.42$. The autocorrelation function of the model matches well with the dominant frequency of the oscillations, but the $\hat{R}_{\mathcal{L}_{2}}(\tau)$ value can hardly match.

\section{Conclusion}
\label{Sec:Conclusions}

We have proposed a data-driven modelling methodology, which consists of hierarchical clustering and network modelling on top of the cluster-based reduced-order model (CROM).
The hierarchical structure is physically consistent with the weakly nonlinear model derived from the mean-field consideration.
The flow field is decomposed into a hierarchy of components, namely the slowly varying mean-flow field, the dominant vortex shedding and the secondary components.
The resulting hierarchical cluster-based reduced-order model (HiCROM) can automate the modelling process based on the representative states and systematically trace the flow dynamics on multiple scales, involving multiple frequencies and multiple attractors.
The cluster-based hierarchical network model (HiCNM) presented in this work is a HiCROM using the directed network to describe the non-trivial transitions between clusters.
Based on the classical CNM, we derived the HiCNM for the transient and post-transient dynamics of the two-dimensional incompressible ``fluidic pinball'', characterized by multiple invariant sets and dynamics, for different Reynolds numbers. 

The considered data set consists of snapshots of the velocity field computed from the direct numerical simulation starting with different initial conditions, which refer to four trajectories: two mirror-conjugated trajectories starting in the vicinity of the symmetric steady solution and the two others starting from the two asymmetric steady solutions.
At the considered Reynolds numbers, all the steady solutions are unstable. 
In this sense, the data set contains the transient and post-transient dynamics involving all the invariant sets of the system. 
The hierarchical modelling is based on the hierarchical clustering under the mean-field consideration.
In the first layer, the first $k$-means$++$ clustering algorithm is applied to the low-pass filtered data and partitions the snapshots into different clusters, as in algorithm~\ref{alg:FilteredKmeans}.
The flow field fluctuations are responsible for the nonlinear mean-field distortions through the Reynolds stress. As a result, the snapshots in different clusters exhibit different states of slowly changing mean-flow field. The network model in the first layer focuses on the global transitions between different invariant sets. 
The clustering result in the first layer will guide the clustering process in the second layer.
The second clustering process in algorithm~\ref{alg:KmeansChildLayer} partitions the snapshots in the same cluster again into sub-clusters, according to the original data.
Based on the sub-clusters, a new network model can be built with a better interpolation of the local dynamics.
The clustering process is similar to the divisive hierarchical clustering, which can continue until each snapshot is a cluster. However, the number of layers depends on the number of characteristic scales in the system, such as the number of coherent frequencies or the fast and slow terms. Hence, two or three layers will be enough to extract the transient and post-transient dynamics out of multiple invariant sets and multiple frequencies in our case.

At $\Rey=80$, the six invariant sets were well identified in the first layer of the HiCNM, including the dynamics around three unstable steady solutions ($\mathcal{C}_1$, $\mathcal{C}_{12}$, $\mathcal{C}_{17}$), one unstable symmetric limit cycle ($\mathcal{C}_7$) and two stable asymmetric limit cycles ($\mathcal{C}_9$, $\mathcal{C}_{11}$).  
The transient dynamics between the multiple invariant sets, and the temporal development of the degrees of freedom associated with the static symmetry breaking, are identified by the model in the first layer.
We further presented the model in the second layer involving the three exact solutions: the destabilization of the symmetric steady solution in cluster $\mathcal{C}_{1}$, the dynamics around the symmetric limit cycle in cluster $\mathcal{C}_{7}$, and the permanent regime on the asymmetric limit cycle in cluster $\mathcal{C}_{9}$.
Compared to a CNM with the same number of clusters, the HiCNM preserves the advantage of automatable modeling, optimizes the cluster distribution, and makes it human-interpretable.

For the quasi-periodic flow regime at $\Rey=105$, the first two layers are identical to the case at $Re=80$.
The HiCNM in the first layer identified the different invariant sets, and the model in the second layer described the local dynamics on the invariant sets. We further introduced the third layer to characterize the new coherent structures at low frequency. 
The sub-division clustering in algorithm~\ref{alg:KmeansChildLayer} was applied on the entering snapshots of the cluster $\mathcal{C}_{6,\, 10}$. 
The low frequency was successfully identified, while the centroids identified the tiny changes of the oscillating jet in the near wake.

At $\Rey=130$, three unstable steady solutions and one chaotic attracting set have been caught in the first layer. The chaotic zone was divided into several clusters. 
In the second layer, we focused on the local structures around the invariant sets. 
We determined the dynamics of the initial transients from the unstable steady solutions in $\mathcal{C}_1$ and $\mathcal{C}_{10}$.
Besides, to preserve the continuity of the data, the second clustering process was applied to the group of clusters in the chaotic regime $\mathcal{C}_{k_1}$, $k_1=4,\dots,7$.
An unstable cycle was identified for the chaotic regime, characterized by random transitions from and to the chaotic clusters with low probability.

Compared to other reduced-order modelling strategies, HiCROM inherits the excellent recognition performance of classical CROM, and provides a universal modelling strategy for identifying transient and post-transient dynamics in a self-supervised manner. 
Multiple transient dynamics can be considered at the same time, which gives a global view of the trajectories between the different invariant sets. Thus, it provides a better understanding of the complex flow dynamics for the multiscale, multi-frequencies and multi-attractors problem. 
To summarize, the HiCNM applied in this work has the following advantages comparing to classical CNM:
\begin{enumerate}[(i) ]
    \item A more robust clustering result with the hierarchical modelling under mean-field consideration, and a better distribution of the clusters.
    \item Better ability to identify the topology on the multi-attractor and multiscale problem. HiCNM identifies transient trajectories between different invariant sets, and locally constructs new CNMs for the different invariant sets if necessary.
    \item No need to find a good compromise between the resolution (the number of clusters) and the network complexity. All clustering algorithms use 10-20 clusters, and the number of clusters depends on the accurate representation needs of the scale involved in the layer. The number of layers depends on the number of characteristic scales in the system. The clusters of different scales or characteristics can be systematically distributed in multiple layers with a clear hierarchy. 
    \item The subdivision is flexible according to the actual needs. We can freely choose one or more object clusters that need to be divided in the next layer to preserve the time continuity and the local characteristics of the data. 
\end{enumerate}
The price is the need for a physical intuition guiding the hierarchical clustering and modelling
by adjusting the design parameters. 
As exemplified in this work, the HiCNM at $\Rey=80$ is fully automated. After reaching the second layer, the HiCNM can well identify the mean flow and the coherent structure with a single frequency. The deeper layers cannot identify other meaningful dynamics. 
At $\Rey=105$, the third layer can identify the secondary frequency with the entering sanpshots in one cluster in the second layer. 
At $\Rey=130$, we merge the asymptotic chaotic clusters in the first layer together, and build a model for them in the second layer to maintain their dynamic continuity. This sacrifices automation but results in a better dynamic representation. We foresee other data-driven methods for these decisions in future work to promote the automation of the hierarchical network model.

In summary, HiCNM provides a flexible and automatable cluster-based modelling framework for complex flow dynamics, 
and shows its ability and applicability to identify transient and post-transient, multi-frequency, multi-attractor dynamics.  
Since the Reynolds decomposition under the mean-field consideration is common for fluid flows, the hierarchical strategy should be extendable to other flows.
Especially for cases that require high-resolution analysis with a large number of clusters, HiCNM greatly simplifies the complexity of the analysis and improves its interpretation.
For high turbulence without frequency boundaries, such as isotropic turbulence, it is reasonable to use a network model first to check whether there is a grouping relationship between clusters, before deciding whether a HiCNM needs to be constructed.
HiCNM is promising for a variety of potential applications, such as topology identification of the state space for complex dynamics, recognition and analysis of the temporal evolution of degrees of freedom associated with different types of instabilities (Hopf, pitchfork, etc.), and feature extraction of the dynamical structure when different spatial/temporal scales are involved.
An alternative direction of this work is the HiCNM-based control, with the aim to find an optimized control strategy from various control laws with multiple scales in different layers.

\section*{Acknowledgements}
Nan Deng appreciates the support of the China Scholarships Council (No.201808070123) during his Ph.D. thesis in the ENSTA Paris of Institut Polytechnique de Paris, 
and numerical supports from the laboratories LISN (CNRS-UPR 3251) and IMSIA (UMR EDF-ENSTA-CNRS-CEA 9219).

This work is supported by a public grant overseen by the French National Research Agency (ANR) under grant `FlowCon' (ANR-17-ASTR-0022), 
the National Natural Science Foundation of China (NSFC) under grant 12172109,
and the Polish Ministry of Science and Higher Education (MNiSW) under the Grant No.: 0612/SBAD/3567.

We appreciate valuable discussions with Guy Cornejo-Maceda, Hao Li, and Fran\c{o}is Lusseyran
and the HIT fluidic pinball team: Bingxi Huang, Wenpeng Li, Yiqing Li, Kexin Zhao, Qixin Lin, Ruixuan Shen, Xin Wang and Shangyan Xie.
Last but not least, we thank the referees for their thoughtful suggestions, which have significantly improved the manuscript.

{\bf Declaration of Interests}. The authors report no conflict of interest.

\appendix
\section{Blockage effect in the fluidic pinball}
\label{Sec:BlockageRatio}

\begin{figure}
 \centerline{
 \includegraphics[width = .55\linewidth]{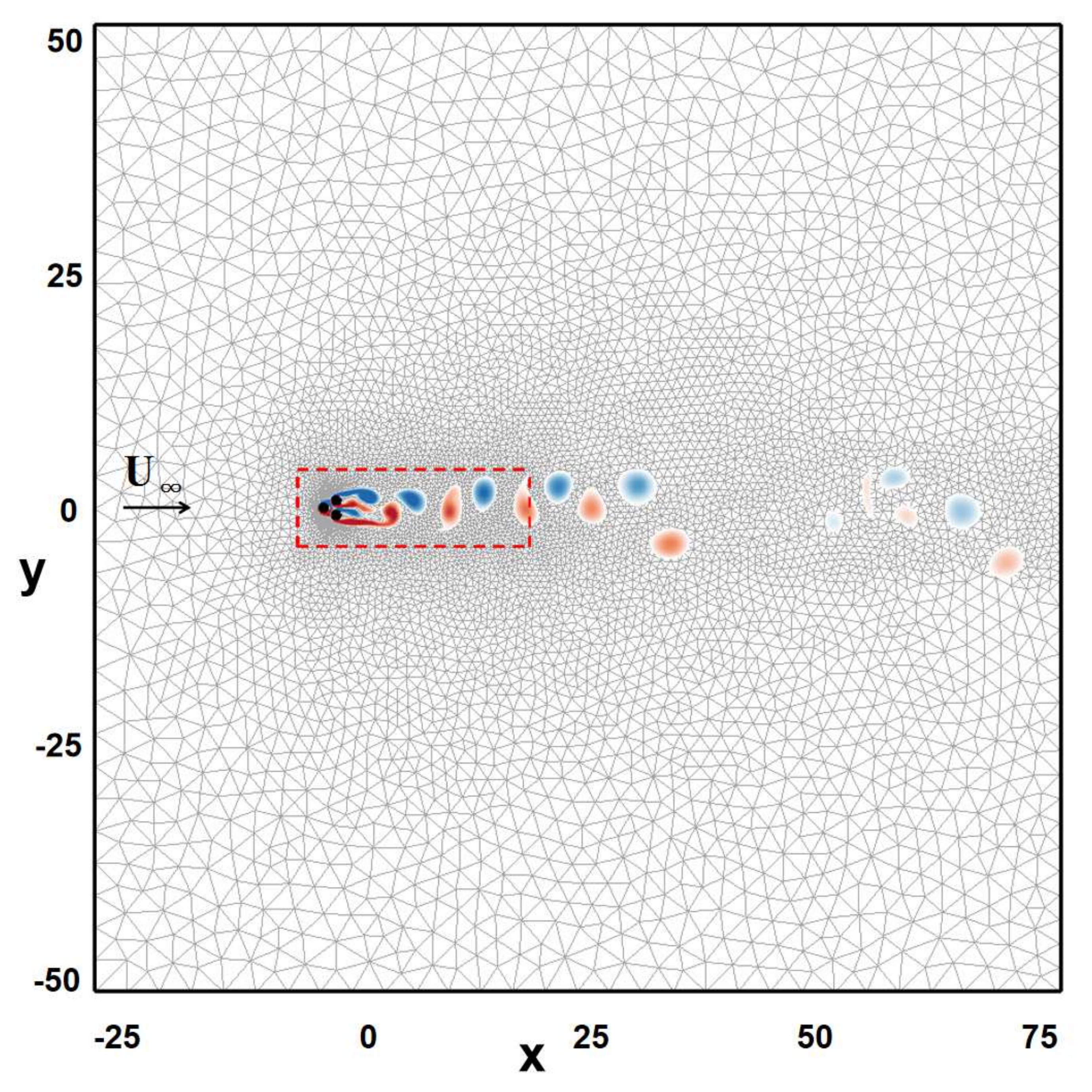}
 }
\caption{Configuration of the fluidic pinball and dimensions of the simulated domain of the blockage ratio $B=0.025$. 
A typical field of vorticity at $\Rey=150$ is represented in color with $[-1.5, 1.5]$. The upstream velocity is denoted $ U_\infty $.  An observation zone of size $[-4D,+20D]\times[-4D,+4D]$ is marked out with a red dashed box.}
\label{Fig:pinball2}
\end{figure}

We run the DNS with an enlarged computational domain with the blockage ratio $B=0.025$, which is bounded by a rectangular box of size $[-25D,+75D]\times[-50D,+50D]$, as shown in figure~\ref{Fig:pinball2}. 
The unstructured grid has 14 831  triangles and 29 961 vertices as compared to the grid in \S~\ref{Sec:Pinball}.

\begin{figure}
\centerline{
\begin{tabular}{cccc}
(a) & \raisebox{-0.5\height}{\includegraphics[width=.45\linewidth]{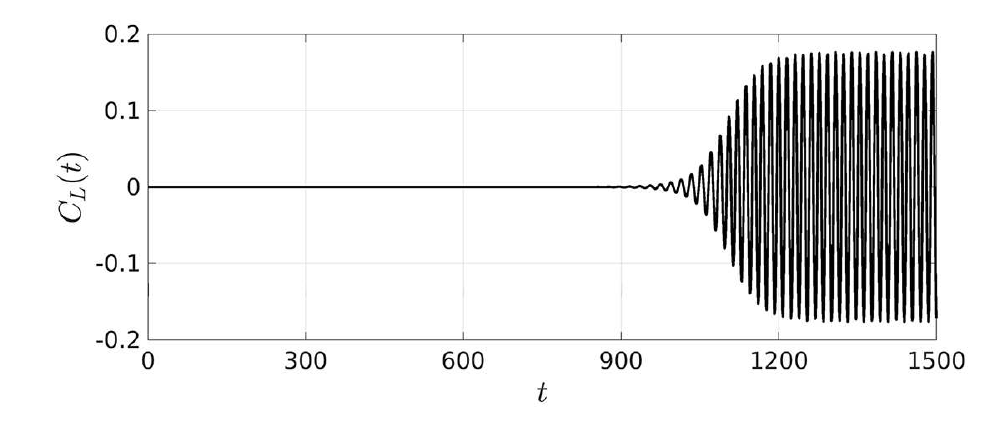}} &  
(b) & \raisebox{-0.5\height}{\includegraphics[width=.45\linewidth]{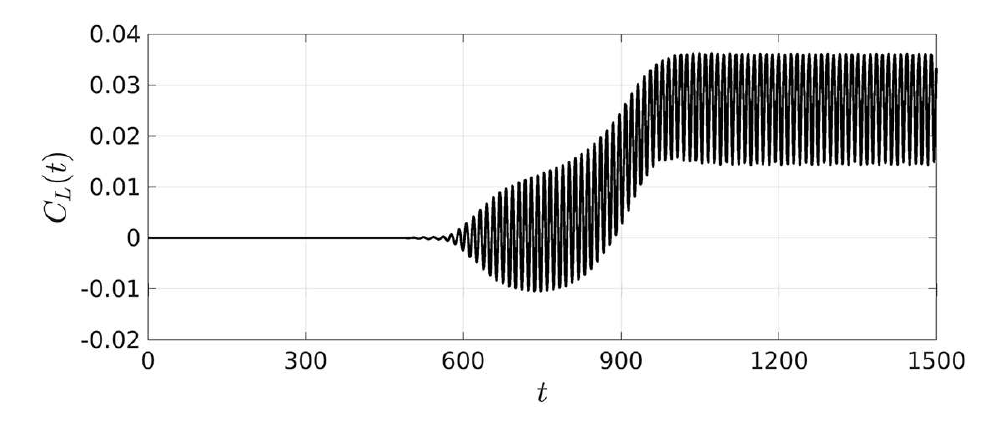}} \\
(c) & \raisebox{-0.5\height}{\includegraphics[width=.45\linewidth]{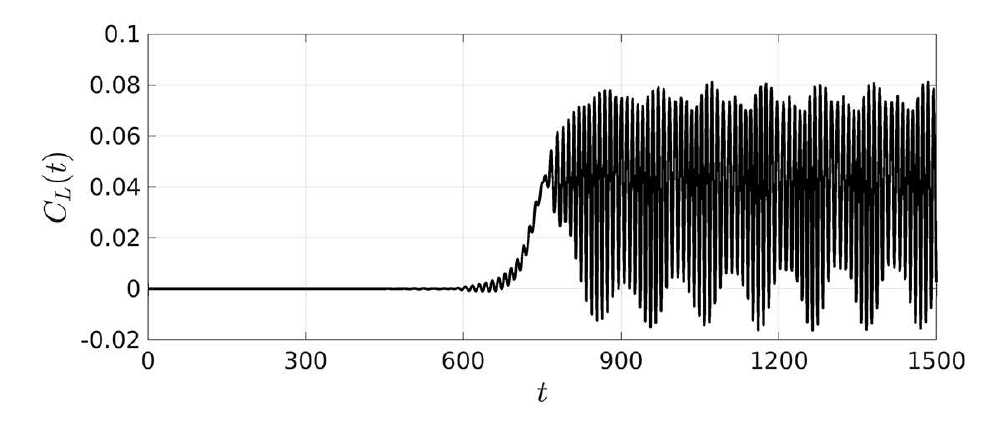}} &
(d) & \raisebox{-0.5\height}{\includegraphics[width=.45\linewidth]{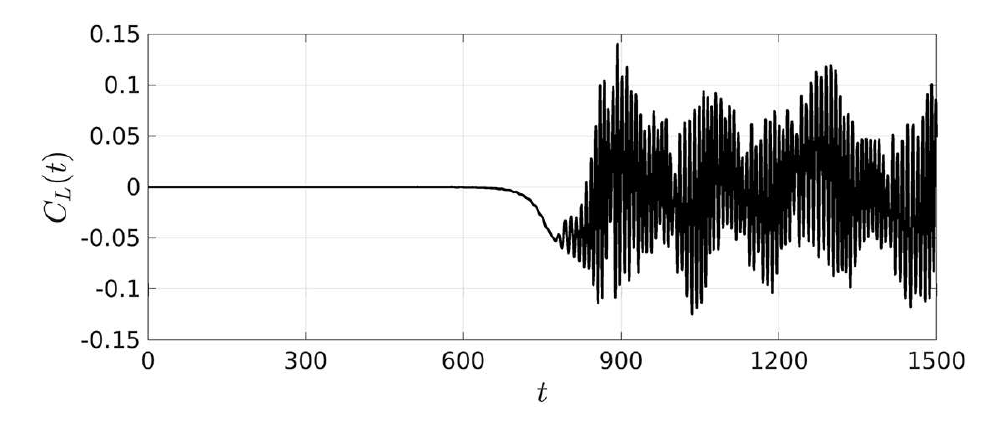} }\\
\end{tabular}
}
\caption{Transient and post-transient dynamics of the fluidic pinball with the enlarged grid, illustrated with the time evolution of the lift coefficients $C_L$ starting with the symmetric steady solution at different Reynolds numbers: $\Rey = 30$ (a) , $90$ (b) , $120$ (c), $150$ (d).}
\label{Fig:CL_NewGrid}
\end{figure}

The vortices in the near wake $0<x<20D$ are concentrated in $|y|<4D$.
Linear stability analysis of the symmetric steady solution indicates that the critical value of the primary Hopf bifurcation does not change $\Rey_1=18$, but the next bifurcations are found for larger Reynolds numbers. 
The pitchfork bifurcation of the symmetric steady solution is changed to $\Rey_2=81$, but the transient and post transient dynamics for different flow regimes remain qualitatively the same, as shown in figure~\ref{Fig:CL_NewGrid}. 

Enlarging the computational domain reduces the blockage effect, but the blockage is practically difficult to suppress or even reduce.
In this work, we are interested in the richness of the dynamics to evaluate our method.
The blockage is not critical as we have similar numerical results for the transient and post-transient dynamics.
How the location of lateral boundaries, as well as upstream and downstream boundaries, affect the bounded flow will be discussed in our future work.

\section{Vertical transitions in the hierarchical network model}
\label{Sec:VT_HiCNM}

In figure~\ref{Fig:H_network_cell}, the snapshots entering into and leaving from $\mathcal{C}_{k_{l-1}}$ are marked out for each trajectory by the characteristic function \eqref{Eqn:CharacteristicFunction2}, and can be used to describe the vertical transitions.

When a cluster in the parent layer $\mathcal{C}_{k_{l-1}}$ is activated, it also activates a horizontal transition through all its subclusters in the child layer.
The ports of entry and exit for the subclusters are indicated by the entering and exiting snapshots.
The entering snapshots belong to the first activated sub-cluster in each trajectory.
The horizontal transition through subclusters ends with the exiting snapshots, and is forced to return to the parent layer.
At the next time step after the exiting snapshots, $\mathcal{C}_{k_{l-1}}$ and all its subclusters will deactivate. 

With an additional condition from the subclusters of the child layer, the characteristic function \eqref{Eqn:CharacteristicFunction2} can be defined as
\begin{subequations}
\label{Eqn:CharacteristicFunction3}
\begin{eqnarray}
  \chi_{o, k_{l-1} \rightarrow k_{l}}^{m} &:= & \begin{cases}
                1, &  \mbox{if }\bm{u}^{m-1} \notin \mathcal{C}_{k_{l-1}} \And \bm{u}^{m} \in \mathcal{C}_{k_{l-1}, k_{l}}, \\ 
                0, &  \mbox{otherwise}.
               \end{cases} \\
  \chi_{e, k_{l-1} \rightarrow k_{l}}^{m} &:= & \begin{cases}
                1, &  \mbox{if }\bm{u}^{m+1} \notin \mathcal{C}_{k_{l-1}}  \And \bm{u}^{m} \in \mathcal{C}_{k_{l-1}, k_{l}}, \\ 
                0, &  \mbox{otherwise}.
               \end{cases}
\end{eqnarray}
\end{subequations}

The number of entering snapshots $n_o$ and of exiting snapshots $n_e$ in each subclusters $\mathcal{C}_{k_1,\cdots, k_{l}}$ read
\begin{equation}
\label{Eqn: NumVTrans}
n_{o,\,k_{l}} = \sum\limits_{m=1}^M\, \chi_{o, k_{l-1} \rightarrow k_{l}}^{m}, \> \> \>
n_{e,\,k_{l}} = \sum\limits_{m=1}^M\, \chi_{e, k_{l-1} \rightarrow k_{l}}^{m},
\end{equation}
where only the final subscript of the subcluster index in the current layer is indicated, and as well in the following.

For the cluster $\mathcal{C}_{k_{l-1}}$, the probability of vertical transition into and out of the child layer $Q_{o,k_{l}}$ and $Q_{e,k_{l}}$, are defined as
\begin{equation}
\label{Eqn: ProbabilityVTrans}
Q_{o,\, k_{l}} = \frac{n_{o,\,k_{l}}}{n_o},   \>    \>   
Q_{e,\, k_{l}} = \frac{n_{e,\,k_{l}}}{n_e}.
\end{equation}
Note that $\sum\limits_{k_{l}=1}^{k_{l}} Q_{o,\,k_{l}} =1$ and $\sum\limits_{k_{l}=1}^{k_{l}} Q_{e,\, k_{l}} =1$.

\subsection*{An example of the hierarchical network model}
\label{SubSubSec:example}

\begin{figure}
 \centerline{
 \includegraphics[width = .9\linewidth]{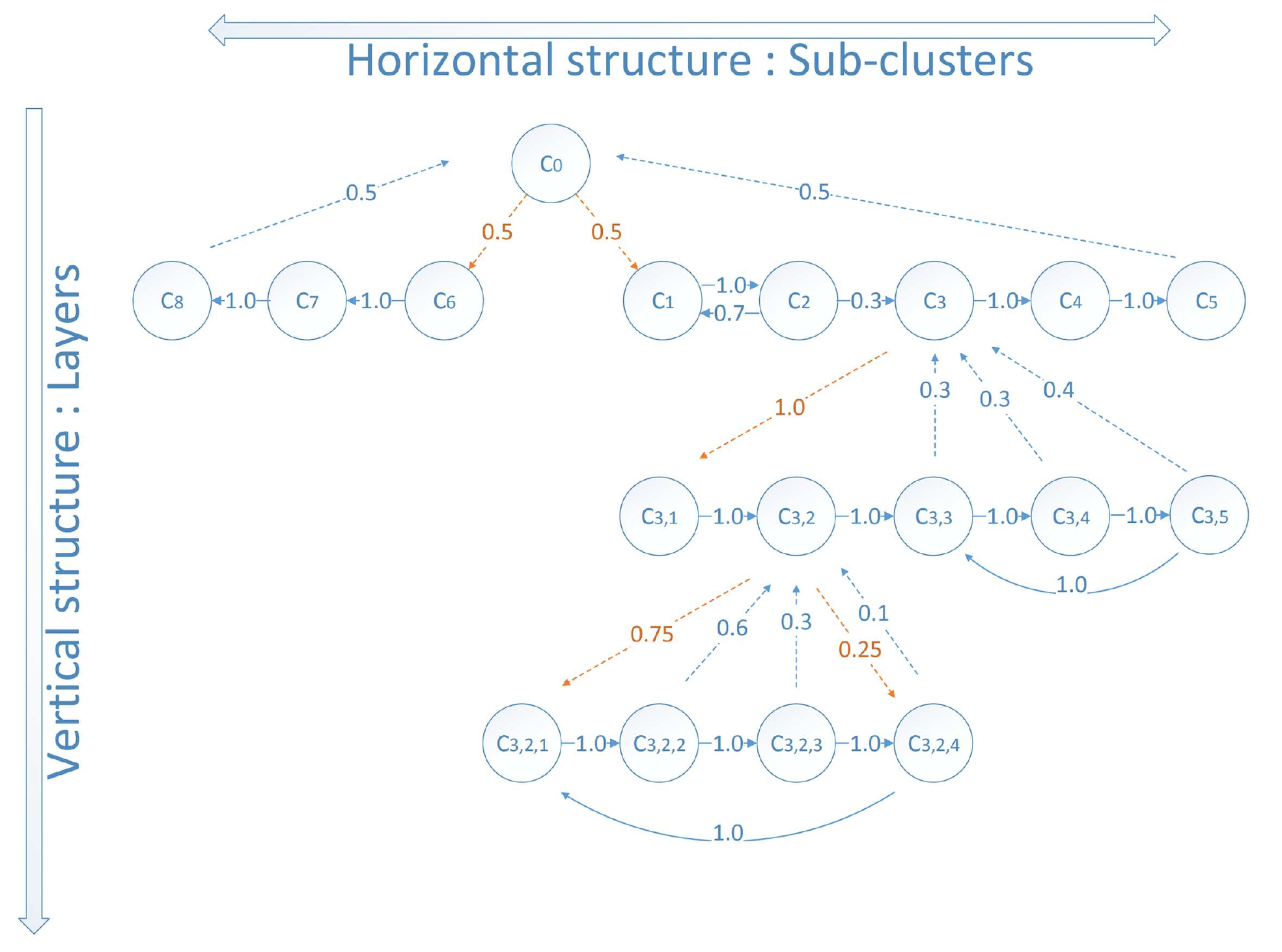}
 }
\caption{An illustration of an hierarchical network model. 
The vertical structure describes the transitions between the parent-child layers with dashed arrows: the red dashed arrows present the possible inlet from the parent layer to the child layer, and the blue dashed arrows denote the outlet back to the parent layer. 
The horizontal structure presents the transition between the subclusters with arrows.
The numbers indicate the possibility of each transition.}
\label{Fig:H_network_illustration}
\end{figure}

An example of HiCNM is illustrated in figure~\ref{Fig:H_network_illustration}. The model is constructed from the full data set containing one or several individual trajectories of discrete state snapshots, each starting from different initial conditions and converging toward possibly different attracting sets. 
From the top to the bottom of this tree structure, only one cluster in each parent layer has been chosen and divided into subclusters in the child layer. 
For the clusters in each layer, we only indicate the sub-cluster number in the current layer.

The full data set is treated as one cluster $\mathcal{C}_0$ on the top, which ensembles all the snapshots.
The clustering algorithm in the first layer has divided the snapshots in this cluster into eight subclusters.
Two entering subclusters $\mathcal{C}_1$ and $\mathcal{C}_6$ are sketched with the same probability, as could be found when the full data set contains two pathways with equal probability in the state space of the system. 
The trajectories starting from the two entering subclusters have no intersecting cluster, and return to $\mathcal{C}_0$ from their respective existing subclusters,  $\mathcal{C}_5$ and $\mathcal{C}_8$, with the same probability. 
The trajectory in layer $\mathcal{L}_1$, $\mathcal{C}_6 \rightarrow \mathcal{C}_7 \rightarrow \mathcal{C}_8$ is a simple one-way transition. The trajectory $\mathcal{C}_1 \rightarrow \dots \rightarrow \mathcal{C}_5$ is more complex, due to a possible return dynamics from  $\mathcal{C}_2$ to $\mathcal{C}_1$.
The clustering algorithm works on the snapshots in cluster $\mathcal{C}_3$ and has divided them into five subclusters in the second layer $\mathcal{L}_2$. 
For the vertical transition, only one entering sub-cluster has been found but with three exiting subclusters, which means that the snapshots leaving cluster $\mathcal{C}_3$ belong to one of these three subclusters. The dynamics between the subclusters is a simple one-way transition, but with a cycle $\mathcal{C}_{3,\, 3} \rightarrow \mathcal{C}_{3,\, 4}  \rightarrow \mathcal{C}_{3,\, 5}  \rightarrow \mathcal{C}_{3,\, 3} $, which indicates a periodic dynamics.
Next the clustering algorithm is applied to the snapshots in cluster $\mathcal{C}_{3,\, 2}$, resulting in four subclusters in the third layer $\mathcal{L}_3$. 
Among the subclusters, there exist two entering subclusters and three exiting subclusters. 
We notice that $\mathcal{C}_{3,\, 2,\, 4}$ can work as either an entering sub-cluster or an exiting sub-cluster.
The reason is that the entering snapshots and the exiting snapshots of  $\mathcal{C}_{3,\, 2}$ belong to the same sub-cluster in the child layer.
A periodic dynamics exists between the subclusters $\mathcal{C}_{3,\, 2,\, 1} \rightarrow \dots \rightarrow \mathcal{C}_{3,\, 2,\, 4} \rightarrow \mathcal{C}_{3,\, 2,\, 1}$.

A full dendrogram, as in figure~\ref{Fig:H_structure}, is also available if all the clusters in the parent layer are divided.
However, in actual practice, it is not necessary to divide every cluster in the state space.
The clustering algorithm in the first layer divides different invariant sets and the transient states into different clusters. A classic CNM is used to describe the transient dynamics between the invariant sets.
In the second layer, the refined dynamics on the invariant sets exhibits new interesting features.
The snapshots are often concentrated close to the stable/unstable invariant sets.
Within the same invariant set, the snapshots have a relatively homogeneous distribution according to certain rules.
The local dynamics are relatively simple and easy to extract because the clustering result depends entirely on the distribution in the state space.
The transient states from an unstable set to a stable set mix the dynamical behavior of different invariant sets.
Hence, the clusters close to invariant sets need to be divided again, namely close to steady solution, to metastable solution, or to the stable solution.
Usually, these clusters have some characteristics, like a large number of snapshots in the cluster or multiple possible transitions to other clusters.

\section{Clustering with POD}
\label{Sec:POD}

The computational cost can be significantly reduced with POD, as a lossless POD can highly compress the flow field data to accelerate the clustering algorithm.
The clustering algorithm can be applied to the compressed data instead of the high-dimensional velocity fields.
In this work, the snapshots of velocity field are pre-processed by a proper orthogonal decomposition (POD), where $\bm{u}_s(\bm{x})$ is the symmetric steady solution at the Reynolds number under consideration.
Compared to the classical POD method, the symmetric steady solution $\bm{u}_s(\bm{x})$ has been used instead of the ensemble-averaged mean flow $\bm{\bar{u}}(\bm{x})$, because our analysis deals with multiple invariant sets and the mean flow is not a Navier-Stokes solution, which has no dynamical relevance.  
The fluctuating flow field can be decomposed on the basis of the POD modes $\bm{u}_i(\bm{x})$, 
\begin{equation}
\bm{u}^m(\bm{x})-\bm{u}_s(\bm{x}) \approx \sum _{i=1}^{N} a_i^m\,\bm{u}_i(\bm{x}),
 \label{Eqn:POD}
\end{equation}
where the $a_i^m$ are the mode amplitudes. A complete basis for the modal decomposition is given when $N=M$ \citep{berkooz1993arfm}. 
For our cluster-based analysis, the number of modes could be reduced to $N=400$ without loss of relevant information. 
The pre-processing algorithm is detailed in algorithm~\ref{alg:POD}. 
\begin{algorithm}[h]  
  \caption{Pre-processing the velocity field by POD}  
      \label{alg:POD}  
  \begin{algorithmic}[1]  
    \Require  
      $\bm{u}^m$: snapshots of velocity field; \quad
      $\bm{u}_s$: symmetric steady solution
    \Ensure  
      $\bm{u}_i$: leading POD modes;\quad
      $a_i^m$: mode amplitudes
    \State compute POD modes $\bm{u}_i,\, i=1,\dots,N$, for the data base $\{\bm{u}^m(\bm{x})\}, \, m=1,\dots,M$, with choosing $\bm{u}_s$ as the base-flow; 
    \State compute the mode amplitudes $a_i^m = \left( \bm{u}^m- \bm{u}_s, \bm{u}_i \right)_{\Omega}$; \State save the leading $N$ POD modes and the corresponding mode amplitudes.
  \end{algorithmic}  
\end{algorithm}
The computational cost for the cluster analysis can be significantly reduced in the POD subspace \citep{kaiser2014jfm, li2020CNM}, thus enabling an accurate compressed sensing of the original datasets.
The centroids of the velocity field based on the POD mode amplitudes now read:
\begin{equation}
\bm{c}_{k} = \bm{u}_s + \frac{1}{n_k} \sum\limits_{m=1}^M  \> \chi_{k}^{m} \sum\limits_{i=1}^{N} a_i^m\bm{u}_i.
\end{equation} 

We note that this POD process is just an option for data compression to speed up the clustering process, which can approximate the data distribution in a POD subspace with high accuracy. We can even apply the cluster-based approach to a feature-based subspace of the flow data, as \citet{nair2019JFM} who applied CROM to a 3D phase space of the drag and lift forces.  

In contrast to the non-linear reconstruction of flows using a POD basis, the cluster-based approach neither decomposes the flow field nor extracts the dominant structures with the most fluctuating energies.
Instead, it gathers similar snapshots and represents them with a linear combination of snapshots within the cluster.  This combination is always in the original data space and there is no projection comparing to a POD model. 
In addition, the modelling process of the cluster-based method is automated based on the data topology in the state space, which is even suitable for multiple invariant sets.
The differences between the cluster-based methods and the POD-based model have been discussed in \citet{kaiser2014jfm, li2020CNM}. As an extension, the HiCNM strategy provides a multiscale solution for describing complex flow with transient dynamics.
\section{Stochastic model for asymptotic regime}
\label{Sec:Stochastic}

In \S~\ref{Sec:HiCROM_PINBALL}, we have applied HiCNMs for the identification and analysis of complex dynamics with multiple scales and multiple invariant sets.
The centroids and cluster index provide a concise representation of the original data-set, and we can reconstruct the flow based on the time evolution of the cluster index.
From the cluster analysis, we use a transition matrix to statistically record the possible motions between clusters, which can be used to predict the evolution of the cluster index. Hence, the dynamics reconstruction can also come from the stochastic model based on the transition matrix.

For a single trajectory, the transient dynamics is fully predictable and converges to the asymptotic dynamics.
A stochastic model for the transient dynamics is not suggested because the clustering result may suffer from the multiscale problem, and some random walks will be mistakenly introduced into the transition matrix, like the destabilizing stage from the steady solution.
In contrast, if there is no multiscale problem, all the transition probabilities should be 1 due to the predictable transient dynamics. Hence, there is no need to build the dynamics from a stochastic model and to discuss the probability distribution.

\begin{figure}
\centerline{
\begin{tabular}{cc}
\includegraphics[width=.6\linewidth]{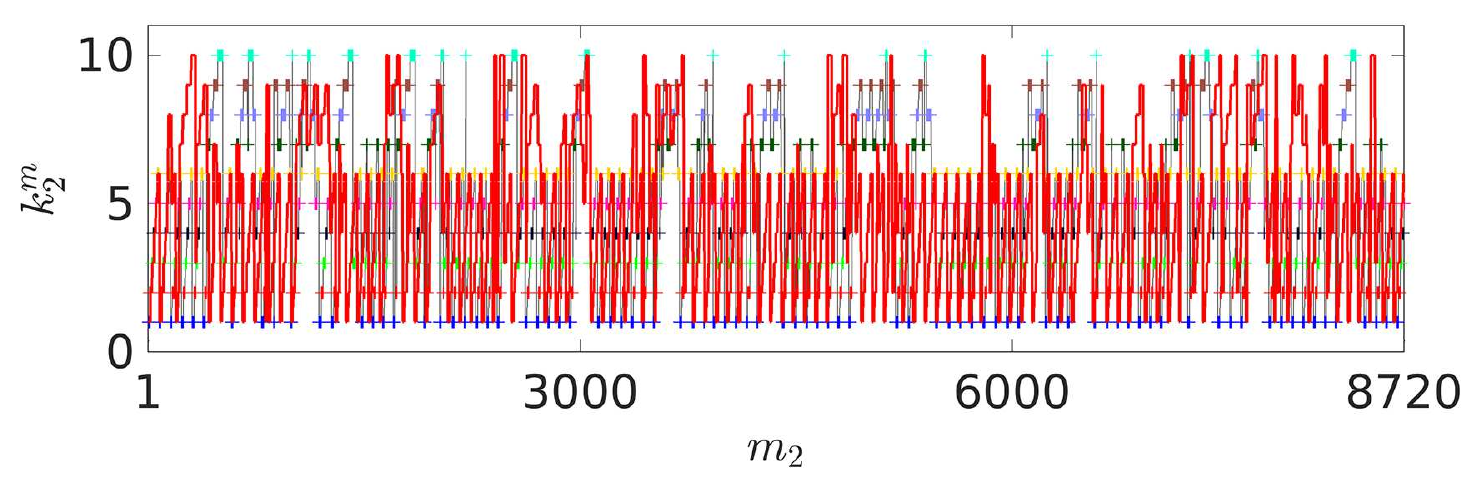} &
\includegraphics[width=.3\linewidth]{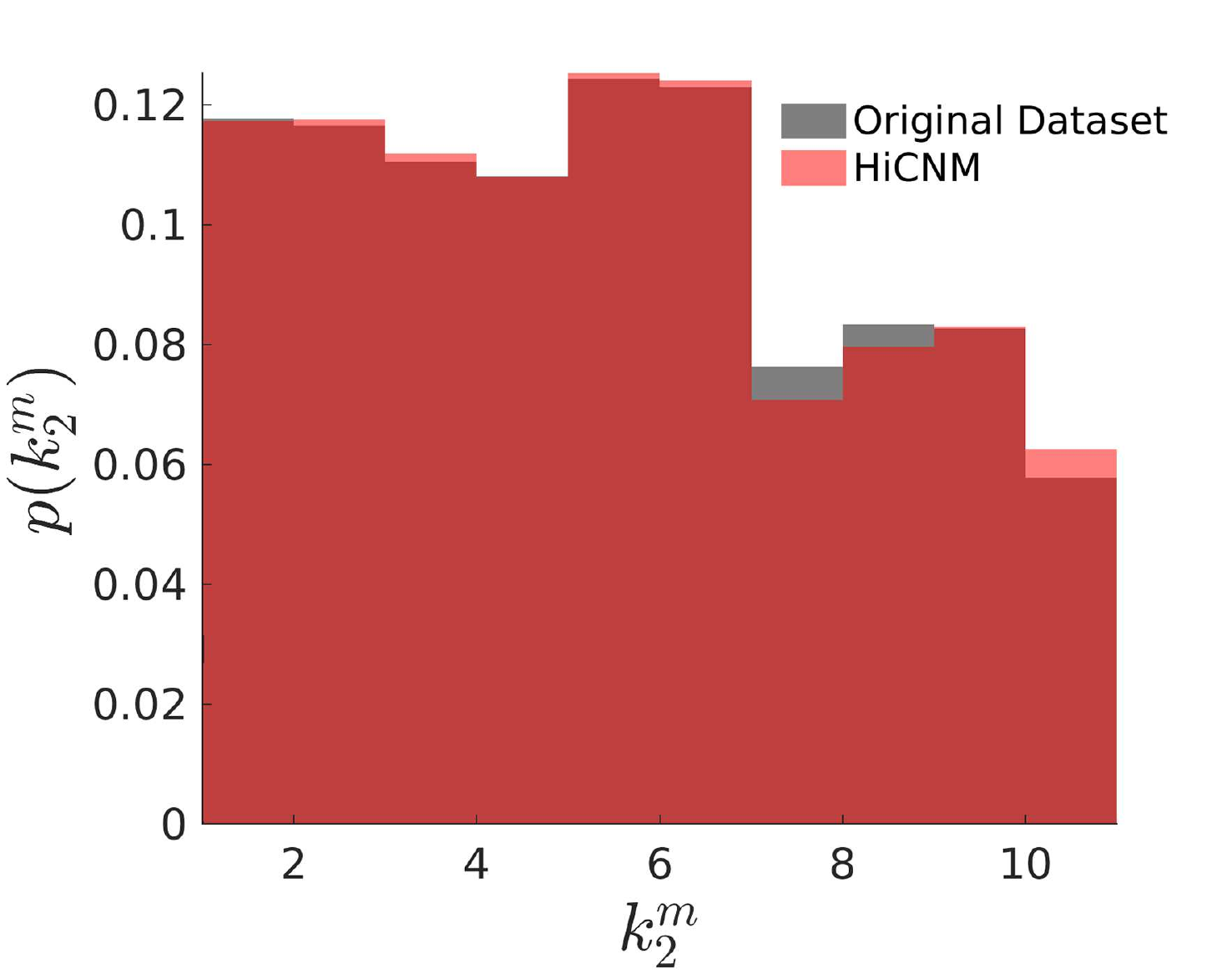} \\
(a) & (b) 
\end{tabular}
}
\caption{Time evolution of the cluster index: the reconstructed dynamics (red curve) with the same initial cluster of the original dynamics (gray curve with colored markers) for $m_2=1, \, \ldots , \, 8720$ (a); and probability distribution of all $m_2$ in figure~\ref{Fig:Matrix130_layer2_CHAOS} by a stochastic model using the transition matrix (b), at $\Rey=130$.}
\label{Fig:Stochastic130}
\end{figure}

The asymptotic limit cycle at $\Rey=80$ and torus at $\Rey=105$ are also fully predictable as there is no random walk. Only for the chaotic dynamics in the asymptotic regime at $\Rey=130$, can the transition matrix be used to construct a stochastic model. 
According to the current state in one of the clusters, it will choose the next destination according to the probability of transitions.
Based on the local dynamics for the chaotic state in the group of clusters $\mathcal{C}_{k_1}$, $k_1 = 4, \ldots, 7$, the reconstructed dynamics and the probability distribution provided by the stochastic model are shown in figure~\ref{Fig:Stochastic130}.
The main cycle $\mathcal{C}_1 \rightarrow \cdots \rightarrow \mathcal{C}_{6} \rightarrow \mathcal{C}_1$ and the random walks to $\mathcal{C}_{k_2}$, $k_2 = 7, \ldots, 10$ have been fairly reproduced, with the appropriate probability distribution.

\bibliographystyle{jfm}
\bibliography{Main}

\end{document}